\documentclass[12pt,a4paper]{article}

\usepackage{amsmath}
\usepackage{amssymb}
\usepackage[nosort]{cite}
\usepackage[hyperref,bulletsep]{collect}
\def\gfxon{\usepackage[final]{graphicx}}

\gfxon

\def\({\left(}
\def\){\right)}

\makeatletter
\let\old@makecaption=\@makecaption
\def\@makecaption{\small\old@makecaption}
\makeatother

\makeatletter \@addtoreset{equation}{section} \makeatother

\makeatletter
\let\old@startsection=\@startsection
\renewcommand{\@startsection}[6]{\old@startsection{#1}{#2}{#3}{#4}{#5}{#6\mathversion{bold}}}
\makeatother

\let\oldPhi=\Phi
\let\oldOmega=\Omega
\let\oldLambda=\Lambda
\let\oldPsi=\Psi
\let\oldGamma=\Gamma
\let\oldDelta=\Delta
\let\oldSigma=\Sigma
\let\oldTheta=\Theta
\let\oldPi=\Pi
\renewcommand{\Phi}{\mathnormal{\oldPhi}}
\renewcommand{\Omega}{\mathnormal{\oldOmega}}
\renewcommand{\Psi}{\mathnormal{\oldPsi}}
\renewcommand{\Gamma}{\mathnormal{\oldGamma}}
\renewcommand{\Sigma}{\mathnormal{\oldSigma}}
\renewcommand{\Delta}{\mathnormal{\oldDelta}}
\renewcommand{\Theta}{\mathnormal{\oldTheta}}
\renewcommand{\Pi}{\mathnormal{\oldPi}}
\renewcommand{\Lambda}{\mathnormal{\oldLambda}}

\newcommand{\indup}[1]{_{\mathrm{#1}}}

\newcommand{\matr}[2]{\left(\begin{array}{#1}#2\end{array}\right)}
\newcommand{\smatr}[4]{\matr{c|c}{#1&#2\\\hline #3&#4}}
\newcommand{\alg}[1]{\mathfrak{#1}}
\newcommand{\grp}[1]{\mathrm{#1}}

\newcommand{\sfrac}[2]{{\textstyle\frac{#1}{#2}}}
\newcommand{\half}{\sfrac{1}{2}}
\newcommand{\quarter}{\sfrac{1}{4}}

\newcommand{\pint}{\makebox[0pt][l]{\hspace{3.4pt}$-$}\int}
\newcommand{\pexp}{\mathrm{P}\exp}

\newcommand{\superN}{\mathcal{N}}

\newcommand{\order}[1]{\mathcal{O}(#1)}

\newcommand{\hateq}{\mathrel{\widehat{=}}}

\newcommand{\trans}{{\scriptscriptstyle\mathsf{T}}}
\newcommand{\strans}{{\scriptscriptstyle\mathsf{ST}}}
\newcommand{\sgrad}{\eta}

\newcommand{\Real}{\mathbb{R}}
\newcommand{\Comp}{\mathbb{C}}

\newcommand{\Integers}{\mathbb{Z}}

\newcommand{\lrbrk}[1]{\left(#1\right)}
\newcommand{\bigbrk}[1]{\bigl(#1\bigr)}
\newcommand{\Bigbrk}[1]{\Bigl(#1\Bigr)}

\newcommand{\bigcomm}[2]{\big[#1,#2\big]}
\newcommand{\comm}[2]{[#1,#2]}

\newcommand{\set}[1]{\{#1\}}

\newcommand{\sset}[2]{\{#1||#2\}}
\newcommand{\bigsset}[2]{\big\{#1\big|\big|#2\big\}}

\newcommand{\mono}{\omega}
\newcommand{\Mono}{\Omega}
\newcommand{\sheetsign}{\varepsilon}
\newcommand{\contour}[1]{\mathcal{#1}}
\newcommand{\resolvsl}[1][]{\makebox[0pt][l]{\hspace{0.06em}$/$}#1G}
\newcommand{\sheetsl}[1][]{\makebox[0pt][l]{\hspace{0.06em}$/$}#1p}
\newcommand{\resolvHsl}[1][]{\makebox[0pt][l]{\hspace{0.15em}$/$}#1H}

\newcommand{\nn}{\nonumber} 
\newcommand{\nln}{\nonumber\\}
\newcommand{\nl}{\nonumber\\&\hspace{-4\arraycolsep}&\mathord{}}

\newcommand{\earel}[1]{\mathrel{}&\hspace{-2\arraycolsep}#1\hspace{-2\arraycolsep}&\mathrel{}}
\newcommand{\eq}{\earel{=}}

\makeatletter
\newcommand{\newop}[2]{\def#1{\mathop{\operator@font #2}\nolimits}}
\newcommand{\newbin}[2]{\def#1{\mathbin{\operator@font #2}}}
\makeatother

\newop{\Re}{Re}
\newop{\Im}{Im}
\newop{\diag}{diag}
\newop{\rank}{rank}
\newop{\Tr}{Tr}
\newop{\tr}{tr}
\newop{\str}{str}
\newop{\sdet}{sdet}
\newop{\sign}{sign}

\def\[{\begin{equation}}
\def\]{\end{equation}}
\def\<{\begin{eqnarray}}
\def\>{\end{eqnarray}}

\makeatletter
\def\mr@ignsp#1 {\ifx\:#1\@empty\else #1\expandafter\mr@ignsp\fi}%
\newcommand{\multiref}[1]{\begingroup
\xdef\mr@no@sparg{\expandafter\mr@ignsp#1 \: }%
\def\mr@comma{}%
\@for\mr@refs:=\mr@no@sparg\do{\mr@comma\def\mr@comma{,}\ref{\mr@refs}}%
\endgroup}
\makeatother

\newcommand{\hypref}[2]{\ifx\href\asklfhas #2\else\href{#1}{#2}\fi}

\newcommand{\secref}[1]{Sec.~\multiref{#1}}

\newcommand{\appref}[1]{App.~\multiref{#1}}

\newcommand{\figref}[1]{Fig.~\multiref{#1}}
\renewcommand{\eqref}[1]{(\multiref{#1})}

\newenvironment{bulletlist}{\begin{list}{$\bullet$}{\leftmargin1.5em\itemsep0pt}}{\end{list}}

\ifx\href\asklfhas\newcommand{\href}[2]{#2}\fi
\newcommand{\arxivno}[1]{\href{http://arxiv.org/abs/#1}{#1}}

\begin{document}

\thispagestyle{empty}
\begin{flushright}\footnotesize
\texttt{\arxivno{hep-th/0502226}}\\
\texttt{ITEP-TH-13/05}\\
\texttt{LPTENS-05/10}\\
\texttt{NSF-KITP-05-12}\\
\texttt{PUTP-2152}\\
\texttt{UUITP-03/05}
\end{flushright}
\vspace{0.3cm}

\renewcommand{\thefootnote}{\fnsymbol{footnote}}
\setcounter{footnote}{0}

\begin{center}
{\Large\textbf{\mathversion{bold}
The Algebraic Curve of\\
Classical Superstrings
on $AdS_5\times S^5$
}\par} \vspace{0.8cm}

\textsc{N.~Beisert$^{a}$, V.A.~Kazakov$^{b,}$\footnote{Membre de
l'Institut Universitaire de France}, K.~Sakai$^b$ and
K.~Zarembo$^{c,}$\footnote{Also at ITEP, Moscow, Russia}} \vspace{5mm}

\textit{$^{a}$ Joseph Henry Laboratories, Princeton University,\\
Princeton, NJ 08544, USA} \vspace{3mm}

\textit{$^{b}$
Laboratoire de Physique Th\'eorique\\
de l'Ecole Normale Sup\'erieure et l'Universit\'e Paris-VI,\\
Paris, 75231, France}\vspace{3mm}

\textit{$^{c}$
Department of Theoretical Physics,\\
Uppsala University, 751 08 Uppsala, Sweden}
\vspace{5mm}

\texttt{nbeisert@princeton.edu}\\
\texttt{kazakov,sakai@lpt.ens.fr}\\
\texttt{konstantin.zarembo@teorfys.uu.se}\par\vspace{0.8cm}

\vfill

\textbf{Abstract}\vspace{5mm}

\begin{minipage}{12.7cm}
We investigate the monodromy of the Lax connection for
classical IIB superstrings on $AdS_5\times S^5$.
For any solution of the equations of motion
we derive a spectral curve of degree $4+4$.
The curve consists purely of conserved quantities,
all gauge degrees of freedom have been eliminated in this form.
The most relevant quantities of the solution,
such as its energy, can be expressed through
certain holomorphic integrals on the curve.
This allows for a classification of finite gap solutions
analogous to the general solution of
strings in flat space.
The role of fermions in the
context of the algebraic curve is clarified.
Finally, we derive a set of integral equations
which reformulates the algebraic curve as
a Riemann-Hilbert problem.
They agree with the planar, one-loop
$\mathcal{N}=4$ supersymmetric gauge theory proving the
complete agreement of spectra in this approximation.
\end{minipage}

\vspace*{\fill}

\end{center}

\newpage
\setcounter{page}{1}
\renewcommand{\thefootnote}{\arabic{footnote}}
\setcounter{footnote}{0}


\section{Introduction and Overview}
\label{sec:Intro}

Strings in flat space have been solved a long time ago.
The solution of the classical equations of motion
is straight-forward and obtained by a Fourier transformation,
or mode decomposition, of the world sheet.
The string is then represented by a collection of independent harmonic
oscillators, one for each mode and orientation in target space.
The oscillators are merely coupled by the Virasoro and level-matching
constraints.
The conserved, physical quantities of the string
are the absolute values of oscillator amplitudes.
Quantization of this system essentially poses no problem.
The harmonic oscillators are excited in quanta and
the amplitudes turn into integer-valued excitation numbers.

Maldacena's conjecture \cite{Maldacena:1998re,Gubser:1998bc,Witten:1998qj}
however brought about special attention on strings in curved target spaces
with `RR-flux', in particular IIB superstrings on $AdS_5\times S^5$.
There, a solution and quantization is much more involved
due to the highly non-linear nature of the string action \cite{Metsaev:1998it}.
A direct quantization of the world sheet theory is furthermore obstructed
by conformal and kappa symmetry which require gauge fixing.
This introduces a number of additional terms and usually
makes the problem intractable.

One path to quantization is related to
the maximally supersymmetric plane-wave background \cite{Blau:2001ne,Blau:2002dy}
and the correspondence to gauge theory \cite{Berenstein:2002jq}.
In this background the solution and quantization
closely resembles its flat space counterpart \cite{Metsaev:2001bj,Metsaev:2002re}.
The full $AdS_5\times S^5$ background may be regarded
as a deformation of plane waves.
Following this idea, one can obtain a
quantum string on $AdS_5\times S^5$
in a perturbation series around plane waves \cite{Callan:2003xr}.
This approach has yielded several important insights into the
quantum nature of the string, but there are drawbacks:
The perturbative expansion is very involved,
the first order is feasible \cite{Parnachev:2002kk,Callan:2004uv,Callan:2004ev,McLoughlin:2004dh},
but beyond there are no definite answers available yet.
Even if this problem might be overcome,
still we would be limited to a certain region of the parameter
space of full $AdS_5\times S^5$ which
is insensitive to global aspects.

Another approach to strings in curved space
is to consider classical solutions,
see e.g.~\cite{Gubser:2002tv,Frolov:2002av,Russo:2002sr,Minahan:2002rc,Tseytlin:2002ny}.
For these solutions with large spins one can show that quantum effects are
suppressed and already the classical solution
yields a good approximation for the full energy.
Even more excitingly, Frolov and Tseytlin discovered that
many of these spinning string solutions
have an expansion which is in qualitative agreement with
the loop expansion of gauge theory \cite{Frolov:2003qc}.
Their conjecture of a quantitative agreement has been confirmed in
several cases in \cite{Beisert:2003xu,Beisert:2003ea}
and many more works since,%
\footnote{Here, as well as in the case
of near plane-wave strings
there are discrepancies starting
at three gauge theory loops \cite{Callan:2003xr,Serban:2004jf}.
This puzzle can also be reformulated as the
question why it works at one and two loops in the first place.
We have little to add on this issue.}
see \cite{Tseytlin:2003ii,Tseytlin:2004cj,Beisert:2004ry,Beisert:2004yq,Zarembo:2004hp}
for reviews of the subject.
Finding exact solutions is not trivial,
the complexity of the functions
increases with the complexity of the solution.
The functions that occur are of algebraic, elliptic or
hyperelliptic type and
many of those which can be expressed using conventional
functions have been found.
While in principle each and every solution
can be found using suitable (unconventional) functions, 
it is impossible to catalog infinitely many of them
in order to understand their generic structure.

Finding the energy spectrum of superstrings on $AdS_5\times S^5$
therefore appears a too difficult problem to be solved explicitly.
Instead one can ask a more moderate question:
How is the spectrum of string solutions organized?
In other words, can we classify string solutions
even though we cannot write them explicitly?
Understanding the classification at the classical level
might be an essential step towards understanding the quantum string.
The classification was started in \cite{Kazakov:2004qf}
for bosonic strings on $\Real\times S^3$
which is a subspace of the full $AdS_5\times S^5$ background.
It was shown that for each solution of the equations of motion
there exists a corresponding hyperelliptic curve.
The key physical data of the solution,
such as the energy and Noether charges,
were identified in the algebraic curve.%
\footnote{This explains, among other things, why the classical energy,
one of these charges, is typically expressed through hyperelliptic functions.
The various integration constants of the classical solution
turn into moduli of the algebraic curve
which appear as parameters to the hyperelliptic functions.
See also \cite{Arutyunov:2003uj,Arutyunov:2003za} for 
a discussion of the moduli of some particular curves.}

At this point one can turn the logic around and
investigate the moduli space of admissible curves,
i.e.~those curves which correspond to some classical solution.
This leads to a solution of the spectral problem
in terms of algebraic curves, which is probably
as close to an explicit solution as it can be.
However, one would have to ensure that
all relevant constraints on the structure of admissible
curves have been correctly identified.
A survey of the moduli space of admissible curves
suggests that this is indeed the case:
There turns out to be one continuous modulus per genus and each
handle of the curve can be interpreted as a particular
string mode. This count matches with strings in flat space,
which has one amplitude per string oscillator.
Although two distinct theories are compared here, one can
expect that the number of local degrees of freedom
of the string should be independent of the background.
We furthermore believe that the (conserved) moduli
of a curve represent a complete set of action variables
for the string.
The moduli space of admissible curves would thus represent
half the phase space of the string model.

Another interesting option is to reformulate the problem of finding
admissible curves as a Riemann-Hilbert problem. This is achieved by
representing the curve as a collection of Riemann sheets connected
by branch cuts. 
The branch cuts are represented by
integrals over contours and densities in the complex plane. The
admissibility conditions turn into integral equations on these
contours and densities. This representation reveals an underlying
scattering problem and the branch cuts represent the fundamental
particles. The integral equations select equilibrium states of the
scattering problem. This can be compared to a direct Fourier
transformation of the string: The Fourier transformation transforms
the equations of motion into equations among the different Fourier
modes. Conceptually, the resulting equations are very similar to the
integral equations. The main difference between the two approaches
is that there are interactions between arbitrarily many Fourier
modes due to the highly non-linear nature of the strings, while the
interactions for the integral equations are only \emph{pairwise}! In
some sense, the algebraic curve can thus be interpreted as a clever
mode decomposition specifically tailored for the particular curved
background.

The pairwise, i.e.~factorized, nature of
the scattering problem leads us to integrability. Indeed, the
algebraic curve was constructed using the Lax connection, a family
of flat connections on the two-dimensional string world sheet. For
sigma models on group manifolds and symmetric coset spaces, such as
$\grp{SU}(2)=S^3$, this connection is well-known
\cite{Zakharov:1973pp}
and related to integrability as well as an infinite set of conserved charges
\cite{Pohlmeyer:1975nb,Luscher:1977rq,Brezin:1979am,Eichenherr:1981sk}
of the two-dimensional theory.
Integrable structures were also found in the AdS/CFT dual
$\superN=4$ gauge theory: The dual of the world-sheet Hamiltonian,
the planar dilatation operator (see~\cite{Beisert:2004ry} for a
review), was shown to be integrable at leading loop order
\cite{Minahan:2002ve,Beisert:2003yb}. Moreover, there are
indications that integrability is not broken by higher-loop effects
\cite{Beisert:2003tq,Beisert:2003ys}. In gauge theory, integrability
enables one to construct a Bethe ansatz \cite{Staudacher:2004tk} to
diagonalize local operators, which are isomorphic to quantum spin
chains. This leads to a set of algebraic equations
\cite{Minahan:2002ve,Beisert:2003yb,Serban:2004jf,Beisert:2004hm}
whose solutions are in one-to-one correspondence to eigenstates of
the dilatation operator. In the limit of states with a large number
of partons, which is at the heart of the spinning-strings
correspondence, the discrete Bethe equations turn into integral
equations \cite{Sutherland:1995aa,Beisert:2003xu}. These are very
similar to the integral equations from the string sigma model. In
fact, it was shown that the higher-loop Bethe equations in the
$\alg{su}(2)$ sector \cite{Serban:2004jf,Beisert:2004hm} match with
the equations from classical string theory on $\Real\times S^3$ up
to two gauge-theory loops \cite{Kazakov:2004qf}. This proves the
equality of energy spectra in this limit and sector. Alternatively,
one can also derive an algebraic curve for gauge theory and compare
it to the one for the sigma model \cite{Kazakov:2004qf}. An
altogether different approach to showing the agreement of spectra
uses coherent states \cite{Kruczenski:2003gt,Kruczenski:2004kw}.

\medskip

The solution of the spectral problem in terms
of algebraic curves has since been extended to
three other subsectors of the full superstring:
Bosonic strings on $AdS_3\times S^1$ \cite{Kazakov:2004nh},
on $\Real\times S^5$ \cite{Beisert:2004ag}
and on $AdS_5\times S^1$ \cite{Schafer-Nameki:2004ik}.
Also some features of the assembly of full $AdS_5$ and full $S^5$
are known \cite{Arutyunov:2004yx}.
In all previous analyses, however, \emph{fermions} have
been excluded.%
\footnote{Within the Frolov-Tseytlin correspondence
fermions have been treated in 
\cite{Mikhailov:2004xw,Hernandez:2004kr}
using the coherent state approach.}
While this is justified (for almost all practical purposes)
at the classical level, they are certainly required
to give a consistent quantum theory.
It is therefore essential to include them, even at the classical level.
This can indeed be done, even though it is a classical setting.%
\footnote{Having fermions in classical \emph{equations}
is not a problem, but we run into difficulties
when we try to find explicit \emph{solutions}, which would
require the introduction of actual Grassmann numbers.}

In the present article we shall derive the solution
of the spectrum of IIB superstrings on $AdS_5\times S^5$
in terms of algebraic curves.
The starting point will be the family of flat connections
found by Bena, Polchinski and Roiban \cite{Bena:2003wd}.
Using its open Wilson loop around the closed-string world sheet,
the so-called monodromy, we can derive an algebraic curve.
As was demonstrated in \cite{Bena:2003wd}
the Lax connection exists prior to gauge fixing.
We therefore do not fix any gauge,
neither of conformal nor of kappa symmetry,
in contrast to \cite{Kazakov:2004qf,Kazakov:2004nh,Beisert:2004ag}
and especially \cite{Arutyunov:2004yx}.
The emergent curve is neither a regular algebraic curve, nor an
algebraic supercurve, i.e.~not a supermanifold.
It almost splits in two parts,
but it is held together by the fermions.
Each part has degree four and corresponds to one of the
$S^5$ and $AdS_5$ coset models.
The bosonic degrees of freedom give rise
to square-root branch points and cuts connecting them.
These appear only within each set of four Riemann sheets.
We shall show that, conversely, the fermions give rise to poles.
Poles come in pairs,
one of them is on the $S^5$-part of the curve,
the other on the $AdS_5$-part while their residues are the same.%
\footnote{The residues are products of two Grassmann-odd numbers.
Therefore, they are Grassmann-even,
but not of zeroth degree, i.e.~they cannot be
represented by common numbers.
This is why fermions can be neglected for almost
all practical purposes.
The derivations are however simplified by ignoring this fact.}
Their position within the algebraic curve
is determined by the bosonic background.
The two parts of the curve are furthermore linked by the
Virasoro constraint:
It relates a set of fixed poles
between the two parts of the curve.
These poles are an important general characteristics of the model
and do not correspond to fermions.

The precise structure of the algebraic curve
and its representation in the form of integral equations
constitute the key information from string theory
for a comparison with gauge theory
\cite{Beisert:2003jj,Beisert:2003yb,Beisert:2004ry,Schafer-Nameki:2004ik}
via the Frolov-Tseytlin proposal.
Using an integral representation for the curve, we
are able to show agreement of the spectra at leading order
in the effective coupling constant.
A more detailed comparison will be performed 
in the follow-up article \cite{Beisert:2005di}.

\bigskip

The structure of this article is as follows:
In \secref{sec:Super} we will investigate the monodromy of the
Lax connection and derive an algebraic curve from it.
The remainder of the section is devoted to finding
the analytic properties of the curve and relating them to
data of the associated string solution.
Then we decouple from the underlying string solution in
\secref{sec:Moduli} and consider the set of admissible curves.
After counting the number of moduli, we shall identify them with
certain integrals on the curve. Their relationship to the global charges
is established.
In the final \secref{sec:Integral} we shall represent the algebraic curve
by means of its branch cuts between the Riemann sheets.
The resulting equations are closely related to the equations
one obtains from spin chains in the thermodynamic limit.
We show that they agree with one-loop gauge theory.
We conclude and give an outlook in \secref{sec:Concl}.
The appendices contain a review of supermatrices
(\appref{sec:SuperAlgebra}), the relation between
coset and vector models
(\appref{sec:Vector}) and explicit but lengthy expressions
related to the full supersymmetric sigma model (\appref{sec:Beauty}).

\section{Supersymmetric Sigma Model}
\label{sec:Super}

We start by investigating the $AdS_5\times S^5$
supersymmetric sigma model on a closed string worldsheet.
First of all we present the sigma model in terms
of its fields, currents and constraints.
Then we review the Lax connection and
its monodromy and show that
the essential physical information (action variables)
is described by an algebraic curve.
The remainder of this section is devoted to
special properties of the curve and
relating them to physical quantities.

The $AdS_5\times S^5$ superspace can be represented as the coset
space of the supergroup $\grp{PSU}(2,2|4)$ over
$\grp{Sp}(1,1)\times\grp{Sp}(2)$. Up to global issues, but
preserving the algebraic structure, we can change the signature of
the target spacetime. Here we will consider the coset
$\grp{PSL}(4|4,\Real)/\grp{Sp}(4,\Real)\times\grp{Sp}(4,\Real)$.
This choice is convenient as we can completely avoid complex conjugation
which may be somewhat confusing, especially in a supersymmetric
setting. See e.g.~\cite{Hatsuda:2004it,Das:2004hy}
for an explicit treatment of the $\grp{PSU}(2,2|4)$ coset model.
The global issues that we should keep in mind are whether
the string can wind around the manifold. For $S^5$ this is certainly the
case, while for $AdS_5$ there should be no windings. Note that
the physical $AdS_5$ is a universal cover and there cannot be windings
around the unfolded time circle. Likewise,
the involved group manifolds are considered to be universal coverings.

\subsection{The Coset Model}
\label{sec:Super.Frame}

The Metsaev-Tseytlin string is a coset space sigma model.
To represent the coset,
we consider a group element $g$ of $\grp{PSL}(4|4,\Real)$ and
two constant $(4|4)\times (4|4)$ matrices%
\footnote{For a short review of
the algebra of supermatrices, cf.~\appref{sec:SuperAlgebra}.}
\[\label{eq:Super.E1E2}
{E}_1=\smatr{E}{0}{0}{0},\qquad {E}_2=\smatr{0}{0}{0}{E},
\]
which break $\grp{PSL}(4|4,\Real)$ to $\grp{Sp}(4,\Real)\times\grp{Sp}(4,\Real)$.
Here, $E$ is an antisymmetric $4\times 4$ matrix%
\footnote{In fact, any $E=-E^\trans$ with
$\varepsilon_{\alpha\beta\gamma\delta}E^{\alpha\beta}E^{\gamma\delta}\neq 0$
would suffice and one could as well pick distinct matrices $E$ for
$E_1$ and $E_2$.}
\[\label{eq:Super.Edef}
E=\matr{rr}{0&+I\\-I&0},
\]
where each entry corresponds to a $2\times 2$ block and $I$ is
the identity matrix. We shall denote
the pseudo-inverses of $E_1,E_2$ by
\[\label{eq:Super.E1E2bar}
{\bar E}_1=\smatr{E^{-1}}{0}{0}{0},\qquad
{\bar E}_2=\smatr{0}{0}{0}{E^{-1}}.
\]
These are defined such that
a product of $E_a$ and $\bar E_b$ is
a projector to the even/odd subspace
if $a=b$ or zero if $a\neq b$.
Finally, let us introduce a grading matrix
\[\label{eq:Super.Grading}
\sgrad=\smatr{+I}{0}{0}{-I}
\]
which will be useful at various places.

The breaking of $\grp{PSL}(4|4,\Real)$ to
$\grp{Sp}(4,\Real)\times\grp{Sp}(4,\Real)$
is achieved as follows:
The matrix $E_1$ is invariant under
$E_1\mapsto h E_1 h^\strans$
for elements $h$ of a subgroup
$\grp{Sp}(4,\Real)\times\grp{SL}(4,\Real)$
of $\grp{PSL}(4|4,\Real)$.
Similarly, $E_2\mapsto hE_2 h^\strans$ is invariant
under a $\grp{SL}(4,\Real)\times \grp{Sp}(4,\Real)$.
The combined map $(E_1,E_2)\mapsto (hE_1h^\strans,hE_2h^\strans)$
leaves $(E_1,E_2)$ invariant precisely for
$h\in \grp{Sp}(4,\Real)\times \grp{Sp}(4,\Real)$.
Thus the element $(gE_1g^\strans,gE_2g^\strans)$
with $g\in \grp{PSL}(4|4,\Real)$
parametrizes the $AdS_5\times S^5$ superspace.%
\footnote{Note that $E_1$ is an antisymmetric supermatrix,
$E_1^\strans=-\sgrad E_1$, while $E_2$ is symmetric,
$E_2^\strans=+\sgrad E_2$. Therefore also $g(E_1\pm E_2)g^\strans$ or
$g(E_1\pm iE_2)g^\strans$ parametrize the coset as we can
disentangle the contributions from $E_1$ and $E_2$ by projecting to
the symmetric and antisymmetric parts.}

We now introduce the supermatrix-valued field
$g(\tau,\sigma)\in \grp{PSL}(4|4,\Real)$ on the worldsheet.
It satisfies $\sdet g=1$
and we identify group elements which are related by an abelian
rescaling, $g\hateq \xi g$.%
\footnote{For $(4|4)\times (4|4)$ supermatrices, $\sdet \xi I=1$ for
any number $\xi$.}
The field $g$ is not necessarily strictly periodic
but
\[\label{eq:Super.Periodicity}
g(\tau,\sigma+2\pi)=g(\tau,\sigma)\,h(\tau,\sigma).
\]
with $h(\tau,\sigma)$ an element of $\grp{Sp}(4,\Real)\times \grp{Sp}(4,\Real)$.
We define the standard $g$-connection $J$ as
\[\label{eq:Super.J}
{J}= -{g}^{-1}d{g}.
\]
It is flat and supertraceless
\[\label{eq:Super.dJ}
d{J}= {J}\wedge {J}\quad\mbox{and}\quad \str J=0
\]
by means of the usual identities and $\sdet g=1$. The algebra
$\alg{psl}(4|4,\Real)$ can be decomposed into four parts obeying a
$\Integers_4$-grading
\cite{Metsaev:1998it,Kallosh:1998zx,Berkovits:1999zq,Roiban:2000yy}.%
\footnote{The $\Integers_4$-grading is directly related
to supertransposing, c.f.~\appref{sec:SuperAlgebra}.}
The connection
decomposes as follows
\[\label{eq:Super.SplitJ}
{J}={H}+{Q}_1+{P}+{Q}_2.
\]
We use the constant supermatrices $E_{1,2},\bar E_{1,2}$ to project
to the various components
\<\label{eq:Super.HQPQ}
{H}\eq
\half {E}_1 {\bar E}_1 \,{J}\,{E}_1 {\bar E}_1
-\half {E}_1 \,{J}^\strans\,{\bar E}_1
+\half {E}_2 {\bar E}_2 \,{J}\,{E}_2 {\bar E}_2
-\half {E}_2 \,{J}^\strans\,{\bar E}_2,
\nln
{Q}_1\eq
\half {E}_1 {\bar E}_1 \,{J}\,{E}_2 {\bar E}_2
+\half {E}_1 \,{J}^\strans\,{\bar E}_2
+\half {E}_2 {\bar E}_2 \,{J}\,{E}_1 {\bar E}_1
-\half {E}_2 \,{J}^\strans\,{\bar E}_1,
\nln
{P}\eq
\half {E}_1 {\bar E}_1 \,{J}\,{E}_1 {\bar E}_1
+\half {E}_1 \,{J}^\strans\,{\bar E}_1
+\half {E}_2 {\bar E}_2 \,{J}\,{E}_2 {\bar E}_2
+\half {E}_2 \,{J}^\strans\,{\bar E}_2,
\nln
{Q}_2\eq
\half {E}_1 {\bar E}_1 \,{J}\,{E}_2 {\bar E}_2
-\half {E}_1 \,{J}^\strans\,{\bar E}_2
+\half {E}_2 {\bar E}_2 \,{J}\,{E}_1 {\bar E}_1
+\half {E}_2 \,{J}^\strans\,{\bar E}_1.
\>
They satisfy the $\Integers_4$-graded Bianchi identities in
\cite{Bena:2003wd}
\<\label{eq:Super.dHQPQ}
d{H}\eq
{H}\wedge {H}
+{Q}_1\wedge {Q}_2
+{P}\wedge {P}
+{Q}_2\wedge {Q}_1,
\nln
d{Q}_1\eq
{H}\wedge {Q}_1
+{Q}_1\wedge {H}
+{P}\wedge {Q}_2
+{Q}_2\wedge {P},
\nln
d{P}\eq
{H}\wedge {P}
+{Q}_1\wedge {Q}_1
+{P}\wedge {H}
+{Q}_2\wedge {Q}_2,
\nln
d{Q}_2\eq
{H}\wedge {Q}_2
+{Q}_1\wedge {P}
+{P}\wedge {Q}_1
+{Q}_2\wedge {H},
 \>
and their supertraces vanish
\[\label{eq:Super.STr}
\str{H}=\str{Q}_1=\str {P}=\str {Q}_2= 0.
\]
Note that $\str{H}=\str{Q}_1=\str {Q}_2=0$ is
satisfied by means of the projections
\eqref{eq:Super.HQPQ} while
$\str{P}=\str{J}=0$ holds due to \eqref{eq:Super.dJ}.

The action of the IIB superstring on $AdS_5\times S^5$ given in
\cite{Metsaev:1998it} in terms of the connections $P,Q_{1,2}$
reads \cite{Roiban:2000yy}
\[\label{eq:Super.Action}
S_{\sigma}=
\frac{\sqrt{\lambda}}{2\pi}\int
\lrbrk{\half \str {P}\wedge {\ast {P}}
-\half \str Q_1\wedge Q_2
+\Lambda\wedge  \str P}.
\]
%
We have introduced the Lagrange multiplier $\Lambda$ to enforce
$\str P=0$. In fact, we cannot remove  the part proportional to the
identity matrix because of the identity $\str I=0$. The equations of
motion read
\<\label{eq:Super.EOM}
0\eq
{P}\wedge {Q}_2
-{\ast{P}}\wedge {Q}_2
+{Q}_2\wedge {P}
-{Q}_2\wedge {\ast{P}},
\nln
d{\ast{P}}\eq
{H}\wedge {\ast{P}}
+{Q}_1\wedge {Q}_1
+{\ast{P}}\wedge {H}
-{Q}_2\wedge {Q}_2
+d\Lambda,
\nln
0\eq
{P}\wedge {Q}_1
+{\ast{P}}\wedge {Q}_1
+{Q}_1\wedge {P}
+{Q}_1\wedge {\ast{P}}.
\>
The appearance of $\Lambda$ in the equations of motion is
related to the projective identification $g\hateq \xi g$.
The equations of motion can also be written as the $g$-covariant conservation of the
global $\alg{psl}(4|4,\Real)$ symmetry current $K$
\[\label{eq:Super.Current}
d{\ast K}
-{J}\wedge {\ast K}
-{\ast K}\wedge {J}=0,
\qquad
K=P+\half {\ast Q_1}-\half {\ast Q_2}-{\ast}\Lambda.
\]
The above equations of motion follow after decomposition into the
$\Integers_4$-graded components. The dependence of $K$ on the
unphysical Lagrange multiplier reflects the ambiguity in the
definition of the abelian part of $K$ in $\alg{psl}(4|4,\Real)$.
In the fixed frame,%
\footnote{We shall distinguish between a moving frame and a fixed frame.
In the moving frame $E$ is a constant matrix and
the fundamental field is $g$. The gauge connection is $D=d-J$.
In the fixed frame the matrix corresponding to $E$ is
$gEg^\strans$. It is not constant but rather the fundamental field.
The gauge connection is trivial, $D=d$.
See \appref{sec:Vector} for a comparison of both formalisms.
We use uppercase and lowercase letters
for the moving and fixed frames, respectively.}
which is related to the moving one by $k=gKg^{-1}$,
the equations for the current are even shorter
\[\label{eq:Super.current}
d{\ast k}=0.
\]
The global symmetry charges are consequently given by
\[\label{eq:Super.ChargesOld}
s=\frac{\sqrt{\lambda}}{2\pi}\oint_\gamma {\ast k}
=\frac{\sqrt{\lambda}}{2\pi}\int_0^{2\pi} d\sigma\, k_\tau.
\]
These do not depend on the form of the path $\gamma$ around the closed loop
and are thus conserved physical quantities.
For later convenience,
we rewrite $s=g(0)Sg^{-1}(0)$ in terms of the
moving-frame current $K$ as follows
\<\label{eq:Super.Charges}
 S \eq\frac{\sqrt{\lambda}}{2\pi}\int_0^{2\pi} d\sigma\,
g^{-1}(0)g(\sigma) K_\tau(\sigma) g^{-1}(\sigma)g(0)
\nln
\eq\frac{\sqrt{\lambda}}{2\pi}\int_0^{2\pi} d\sigma\,
\lrbrk{\pexp \int_{0}^{\sigma}d\sigma' J_\sigma(\sigma')}^{-1} K_\tau(\sigma)
\lrbrk{\pexp \int_{0}^{\sigma}d\sigma' J_\sigma(\sigma')}
.
\>
Here, as for the remainder of the article,
the path ordering symbol $\mathrm{P}$ puts the
values of $\sigma$ in decreasing order from left to right.

In addition to the equation of motion, the Virasoro constraints
following from variation of the world-sheet metric (which appears
only within the dualization $\ast$) are given by
\[\label{eq:Super.Virasoro}
\str P_\pm^2 =0.
\]
Here we have introduced the light-cone coordinates
\[\label{eq:Super.Lightcone}
\sigma_\pm=\half(\tau\pm\sigma), \qquad
\partial_\pm=\partial_\tau\pm\partial_\sigma,\qquad
P_\pm=P_\tau\pm P_\sigma.
\]
%

\subsection{Lax Connection and Monodromy}
\label{sec:Super.Lax}

A family of flat connections $a(\kappa)$ for the superstring on
$AdS_5\times S^5$ was derived in \cite{Bena:2003wd}.%
\footnote{For complex values of $\kappa$, there is only
one family of flat connections.
The other family mentioned in \cite{Bena:2003wd} is trivially
obtained by replacing $\kappa$ with $i\pi-\kappa$.}
This was expressed in the fixed frame,
which is related to moving one by $j=gJg^{-1}$ and similarly for $H,Q_1,P,Q_2$.
The Lax connection is given by
\<\label{eq:Super.LaxBPR}
a(\kappa)\eq \alpha(\kappa)\,
p+\beta(\kappa)\, ({\ast p}-\Lambda) +\gamma(\kappa)\, (q_1+q_2)
+\delta(\kappa)\, (q_1-q_2) \>
with
\[\label{eq:Super.LaxBPRCoeff}
\begin{array}[b]{rclcrcl}
\alpha(\kappa)\eq-2\sinh^2 \kappa, &&
\gamma(\kappa)\eq 1- \cosh \kappa,
\\[3pt]
\beta(\kappa)\eq  2\sinh \kappa\cosh \kappa, &&
\delta(\kappa)\eq\sinh \kappa.
\end{array}
\]
We will employ a more convenient parametrization by setting
$z=\exp\kappa$. The coefficient functions become
\[\label{eq:Super.LaxCoeff}
\begin{array}[b]{rclcrcl}
\alpha(z)\eq 1 - \half z^2 - \half z^{-2}, &&
\gamma(z)\eq 1 - \half z- \half z^{-1} ,
\\[3pt]
\beta(z)\eq \half z^2 -\half z^{-2} , &&
\delta(z)\eq \half z-\half z^{-1}.
\end{array}
\]
We would now like to transform the connection to the moving frame using
$J=g^{-1}jg$ and compute
\<\label{eq:Super.LaxTrans}
d-A(z) \eq
g^{-1}\bigbrk{d+a(z)}g =
d-J+g^{-1}a(z)g
\\\nn\eq
d-H+ (\alpha-1)\, P
+\beta\, ({\ast P}-\Lambda)
+(\gamma-1)\, (Q_1+Q_2)
+\delta\, (Q_1-Q_2),
\>
where the Lax connection reads
\[\label{eq:Super.Lax}
A(z)=
H+
\bigbrk{\half z^2+\half z^{-2}} P
+\bigbrk{-\half z^2+\half z^{-2}}\lrbrk{ {\ast P}-\Lambda}
+z^{-1} Q_1
+z\,Q_2.
\]
As was shown in \cite{Bena:2003wd}, it satisfies the flatness
condition
\[\label{eq:Super.LaxSqr}
\bigbrk{d-A(z)}^2=0
\]
by means of the equations of motion. It is also traceless
for obvious reasons, $\str A(z)=0$.

\begin{figure}\centering
\includegraphics{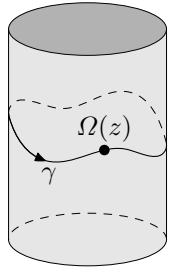}
\caption{The monodromy $\Mono(z)$ is the open Wilson loop
of the Lax connection $A(z)$ around the string.}
\label{fig:Monodromy}
\end{figure}
As emphasized in \cite{Kazakov:2004qf,Beisert:2004ag},
an important object for the solution of the spectral problem
is the open Wilson loop of the Lax connection around the closed string.
It is given by
\[\label{eq:Super.Mono0}
\Mono_0(z)=\pexp\int_0^{2\pi} d\sigma \,A_\sigma(z)
\simeq
\pexp\oint A(z).
\]
The \emph{monodromy} which is defined as%
\footnote{For $z=1$ the Lax connection $A(z)=J$ is the gauge connection.
The additional factor $\Mono_0^{-1}(1)=g(0)^{-1}g(2\pi)=h(0)$
therefore transforms the monodromy back to the tangent space at $\sigma=0$.}
\[\label{eq:Super.Mono}
\Mono(z)=\Mono_0^{-1}(1)\,\Mono_0(z)
\]
is independent of the path $\gamma$ around the closed string;
it merely depends on the point $\gamma(2\pi)=\gamma(0)$
where the path is cut open.
More explicitly, a shift of $\gamma(0)$ leads to
a similarity transformation ($\simeq$), see e.g.~\cite{Beisert:2004ag}.
Therefore, the eigenvalues of $\Mono(z)$ are invariant, physical quantities.
Note that we did not specify any particular gauge of conformal or
kappa symmetry.
Under kappa symmetry the Lax connection transforms by
conjugation \cite{Berkovits:2004jw} and consequently leaves
the eigenvalues invariant as well.
For definiteness we define $\Mono(z)$ through
the path $\sigma\in [0,2\pi]$ at $\tau=0$.
Also note that $\str A(z)=0$ leads to $\sdet \Mono(z)=1$.

In the Hamiltonian formulation, the eigenvalues of the monodromy
represent action variables of the sigma model.%
\footnote{See \cite{Das:2004hy} for an investigation of
the Poisson brackets of the monodromy.}
We have a one-parameter family of them and it is not inconceivable that they
form a complete set. So we might have a sufficient amount of
information to fully characterize the class of solution.
The time-dependent angle variables and all gauge degrees of freedom are
completely projected out in the eigenvalues of $\Mono(z)$. This is a
very good starting point for a quantum theory:
For quantum eigenstates we can measure all the action variables exactly
but information of the angle variables is obscured by
the uncertainty principle.

\subsection{The Algebraic Curve}
\label{sec:Super.Curve}

The physical information of the monodromy matrix is its conjugation
class. Let $u(z)$ diagonalize $\Mono(z)$ as follows
\[\label{eq:Super.MonoDiag}
u(z)\Mono(z)u^{-1}(z)=
\diag
\bigsset{e^{i\tilde p_1(z)},e^{i\tilde p_2(z)},e^{i\tilde p_3(z)},e^{i\tilde p_4(z)}}
        {e^{i\hat p_1(z)},e^{i\hat p_2(z)},e^{i\hat p_3(z)},e^{i\hat p_4(z)}}.
\]
Note that the eigenvalues $e^{i\tilde p_k}$ and $e^{i\hat p_l}$
corresponding to the two gradings are distinguishable,
they cannot be interchanged by a (bosonic) similarity transformation.
We can associate $\tilde p_k$ to $S^5$ while
$\hat p_k$ corresponds to $AdS^5$.
In contrast, we may freely interchange eigenvalues within each set of four.
Unimodularity, $\sdet\Mono(z)=1$, translates to the
condition
\[\label{eq:Super.Unimodular}
\tilde p_1+\tilde p_2+\tilde p_3+\tilde p_4
-\hat p_1-\hat p_2-\hat p_3-\hat p_4
\in 2\pi \Integers.
\]

The monodromy \eqref{eq:Super.Mono} depends analytically on the
spectral parameter $z$ by definition except at the singular points
$z=0$ and $z=\infty$. This however does
not imply that also the eigenvalues
$\sset{e^{i\tilde p_k}}{e^{i\hat p_k}}$ enjoy the same property.

Let us first consider a point $\tilde z_a$ where
two eigenvalues $e^{i\tilde p_k},e^{i\tilde p_l}$
corresponding to the $S^5$-part of the sigma model degenerate.
The restriction of $\Mono(z)$ to the subspace of the two corresponding
eigenvalues then takes the general form
\[\label{eq:Super.22Bosonic}
\Gamma=\matr{cc}{a&b\\c&d}
\]
with some coefficients $a,b,c,d$ depending analytically on $z$.
Its eigenvalues are given by the general formula
\[\label{eq:Super.22BosonicEV}
\gamma_{1,2}=\frac{1}{2}\lrbrk{a+d\pm \sqrt{(a-d)^2+4bc}}.
\]
At $z=\tilde z_a$ the combination
$f=(\gamma_1-\gamma_2)^2=(a-d)^2+4bc=(\Tr\Gamma)^2-2\Tr\Gamma^2$
under the square root vanishes,
$f(\tilde z_a)=0$.
In the generic case, one can expect $f'(\tilde z_a)\neq 0$.
This implies the well-known fact that crossing of eigenvalues
usually gives rise to a square-root singularity:
\[\label{eq:Super.BranchS5}
e^{i\tilde p_{k,l}(z)}= e^{i\tilde p_{k}(\tilde z_a)}
\lrbrk{1\pm\tilde\alpha_a\sqrt{z-\tilde z_a}+\order{z-\tilde z_a}}.
\]
Similarly, coincident $AdS_5$-eigenvalues
$e^{i\hat p_k}$ and $e^{i\hat p_l}$ at $\hat z_a$ lead
to square-root singularities
\[\label{eq:Super.BranchAdS5}
e^{i\hat p_{k,l}(z)}= e^{i\hat p_{k}(\hat z_a)}
\lrbrk{1\pm\hat\alpha_a\sqrt{z-\hat z_a}+\order{z-\hat z_a}}.
\]

The behavior around a point $z^{\ast}_a$ where eigenvalues of
opposite gradings, $e^{i\tilde p_k}$ and $e^{i\hat p_l}$, coincide is
quite different: Consider the submatrix of $\Mono(z)$ on the
subspace of the two associated eigenvectors
\[\label{eq:Super.22Super}
\Gamma=\smatr{a}{b}{c}{d}
\]
The eigenvalues of this supermatrix $\Gamma$ are given by
\[\label{eq:Super.22SuperEV}
\gamma_1=\frac{bc}{d-a}+a\,,\qquad \gamma_2=\frac{bc}{d-a}+d\,,
\]
where again $a,b,c,d$ are given by analytic functions in $z$.
At $z=z^{\ast}_a$, the combination $f=a-d=\gamma_1-\gamma_2=\str \Gamma$
in the denominator is zero by definition, $f(z^\ast_a)=0$.
Generically, we cannot however expect that also the numerator $bc$ vanishes
and therefore we find a pole singularity at $z^{\ast}_a$
\[\label{eq:Super.FermiSing}
e^{i\tilde p_{k}(z)}=
e^{i\makebox[0pt][l]{\hspace{0.06em}$\scriptstyle/$}\tilde p_k(z^{\ast}_a)}
\lrbrk{\frac{\alpha^\ast_a}{z-z^{\ast}_a}+1+\order{z-z^{\ast}_a}}
= e^{i\hat p_{l}(z)}.
\]
Note that the residue $\alpha^\ast_a$
as well as the regular part
$e^{i\makebox[0pt][l]{\hspace{0.06em}$\scriptstyle/$}p(z^{\ast}_a)}$
of $e^{ip(z)}$ at $z=z^\ast_a$ are the same for both eigenvalues.%
\footnote{It might be worthwhile to point out that
$\alpha^\ast_a=-bc$ is the product of two Grassmann-odd quantities
and thus, in principle, cannot be an ordinary number. It satisfies a
nilpotency condition $(\alpha^\ast_a)^2=0$ which, however, does not
quite make it trivial. When quantizing the string, these factors
give rise to fermionic excitations due to quantum 
$\hbar \sim 1/L$ effects. This effect can already be seen in 
the one-loop spin chain for the $\superN=4$ gauge theory 
\cite{Beisert:2003yb} where there are no nilpotent objects.} 
We thus learn that the set of eigenvalues of
$\Mono(z)$ depends analytically on $z$ except at a set of points
$\set{0,\infty,\tilde z_a,\hat z_a,z^{\ast}_a}$. Let us assume that
there are only finitely many singularities of this kind. The cases
of an infinite number of singularities can hopefully be viewed as
limits of this finite setting. A unique labelling of eigenvalues
cannot be achieved globally, because a full circle around one of the
square-root singularities $\tilde z_a,\hat z_a$ will result in an
interchange of the two eigenvalues associated to the singularity.
Therefore we need to introduce several branch cuts
$\tilde{\contour{C}}_a$ and $\hat{\contour{C}}_a$ in $e^{i\tilde
p_k(z)}$ and $e^{i\hat p_k(z)}$, respectively, which connect the
square-root singularities. The functions $\tilde p_k(z)$ and $\hat
p_k(z)$ are therefore analytic except at $\set{0,\infty,\tilde
{\contour{C}}_a,\hat {\contour{C}}_a,z^{\ast}_a}$. Alternatively, we
could view $e^{i\tilde p(z)}$ and $e^{i\hat p(z)}$ as one function
on suitable four-fold coverings $\tilde{\mathbb{M}}$ and
$\hat{\mathbb{M}}$ of $\bar\Comp$. In that case, the functions
$e^{i\tilde p(z)}$ and $e^{i\hat p(z)}$ are analytic except at
$\set{0,\infty,\tilde z_a,\hat z_a,z^{\ast}_a}$. At $z^{\ast}_a$
both functions $e^{i\tilde p(z)}$ and $e^{i\hat p(z)}$ have poles
with equal residues and regular parts. Finally, at $0$ and $\infty$
there are essential singularities of the type
$e^{i\alpha_0/z^2},e^{i\alpha_\infty z^2}$.

Except for the last two singularities,
the functions $e^{i\hat p(z)},e^{i\tilde p(z)}$ would satisfy all requirements
for algebraic curves.
In order to turn the essential singularities at $\set{0,\infty}$ into
regular singularities,
we take the logarithmic derivative of the eigenvalues.
Let us define the matrix $Y(z)$ according to
\[\label{eq:Super.EVLogDer}
u(z)Y(z)u^{-1}(z)= -iz\,\frac{\partial}{\partial z}\log
\bigbrk{u(z)\Mono(z)u^{-1}(z)},
\]
where $u(z)$ diagonalizes $\Mono(z)$.
In other words, the eigenvalues of $Y(z)$ are the logarithmic
derivatives of the eigenvalues of $\Mono(z)$. The corresponding
eigenvectors are the same.
We can now reduce $Y(z)$ to the following expression
\[\label{eq:Super.DiagMatr}
Y(z)=\Mono^{-1}(z)\bigbrk{-iz\Mono'(z)+\comm{U(z)}{\Mono(z)}},
\qquad U(z)=-izu^{-1}(z) u'(z).
\]
As $\Mono(z)$ is non-zero and its only singularities are at $\set{0,\infty}$,
any further singularities can only originate from $U(z)$.
The diagonalization matrix $u(z)$ has square roots and branch cuts.
It appears that all the branch points of $u(z)$
are turned into single poles in $U(z)$.%
\footnote{This may require a special matrix $u(z)$. The point is
that one can redefine $u(z)\mapsto a(z)u(z)$ with any diagonal
matrix $a(z)$. This is a possible source of non-analyticity in
$U(z)$, which however drops out in $\comm{U(z)}{\Mono(z)}$.}
Consequently, $U(z)$ has poles at
$\set{\tilde z_a,\hat z_a,z^{\ast}_a}$, but all the branch cuts are
removed. Therefore $Y(z)$ is single-valued and analytic on  the
complex plane except at the singularities
$\bar\Comp\backslash\set{0,\infty,\tilde z_a,\hat z_a,z^{\ast}_a}$.

Now we can read off the
eigenvalues $\tilde y(z),\hat y(z)$ of $Y(z)$ from its
characteristic function $F(y,z)$
\[\label{eq:Super.Curve}
F(\tilde y(z),z)=0,\qquad F(\hat y(z),z)=\infty
\]
with
\[\label{eq:Super.CurveFunction}
F(y,z)=\frac{\tilde F_4(z)}{\hat F_4(z)}\,\sdet \bigbrk{y-Y(z)}
=\frac{\tilde F(y,z)}{\hat F(y,z)}\,.
\]
We have included polynomial prefactors $\tilde F_4(z),\hat F_4(z)$
in the definition of $F=\tilde F/\hat F$ which clearly
do not change the algebraic curve.
The purpose of the prefactors is
to remove the poles originating from $U(z)$.
The roots of these prefactors are thus given by the singularities
$\set{\tilde z_a,z^{\ast}_a}$ or $\set{\hat z_a,z^{\ast}_a}$,
respectively. They enable us to write
both $\tilde F$ and $\hat F$ as polynomials,
not only in $y$ (obvious), but also in $z$.

As $Y(z)$ has only pole-type singularities
at $\set{0,\infty,\tilde z_a,\hat z_a,z^{\ast}_a}$,
the above equation defines two algebraic curves $\tilde y(z)$ and $\hat y(z)$
on the Riemann surfaces $\tilde{\mathbb{M}}$ and $\hat{\mathbb{M}}$,
respectively. We can even unite the two curves into one
curve $y(z)=\sset{\hat y(z)}{\tilde y(z)}$ on
$\mathbb{M}=\tilde{\mathbb{M}}\cup \hat{\mathbb{M}}$.
At the points $\set{\tilde z_a},\set{\hat z_a}$,
the functions $\tilde y(z),\hat y(z)$
have inverse square-root singularities.
At $\set{z^\ast_a}$ both functions $\tilde y(z),\hat y(z)$
have double poles with equal coefficients.
Similarly, at $\set{0,\infty}$,
there are singularities of the type
$-2\alpha_0/z^2,2\alpha_\infty z^2$.
Finally, there are no single poles anywhere,
because they would lead to a singular matrix $\Mono$,
which cannot happen.

\subsection{The Central Element}
\label{sec:Super.Central}

Consider the local transformation
\[\label{eq:Super.Ident}
g(\tau,\sigma)\mapsto \xi(\tau,\sigma)\, g(\tau,\sigma)
\]
with $\xi$ a number-valued field which is nowhere zero.
Here we would like to demonstrate that this transformation
does not have any physical effect.
First of all, it changes the current $J$ by $J\mapsto
J-\xi^{-1}d\xi$, but note that $\str J=0$ remains true due to $\str
I=0$. The transformation can now be easily seen to affect only
the $P$-component of $J$
\[\label{eq:Super.IdentP}
P\mapsto P-\xi^{-1}d\xi.
\]
In the equations of motion \eqref{eq:Super.EOM}
the variation drops out
when the Lagrange multiplier shifts accordingly
\[\label{eq:Super.IdentLambda}
\Lambda\mapsto \Lambda-\xi^{-1}{\ast d}\xi-i\,d\zeta-i\upsilon\, d\sigma.
\]
The additional transformation parameters are the field
$\zeta(\tau,\sigma)$ and the constant $\upsilon$. We cannot include
$\upsilon$ in $\zeta$ as $\zeta\to\zeta+\upsilon\,\sigma$ as $\zeta$ would
not be periodic. The action is also invariant except for the term
proportional to $\upsilon$. This actually leads to a change of the
global charges \eqref{eq:Super.Charges}
\[\label{eq:Super.IdentCharge}
S\mapsto
S-\frac{\sqrt{\lambda}}{2\pi}\oint d\zeta
-\upsilon\frac{\sqrt{\lambda}}{2\pi}\oint d\sigma
=
S
-\sqrt{\lambda}\,\upsilon.
\]
This change of the central element of $S$ is unphysical because the
global symmetry is merely $\grp{PSU}(2,2|4)$, not $\grp{SU}(2,2|4)$.

The family of flat connections changes up to a central gauge
transformation
\[\label{eq:Super.IdentLax}
A(z)\mapsto A(z)
-(\half z^2+\half z^{-2})\,\xi^{-1}d\xi
-(\half z^2-\half z^{-2})(i\,d\zeta+i\upsilon\, d\sigma).
\]
As this is an abelian shift, it completely factorizes from
the monodromy and we get
\[\label{eq:Super.IdentMono}
\Mono(z)\mapsto \Mono(z)\, \exp \int_0^{2\pi}d\sigma
\Bigbrk{(1-\half z^2-\half z^{-2})\,\xi^{-1}\partial_\sigma\xi -
(\half z^2-\half z^{-2})(i\partial_\sigma\zeta+i\upsilon)}.
\]
The first term measures the winding number of $\xi$ around $0$ when
going once around the string.
This winding affects both the $AdS_5$
and $S^5$ parts of $g$. However, in the physical setting, the
background is a universal cover and windings around the time-circle
of $AdS_5$ are not permitted. Therefore the term involving $\xi$
does not contribute. Also the term involving $\zeta$ vanishes
because $\zeta$ is periodic. We end up with
\[\label{eq:Super.IdentMonoSimp}
\Mono(z)\mapsto \Mono(z)\, \exp \bigbrk{-i\pi\upsilon (z^2-z^{-2})}.
\]
The factor is abelian and does not change the eigenvectors.
We thus find
\[\label{eq:Super.IdentCurve}
Y(z)\mapsto Y(z)-2\pi \upsilon(z^2+z^{-2}).
\]
This means that we can shift the curve $y(z)$
by a term proportional to $(z^2+z^{-2})$
as long as the factor of proportionality is the same for all sheets.

\subsection{Symmetry}
\label{sec:Super.Sym}

Let us introduce the (antisymmetric) supermatrix
\[\label{eq:Super.ConjMat}
C=E_1- i E_2.
\]
Then there is another useful way of expressing \eqref{eq:Super.HQPQ}
\<\label{eq:Super.HQPQConj}
{H}\eq
\quarter J-\quarter C\,J^\strans\,C^{-1}+\quarter \eta \,J\,\eta-\quarter \eta\, C\,J^\strans\,C^{-1}\,\eta,
\nln
{Q}_1\eq
\quarter J-\sfrac{i}{4} C\,J^\strans\,C^{-1}-\quarter \eta \,J\,\eta+\sfrac{i}{4} \eta\, C\,J^\strans\,C^{-1}\,\eta,
\nln
{P}\eq
\quarter J+\quarter C\,J^\strans\,C^{-1}+\quarter \eta \,J\,\eta+\quarter \eta\, C\,J^\strans\,C^{-1}\,\eta,
\nln
{Q}_2\eq
\quarter J+\sfrac{i}{4} C\,J^\strans\,C^{-1}-\quarter \eta \,J\,\eta-\sfrac{i}{4} \eta\, C\,J^\strans\,C^{-1}\,\eta,
\>
where $\eta$ is the grading matrix \eqref{eq:Super.Grading}.
This form reveals that a conjugation of the four components $H,Q_1,P,Q_2$
of $J$ with $C$ is equivalent to their supertranspose
up to a sign determined by their  grading under $\Integers_4$
\< \label{eq:Super.Conj}
C^{-1}\,{H}\,C\eq-{H}^\strans ,
\nln
C^{-1}\,{Q}_1\,C\eq - i{Q}_1^\strans ,
\nln
C^{-1}\,{P}\,C\eq+{P}^\strans ,
\nln
C^{-1}\,{Q}_2\,C\eq + i{Q}_2^\strans.
\>
When we apply this conjugation to the flat connections we obtain
\<\label{eq:Super.ConjLax}
C^{-1} A(z) C \eq -H^\strans +\bigbrk{\half z^2+\half z^{-2}}
P^\strans +\bigbrk{-\half z^2+\half z^{-2}} {\ast P}^\strans
-iz^{-1} Q_1^\strans +iz\,Q_2^\strans
\nln\eq -A^\strans(-iz).
\>
This, in turn, implies a symmetry relation
for the monodromy $C^{-1} \Mono(z)\, C = \Mono^{-\strans}(-iz)$.
The inverse is due to the overall sign in \eqref{eq:Super.ConjLax}
and the transpose puts the Wilson loop in the original path ordering.
In other words%
\footnote{The contribution from $\Mono^{-1}_0(1)=h(0)$ can be
seen to cancel, because $h\in\grp{Sp}(4,\Real)\times\grp{Sp}(4,\Real)$.}
\[\label{eq:Super.ConjMonoSimp}
\Mono(iz) = C\,\Mono^{-\strans}(z)\, C^{-1}
\]
is related to $\Mono(z)$ by conjugation, inversion and
supertranspose.
This translates to the following symmetry of $Y(z)$ and $U(z)$
\[\label{eq:Super.ConjLogDer}
Y(iz)=-C\, Y^\strans(z)\,C^{-1},\qquad U(iz)=-C\, U^{\strans}(z)\, C^{-1}.
\]
In particular, the characteristic function has the symmetry
\[\label{eq:Super.ConjCurve}
F(y,iz)=\frac{\tilde F_4(iz)}{\hat F_4(iz)}\,
\sdet\bigbrk{y-Y(iz)}=\frac{\hat F_4(z)}{\tilde F_4(z)}\,\sdet\bigbrk{y+Y(z)} =
F(-y,z).
\]
It therefore depends analytically only on the combinations $z^4$, $yz^2$, $y^2$.
In other words, $y(iz)=-y(z)$
and consequently $p(iz)=-p(z)+2\pi \Integers$
with some permutation of the sheets.

To determine the permutation, let us consider
the action on the diagonalized matrix
\[\label{eq:Super.ConjDiag}
Y\indup{diag}(iz)=
-C\indup{diag}(z) Y^{\strans}\indup{diag}(z) \,C\indup{diag}^{-1}(z)
\quad\mbox{with}\quad
C\indup{diag}(z)=u(iz)\,C\,u^{\strans}(z).
\]
As both $Y\indup{diag}(iz)$ and
$Y\indup{diag}^{\strans}(iz)=Y\indup{diag}(iz)$ are
diagonal, $C\indup{diag}(z)$ must be a permutation matrix
and thus constant (up to branch cuts).
In particular, we should investigate the fixed points of $z\mapsto iz$;
these are the singular points $\set{0,\infty}$.
At these points, $C\indup{diag}(z)$ as
defined in \eqref{eq:Super.ConjDiag}
must approach an antisymmetric matrix related to $C$.
As it is constant, it must always be an
antisymmetric permutation matrix
which acts non-trivially with period $2$.
We therefore find that the eigenvalues obey the symmetry
\[\label{eq:Super.ConjEigen}
\tilde y_k(iz)=-\tilde y_{k'}(z),
\qquad
\hat y_k(iz)=-\hat y_{k'}(z)
\]
where we are free to choose the following permutation of sheets
\[\label{eq:Super.Perm}
k'=(2,1,4,3)\qquad\mbox{for}\qquad k=(1,2,3,4).
\]
For the quasi-momentum we find
\[\label{eq:Super.ConjQuasi}
\tilde p_k(iz)=2\pi m\sheetsign_k-\tilde p_{k'}(z),
\qquad
\hat p_k(iz)=-\hat p_{k'}(z).
\]
Here we have introduced
\[\label{eq:Super.Epsi}
\sheetsign_k=(+1,+1,-1,-1)\qquad\mbox{for}\qquad k=(1,2,3,4).
\]
The constant shift $2\pi m$ in $\tilde p_k(iz)$ is related
to winding around $S^5$.
It must be absent for the $AdS_5$ counterpart
$\hat p_k(iz)$ because there cannot be windings in the time direction.

\begin{figure}\centering
\includegraphics{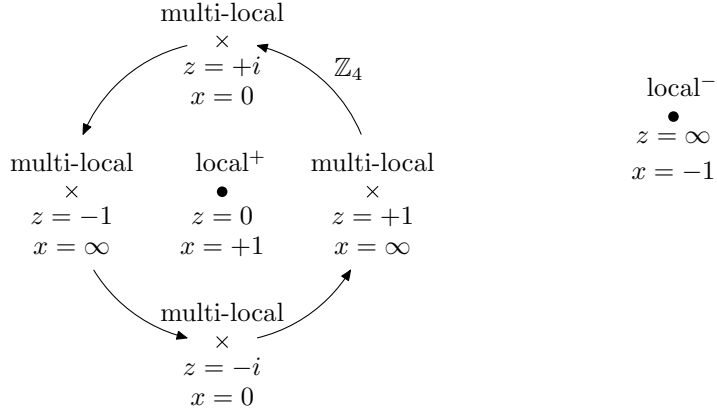}
\caption{Special points of the quasi-momenta. 
The expansion around $z=0,\infty$ yields one sequence
of local charges each,
see \protect\secref{sec:Super.Local}.
At $z=\pm 1,\pm i$ one finds
the Noether charges, discussed in \protect\secref{sec:Super.Global},
and multi-local charges.
All other points are related to non-local charges.}
\label{fig:Charges}
\end{figure}

Finally, we see that $y$ must depend analytically on $z^2$.
We can thus introduce the variable $x$ defined by
\[\label{eq:Super.XvsZ}
x=\frac{1+z^2}{1-z^2}\,,\qquad
z^2=\frac{x-1}{x+1}\,,
\]
which is precisely the variable commonly used for bosonic
sigma models as in \cite{Kazakov:2004qf,Beisert:2004ag}.
The points associated to local and global charges,
discussed in the following subsections, and the
symmetry are related as follows,
see also \figref{fig:Charges}
\[\label{eq:Super.XvsZpoints}
\begin{array}[b]{rclcrcl}
x\eq \infty &\Leftrightarrow& z\eq \pm 1,
\\[3pt]
     x\eq 0 &\Leftrightarrow& z\eq \pm i,
\\[3pt]
     x\eq +1&\Leftrightarrow& z\eq 0,
\\[3pt]
     x\eq -1&\Leftrightarrow& z\eq \infty,
\\[3pt]
     x\earel{\mapsto}1/x &\Leftrightarrow& z \earel{\mapsto}iz.
\end{array}
\]
Note the relation of differentials
\[\label{eq:Super.XvsZdiff}
\frac{dx}{1-1/x^2}=\frac{dz}{z}=d\kappa\,,
\]
where $\kappa=\log z$ is the spectral parameter
used in \cite{Bena:2003wd}.

\subsection{Local Charges}
\label{sec:Super.Local}

At the points $z=0,\infty$ the expansion of the Lax connection
is singular
\[\label{eq:Super.LocalLax}
A(\epsilon^{\pm 1})=
\half \epsilon^{-2}(P\pm{\ast P}\mp \Lambda)
+\epsilon^{-1}Q_{1,2}+
H
+\epsilon\,Q_{2,1}
+\half \epsilon^{2}(P\mp{\ast P}\pm \Lambda)
.
\]
The expansion of the quasi-momentum $p(z)$ at these points
is thus related to local charges.
As was shown in, e.g., \cite{Beisert:2004ag}, in the
absence of the fermionic contributions $Q_{1,2}$,
the leading coefficient of $p(z)$ in $\epsilon$ is
directly related to eigenvalues of the
leading contribution to $A_\sigma$.
Let us repeat the argument for the point $z=0$.
Consider the transformed connection $\bar A(z)$ in
the $\sigma$-direction given by
\[\label{eq:Super.LocalLaxDiag}
\partial_\sigma-\bar A(z)=T(z)\bigbrk{\partial_\sigma-A_\sigma(z)}T^{-1}(z).
\]
Here $T(z)$ and $A(z)$ are given by their expansion in $z$
\[\label{eq:Super.LocalExpand}
T(z)=\sum_{r=0}^\infty z^r T_r,\qquad
\bar A(z)=\sum_{r=-2}^\infty z^r \bar A_r.
\]
We demand that $T_0$ diagonalizes the leading term
\[\label{eq:Super.LocalLaxLead}
\bar A_{-2}=\half T_0P_+T_0^{-1}+\half \Lambda_\sigma
=\diag (\tilde \alpha_1,\tilde \alpha_2,\tilde \alpha_3,\tilde \alpha_4||
\hat \alpha_1,\hat \alpha_2,\hat \alpha_3,\hat \alpha_4).
\]
Since $P$ satisfies $CP^{\strans}C^{-1}=P$, c.f.~\eqref{eq:Super.Conj},
its eigenvalues must be doubly degenerate,
$\tilde \alpha_1=\tilde \alpha_2$, $\tilde \alpha_3=\tilde \alpha_4$,
$\hat \alpha_1=\hat \alpha_2$, $\hat \alpha_3=\hat \alpha_4$.
Furthermore, $P_+$ satisfies the Virasoro constraint $\str P_+^2=0$.
This requires $\tilde \alpha_1=\hat \alpha_1$, $\tilde \alpha_3=\hat \alpha_3$.
The abelian shift by $\Lambda_\sigma$ is compatible with this construction
and we find \cite{Arutyunov:2004yx}
\[\label{eq:Super.LocalLaxLeadFinal}
\bar A_{-2}=\diag (\alpha,\alpha,\beta,\beta||\alpha,\alpha,\beta,\beta)
=\matr{cc}{\alpha I&0\\0&\beta I}.
\]
Here we have introduced a $(2|2)\times(2|2)$ block decomposition
of the $(4|4)\times(4|4)$ supermatrix,
i.e.~each block is a supermatrix.
As the eigenvalues $\alpha$ and $\beta$ are (generically) distinct,
we can use the $T_r(z)$
to bring $\bar A_{r-2}(z)$ to a block-diagonal form
\[\label{eq:Super.LocalLaxBlock}
\bar A_r=\matr{cc}{a_r&0\\0&b_r}\qquad\mbox{or}\qquad
\bar A(z)=\matr{cc}{a(z)&0\\0&b(z)}.
\]

When this is done order by order in perturbation theory,
the resulting $a_r$ and $b_r$ are local
combinations of the fields.
However, the diagonalization of the Lax connection is not yet complete
and a complete diagonalization will lead non-local results.
Still we can obtain local charges:
Although the open Wilson loop $\Mono$ is in general non-local,
its superdeterminant is the exponential of a local charge.
Here $\sdet\Mono=1$ is trivial, but
we can consider only one block of
$T(2\pi)\Mono T(0)^{-1}$
\[\label{eq:Super.LocalMonoBlock}
\mono(z)=\lrbrk{\pexp\int_0^{2\pi}a(1)}^{-1} \lrbrk{\pexp\int_0^{2\pi} a(z)}.
\]
Then $\sdet \mono(z)=\exp iq(z)$ with
\[\label{eq:Super.LocalGenerator1}
q(z)=-i\int_{0}^{2\pi}d\sigma\,\bigbrk{\str a(z)-\str a(1)}.
\]
The expression for the other block involving
$b(z)$ is in fact equivalent due to $\str a+\str b=\str \bar A=0$.
The expansion of $q(z)$ into $q_r$ gives a sequence of local
charges. The term $q_{-2}$ vanishes because $a_{-2}$ is
proportional to the identity. We can also perform a similar
construction around $z=\infty$ leading to similar charges and
thus we have found two infinite sequences of local charges. 
Let us express $q(z)$ through the quasi-momentum $p(z)$.
As $\exp iq$ is the superdeterminant of the block $\mono$ of
$T(2\pi)\Mono T(0)^{-1}$ we can also write $q(z)$ as a sum over a
half of the quasi-momenta
\[\label{eq:Super.LocalGenerator2}
q(z)=\tilde p_1(z)+\tilde p_2(z)-\hat p_1(z)-\hat p_2(z).
\]
Using \eqref{eq:Super.Epsi} we write the generator 
of local charges in the concise form
\[\label{eq:Super.LocalGenerator}
q(z)=\sum_{k=1}^4 \sheetsign_k\bigbrk{\half\tilde p_k(z)-\half\hat p_k(z)}.
\]
Expanded around $z=0,\infty$ it yields the conserved local charges.
In \appref{sec:Local} we will construct the first of these charges.
Note that besides the local charges there is a larger set of
conserved non-local charges.

\subsection{Singularities}
\label{sec:Super.Sing}

We would like to understand the singular behavior
of the quasi-momentum $p(z)$ at $z=0$ better.
In the bosonic case we would be finished
after the semi-diagonalization of the previous section
because all singular terms have been diagonalized
and can be integrated up.
In the supersymmetric case, the remaining singular term $a_{-1}$
is not diagonal and might lead to further singularities at $z=0$.
Here we will show that this does not happen.
The difficulty of the proof is that
any attempt to diagonalize further would lead to non-local terms.

We shall start with one block $a(z)$
of the semi-diagonalized connection $\bar A(z)$.
Let us investigate the logarithm of the monodromy $\mono(z)$
and expand near $z=0$
\footnote{For convenience we omit contributions
from the second term in $\mono$;
they do not change the principal result.}
\[\label{eq:Super.SingWilson}
\log \mono(z)=\int_0^{2\pi}d\sigma\,a(\sigma,z)
+\int_0^{2\pi}d\sigma\int_0^{\sigma}d\sigma'\,
\half \bigcomm{a(\sigma,z)}{a(\sigma',z)}
+\ldots
\]
The further terms involve nested commutators of $a(z)$ at
various points $\sigma$.
The term $a_{-2}=\alpha I$ is abelian and thus contributes only to
the first term. This is not necessarily the most singular term,
as $a_{-1}$ may appear many times within the nested commutators.
To resolve this problem we note that the
full Lax connection obeys
the $\Integers_4$-symmetry relation \eqref{eq:Super.Conj}.
This reduces to a similar relation for the block $a(z)$
\[\label{eq:Super.SignSym}
c\,a^{\strans}(z)\,c^{-1}=-a(iz)\qquad\mbox{or}\qquad
c\,a^{\strans}_r\,c^{-1}=i^{2+r} a_r.
\]
where $c=e_1-ie_2$ with $e_{1,2}$ as in
\eqref{eq:Super.E1E2}, but with $e$ being a $2\times 2$
instead of a $4\times 4$ antisymmetric matrix.
This means that $a_r$ has $\Integers_4$-grading $r$.
Note that the grading is obeyed by commutators,
i.e.~when $x$ and $y$ have gradings $r$ and $s$,
respectively, the commutator $\comm{x}{y}$ has grading $r+s$.
Now consider two $(2|2)\times (2|2)$ supermatrices $x,y$ of grading $-1$.
Then it can be shown (explicitly) that their commutator
$\comm{x}{y}$ is proportional to the identity matrix $I$.
It therefore drops out of any further commutators
and nested commutators can never produce terms of grading
less than $-2$.
Furthermore, all terms of grading $-2$ are proportional to the identity.
The grading coincides with the power of $z$ and we find
\[\label{eq:Super.SingLog}
\log \mono(z)=d_{-2} z^{-2}I+d_{-1} z^{-1}+\order{z^0}
\]
with $d_{-2}$ a number and $d_{-1}$ a matrix of grading $-1$.
To finally diagonalize $\log\mono(z)$
we first use a matrix $\exp(t_{-1} z^{-1})$ which, using
the Baker-Campbell-Hausdorff identity
and for the same reasons as above,
removes the term $d_{-1}$ without lifting
the degeneracy of double poles or creating even higher poles.
Afterwards $\mono(z)$ can be diagonalized perturbatively.
Of course, all of the above holds true for the other block.
We assemble the two blocks and find for the quasi-momentum
\[\label{eq:Super.SingQuasiP}
\tilde p_k(z)\sim \hat p_k(z)\sim \lrbrk{\alpha_0+\sheetsign_k\beta_0}z^{-2}+\order{z^{-1}}
\]
with some coefficients $\alpha_0,\beta_0$
not directly related to $\alpha,\beta$.
We have thus proved that the structure
of residues found in \cite{Arutyunov:2004yx} is
not affected by the fermions.
Note that this distribution on the $p_k$
is compatible with the permutation of sheets
in \secref{sec:Super.Sym}.
Similarly, at $z=\infty$ the expansion of the quasi-momenta is given by
\[\label{eq:Super.SingQuasiM}
\tilde p_k(z)\sim \hat p_k(z)\sim \lrbrk{\alpha_\infty+\sheetsign_k\beta_\infty}z^{2}+\order{z}.
\]
%

\subsection{Global Charges}
\label{sec:Super.Global}

At $z=1$ the expansion of the Lax connection
\[\label{eq:Super.GlobalLax}
A(1+\epsilon)=J-2\epsilon\, {\ast K}+\order{\epsilon^2}
\]
is related to the $\alg{psu}(2,2|4)$ Noether current.
The expansion of the monodromy yields
\[\label{eq:Super.GlobalMono}
\Mono(1+\epsilon)=
I-2\epsilon
\int_0^{2\pi} d\sigma\,
\lrbrk{\pexp \int_{0}^{\sigma}d\sigma' J_\sigma(\sigma')}^{-1} K_\tau(\sigma)
\lrbrk{\pexp \int_{0}^{\sigma}d\sigma' J_\sigma(\sigma')}
+\order{\epsilon^2}
\]
which equals
\[\label{eq:Super.GlobalMonoSimp}
\Mono(1+\epsilon)=I-\epsilon\,\frac{4\pi\,S}{\sqrt{\lambda}}
+\order{\epsilon^2}
\]
by means of \eqref{eq:Super.Charges}.
Not only $z=+1$, but also $z=-1$ and $z=\pm i$ are related to
the global charges,
as can be seen from the symmetry discussed in \secref{sec:Super.Sym}.
The higher orders in the expansion yield
multi-local charges. 
These are the Yangian generators discussed 
in \cite{Bena:2003wd,Hatsuda:2004it,Berkovits:2004xu,Das:2004hy}.

The expansion of the quasi-momenta $\tilde p_k(z)$ associated to $S^5$ at $z=1$ is
\cite{Beisert:2004ag}
\<\label{eq:Super.GlobalSheetsS5}
\tilde p_1(1+\epsilon)\eq
-\epsilon\,\frac{4\pi}{\sqrt{\lambda}}
\lrbrk{+\sfrac{3}{4}\tilde r_1+\sfrac{1}{2}\tilde r_2+\sfrac{1}{4}\tilde r_3+\sfrac{1}{4}r^\ast}+\ldots\,,
\nln
\tilde p_2(1+\epsilon)\eq
-\epsilon\,\frac{4\pi}{\sqrt{\lambda}}
\lrbrk{-\sfrac{1}{4}\tilde r_1+\sfrac{1}{2}\tilde r_2+\sfrac{1}{4}\tilde r_3+\sfrac{1}{4}r^\ast}+\ldots\,,
\nln
\tilde p_3(1+\epsilon)\eq
-\epsilon\,\frac{4\pi}{\sqrt{\lambda}}
\lrbrk{-\sfrac{1}{4}\tilde r_1-\sfrac{1}{2}\tilde r_2+\sfrac{1}{4}\tilde r_3+\sfrac{1}{4}r^\ast}+\ldots\,,
\nln
\tilde p_4(1+\epsilon)\eq
-\epsilon\,\frac{4\pi}{\sqrt{\lambda}}
\lrbrk{-\sfrac{1}{4}\tilde r_1-\sfrac{1}{2}\tilde r_2-\sfrac{3}{4}\tilde r_3+\sfrac{1}{4}r^\ast}+\ldots\,.
\>
Here, $[\tilde r_1,\tilde r_2,\tilde r_3]$ are the
the Dynkin labels of $\grp{SU}(4)$ related to
the spins of $\grp{SO}(6)$ by
$\tilde r_1=J_2-J_3$,
$\tilde r_2=J_1-J_2$,
$\tilde r_3=J_2+J_3$.
The label $r^\ast$ is an unphysical label
related to the $\grp{U}(1)$ hypercharge.
It transforms under the transformation
described in \secref{sec:Super.Central}
as $r^\ast\mapsto r^\ast + \upsilon \sqrt{\lambda}$\,.
Similarly, the expansion for $\hat p_k(z)$ associated to $AdS_5$ reads
\cite{Arutyunov:2004yx}
\<\label{eq:Super.GlobalSheetsAdS5}
\hat p_1(1+\epsilon)\eq
\epsilon\,\frac{4\pi}{\sqrt{\lambda}}
\lrbrk{+\sfrac{3}{4}\hat r_1+\sfrac{1}{2}\hat r_2+\sfrac{1}{4}\hat r_3-\sfrac{1}{4}r^\ast}+\ldots\,,
\nln
\hat p_2(1+\epsilon)\eq
\epsilon\,\frac{4\pi}{\sqrt{\lambda}}
\lrbrk{-\sfrac{1}{4}\hat r_1+\sfrac{1}{2}\hat r_2+\sfrac{1}{4}\hat r_3-\sfrac{1}{4}r^\ast}+\ldots\,,
\nln
\hat p_3(1+\epsilon)\eq
\epsilon\,\frac{4\pi}{\sqrt{\lambda}}
\lrbrk{-\sfrac{1}{4}\hat r_1-\sfrac{1}{2}\hat r_2+\sfrac{1}{4}\hat r_3-\sfrac{1}{4}r^\ast}+\ldots\,,
\nln
\hat p_4(1+\epsilon)\eq
\epsilon\,\frac{4\pi}{\sqrt{\lambda}}
\lrbrk{-\sfrac{1}{4}\hat r_1-\sfrac{1}{2}\hat r_2-\sfrac{3}{4}\hat r_3-\sfrac{1}{4}r^\ast}+\ldots\,.
\>
The Dynkin labels $[\hat r_1,\hat r_2,\hat r_3]$
of $\grp{SU}(2,2)$
are related to the spins of $\grp{SO}(2,4)$ by
$\hat r_1=S_1-S_2$,
$\hat r_2=-E-S_1$,
$\hat r_3=S_1+S_2$.

\subsection{Bosonic $AdS_5\times S^5$, $\Real\times S^5$ and $AdS_5\times S^1$ Sectors}
\label{sec:Super.Bosonic}

The restriction to the classical bosonic string on
$AdS_5\times S^5$ \cite{Arutyunov:2004yx},
$\Real\times S^5$ \cite{Beisert:2004ag} and
$AdS_5\times S^1$ \cite{Schafer-Nameki:2004ik}
is straight-forward:
First of all we remove all possible fermionic poles.
This implies $K^\ast=0$ but we can also set $r^\ast=B=0$
and obtain the bosonic string on $AdS_5\times S^5$.
Then the expansion at $z=1$
\eqref{eq:Super.GlobalSheetsS5,eq:Super.GlobalSheetsAdS5}
as well as the structure of poles at $z=0,\infty$,
c.f.~\secref{sec:Super.Sing},
agrees with \cite{Arutyunov:2004yx}
under the change of spectral parameter
\eqref{eq:Super.XvsZ}.

In the next step we either reduce $AdS_5$ to $\Real$ or
$S^5$ to $S^1$. The isometry groups of both factors $\Real$ and $S^1$
are abelian. For the monodromy corresponding to this factor we can therefore
remove the path ordering
\[\label{eq:Super.BosonicMono}
\Mono(z)=
\lrbrk{\pexp\oint A(1)}^{-1}
\lrbrk{\pexp\oint A(z)}
=\exp\oint \bigbrk{A(z)-A(1)},
\]
We now substitute $A(z)$ from \eqref{eq:Super.Lax}
with $H=Q_1=Q_2=0$, $P=-g^{-1} dg$
and solve $\Mono(z)=e^{ip(z)}$ for the quasi-momentum
\[
p(z)=
i(1-\half z^2+\half z^{-2})\oint g^{-1}dg
-i(-\half z^2+\half z^{-2})\oint g^{-1}{\ast} dg.
\]
The first integral represents the winding number $m$,
it must vanish for $\Real$ and can be non-trivial
for $S^1$. The second integral represents the
global charge, it is proportional to the energy $E$ for $\Real$ and
to the spin $J$ for $S^1$.
By comparing to \eqref{eq:Super.GlobalSheetsAdS5} we find that
in the case of $\Real\times S^5$ the
full quasi-momentum for $AdS_5$ is given by
\[
\hat p_k(z)=
\sheetsign_k\, \frac{\pi\,E}{\sqrt{\lambda}}\, \bigbrk{-\half z^2+\half z^{-2}}.
\]
When the residues at $z=0,\infty$ are matched between
$\tilde p_k$ and $\hat p_k$ we find perfect agreement with
\cite{Beisert:2004ag}.
Equivalently in the case of $AdS_5\times S^1$ the full
quasi-momentum for $S^5$
is obtained by comparing to \eqref{eq:Super.GlobalSheetsS5}%
\footnote{The integral of $g^{-1}dg$ yields
odd multiples of $i\pi$ when $g(2\pi)=-g(0)$,
which is an allowed case.}
\[
\tilde p_k(z)=
\sheetsign_k\, \frac{\pi\,J}{\sqrt{\lambda}}\, \bigbrk{-\half z^2+\half z^{-2}}
+\sheetsign_k \pi m\bigbrk{1-\half z^2-\half z^{-2}}.
\]
Again, after matching the residues,
this is in agreement with \cite{Schafer-Nameki:2004ik}.

\section{Moduli of the Curve}
\label{sec:Moduli}

In this section we investigate the moduli space of admissible curves.
Admissible curves are algebraic curves which
satisfy all the properties derived in the previous section and
which can thus arise from a classical string configuration
on $AdS_5\times S^5$.
For a fixed degree of complexity of the solution,
which manifests as the genus of the curve,
we count the number of degrees of freedom
for admissible curves.
Although it is not obvious that all
admissible curves indeed represent string solutions
(in other words that we have identified all relevant
properties of admissible curves)
we see that this number agrees with strings in flat space.
We take this as evidence that our classification of
string solutions in terms of admissible curves is complete.
We finally identify the discrete parameters and continuous moduli
with certain cycles on the curve and interpret them.
For the comparison to gauge theory
we investigate the
Frolov-Tseytlin limit of the algebraic curve
corresponding to a loop expansion in gauge theory.

\subsection{Properties}
\label{sec:Moduli.Prop}


\begin{figure}\centering
\includegraphics{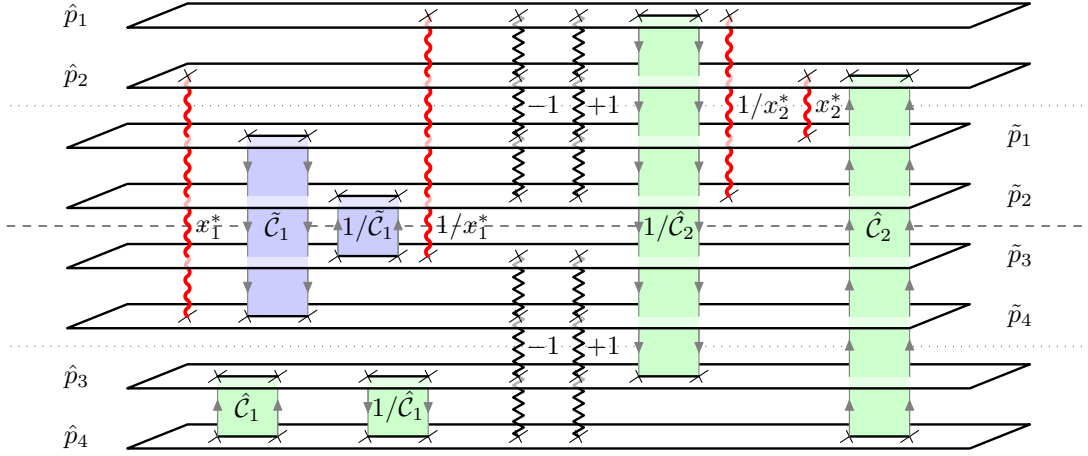}
\caption{Some configuration of cuts and poles for the sigma model.
Cuts $\tilde{\contour{C}}_a$ between the sheets $\tilde p_k$
correspond to $S^5$ excitations and likewise
cuts $\hat{\contour{C}}_a$ between the sheets $\hat p_k$
correspond to $AdS^5$ excitations.
Poles $x^\ast_a$ on sheets $\tilde p_k$ and $\hat p_l$
correspond to fermionic excitations.
The dashed line in the middle is related to physical excitations,
cuts and poles which cross it contribute to the total momentum, 
energy shift and local charges.}
\label{fig:sheetssigma}
\end{figure}
Let us collect the analytic properties of the quasi-momentum
\[\label{eq:Moduli.Sheets}
p(x)=\bigsset{\tilde p_1(x),\tilde p_2(x),\tilde p_3(x),\tilde p_4(x)}
{\hat p_1(x),\hat p_2(x),\hat p_3(x),\hat p_4(x)},
\]
see \figref{fig:sheetssigma} for an illustration.
All sheet functions $\tilde p_k(x)$ and $\hat p_l(x)$
are analytic almost everywhere. The singularities are as follows:
\begin{bulletlist}
\item
At $x=\pm 1$ there are single poles,
c.f.~\secref{sec:Super.Local}.
The four sheets $\tilde p_{1,2}(x),\hat p_{1,2}(x)$
all have equal residues;
the same holds for the remaining four sheets
$\tilde p_{3,4}(x),\hat p_{3,4}(x)$.

\item
Bosonic degrees of freedom are represented
by branch cuts $\set{\tilde{\contour{C}}_a}$, $a=1,\ldots,2\tilde A$
and $\set{\hat{\contour{C}}_a}$, $a=1,\ldots,2\hat A$.
The cut $\tilde{\contour{C}}_a$ connects the sheets
$\tilde k_a$ and $\tilde l_a$ of $\tilde p'(x)$.
Equivalently, $\hat{\contour{C}}_a$ connects the sheets
$\hat k_a$ and $\hat l_a$ of $\hat p'(x)$.
At both ends of the branch cut,
$\tilde x^\pm_a$ or $\hat x^\pm_a$,
there is a square-root singularity on both sheets.

\item
Fermionic degrees of freedom are represented by poles
at $\set{x^\ast_a}$, $a=1,\ldots,2A^\ast$.
The pole $x^\ast_a$ exists on the sheets
$k^\ast_a$ of $\tilde p(x)$ and
$l^\ast_a$ of $\hat p(x)$ with
equal residue.

\end{bulletlist}\pagebreak[2]

Further properties are:
\begin{bulletlist}
\item
For definiteness, we assume the quasi-momentum to approach zero
at $x=\infty$ on all sheets, c.f.~\secref{sec:Super.Global}
\[\label{eq:Moduli.Asymptotics}
\tilde p(x)=\order{1/x},\qquad
\hat p(x)=\order{1/x}.
\]

\item
The quasi-momentum obeys the symmetry $x\mapsto 1/x$,
see \secref{sec:Super.Sym}, as follows
\[\label{eq:Moduli.Sym}
\tilde p_{k}(1/x)=-\tilde p_{k'}(x)+2\pi m \sheetsign_k,\qquad
\hat p_{k}(1/x)= -\hat p_{k'}(x).
\]
We use the permutation $k'$ of $k$
and a sign $\sheetsign_k$ for each sheet $k=(1,2,3,4)$
as defined in \eqref{eq:Super.Perm,eq:Super.Epsi}
\[\label{eq:Moduli.SymDefs}
k'=(2,1,4,3),
\qquad
\sheetsign_k=(+1,+1,-1,-1).
\]
The branch cuts and poles must respect the symmetry.
We therefore consider the cut
$\tilde{\contour{C}}_{\tilde A+a}=1/\tilde{\contour{C}}_{a}$ to be the image of
$\tilde{\contour{C}}_{a}$.
The independent cuts are thus labelled by $a=1,\ldots,\tilde A$.
Similarly for $AdS_5$-cuts $\hat{\contour{C}}_{a}$ and
fermionic poles $x^\ast_a$
\footnote{
Within sums a self-symmetric cut will be counted with weight $1/2$.}
\[\label{eq:Moduli.SymCuts}
\tilde{\contour{C}}_{\tilde A+a}=1/\tilde{\contour{C}}_{a},
\qquad
\hat{\contour{C}}_{\hat A+a}=1/\hat{\contour{C}}_{a},
\qquad
x^\ast_{A^\ast+a}=1/x^\ast_{a}.
\]
Note that there is an arbitrariness of which cuts
are considered fundamental
and which are their images under the symmetry.
E.g.~we might replace $\tilde{\contour{C}}_a$ by
$1/\tilde{\contour{C}}_a$ which effectively interchanges
$\tilde{\contour{C}}_a$ and $\tilde{\contour{C}}_{\tilde A+a}$
without changing the curve.

\item
The unimodularity condition \eqref{eq:Super.Unimodular}
together with \eqref{eq:Moduli.Asymptotics} translates to
\[\label{eq:Moduli.Unimodular}
\tilde p_1+\tilde p_2+\tilde p_3+\tilde p_4=
\hat p_1+\hat p_2+\hat p_3+\hat p_4.
\]

\item
A common shift of all sheets
\[\label{eq:Moduli.Shift}
\tilde p(x)\mapsto \tilde p(x)-\frac{4\pi\upsilon}{1-1/x^2}\,,\qquad
\hat p(x)\mapsto \hat p(x)-\frac{4\pi\upsilon}{1-1/x^2}
\]
is considered unphysical,
c.f.~\secref{sec:Super.Central}.

\end{bulletlist}
For the cuts and poles we define several cycles and periods,
c.f.~\figref{fig:cycles}:
\begin{figure}[t]
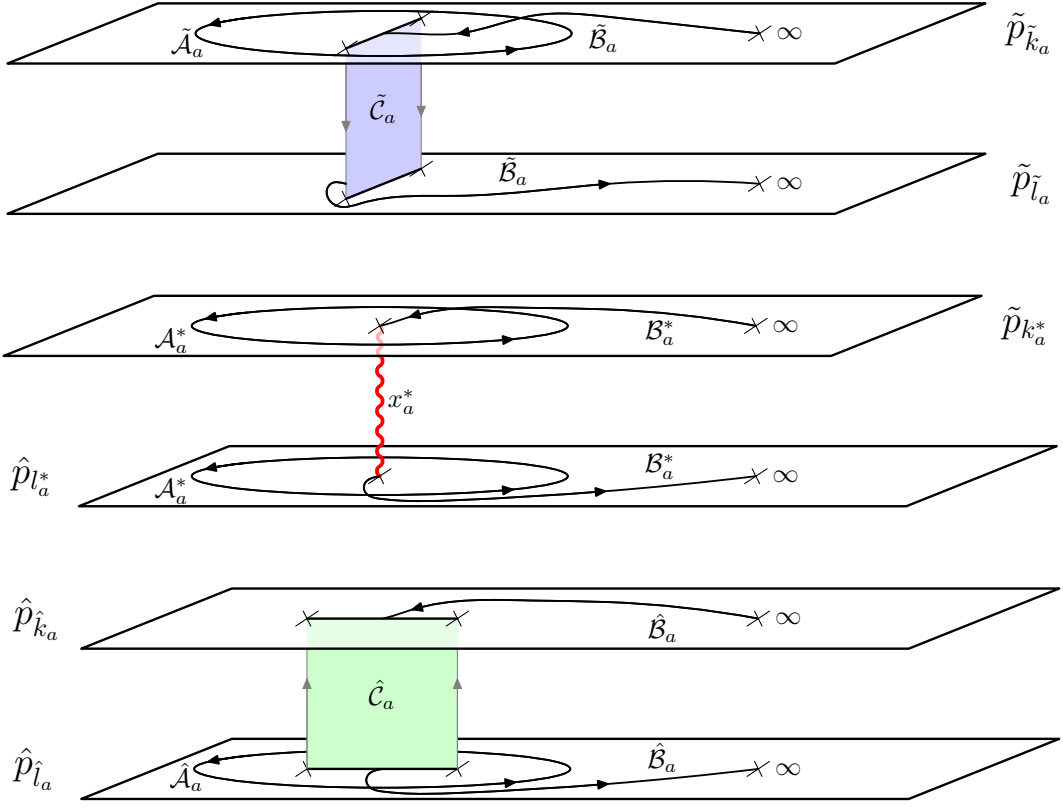
\centering
\includegraphics{bksz.Cycles.S5.eps}%
\vspace{1cm}

\includegraphics{bksz.Cycles.Fermi.eps}%
\vspace{1cm}

\includegraphics{bksz.Cycles.AdS5.eps}

\caption{Cycles for $S^5$-cuts (top), fermionic poles (middle) and
$AdS_5$-cuts (bottom). Generically, $S^5$-cuts are along aligned in the imaginary
direction while $AdS_5$-cuts are along the real axis.}
\label{fig:cycles}
\end{figure}
\begin{bulletlist}
\item
We define the cycles
$\tilde{\contour{A}}_a,\hat{\contour{A}}_a$
which surround the cuts
$\tilde{\contour{C}}_a,\hat{\contour{C}}_a$, respectively.
The cuts, which connect the branch points
$\set{\tilde x^\pm_a},\set{\hat x^\pm_a}$,
have been arranged in such a way that
\[\label{eq:Moduli.ACycleBos}
\oint_{\tilde{\contour{A}}_a}d\tilde p = 0,\qquad
\oint_{\hat{\contour{A}}_a}d\hat p = 0.
\]
This can be achieved by a reorganization of cuts
which corresponds to a
$\grp{Sp}(2\tilde A,\Integers)$ or $\grp{Sp}(2\hat A,\Integers)$
transformation, respectively \cite{Kazakov:2004qf}.

\item
We define the cycle $\contour{A}^\ast_a$
which surrounds the fermionic pole $x^\ast_a$.
There are no logarithmic singularities
at $x^\ast_a$
\[\label{eq:Moduli.ACycleFerm}
\oint_{\contour{A}^\ast_a}d\tilde p = \oint_{\contour{A}^\ast_a}d\hat p=0.
\]
At the singular points $x=\pm 1$ there are no
logarithmic singularities either
\[\label{eq:Moduli.ACycleSing}
\oint_{\pm 1}d\tilde p_k = \oint_{\pm 1}d\hat p_k = 0.
\]

\item
We define periods
$\tilde{\contour{B}}_a,\hat{\contour{B}}_a$
which connect
$x=\infty$ on sheet $\tilde k_a,\hat k_a$ to
$x=\infty$ on sheet $\tilde l_a,\hat l_a$
through the cuts
$\tilde{\contour{C}}_a,\hat{\contour{C}}_a$, respectively,
see \figref{fig:cycles}.
These must be integral
\[\label{eq:Moduli.BCycleBos}
\int_{\tilde{\contour{B}}_a}d\tilde p=2\pi \tilde n_a,\qquad
\int_{\hat{\contour{B}}_a}d\hat p=2\pi \hat n_a,
\]
because the monodromy at both ends of the B-period is trivial,
$\Mono(\infty)=I$.
Together with the asymptotic behavior
\eqref{eq:Moduli.Asymptotics} and single-valuedness
\eqref{eq:Moduli.ACycleBos,eq:Moduli.ACycleFerm,eq:Moduli.ACycleSing}
this implies that $\tilde p(x),\hat p(x)$ must jump by $2\pi \tilde
n_a,2\pi \hat n_a$ when passing through the cut
$\tilde{\contour{C}}_a,\hat{\contour{C}}_a$, respectively. This is
written as the equivalent condition
\<\label{eq:Moduli.DiscontBos}
\sheetsl[\tilde]_{\tilde l_a}(x)
-\sheetsl[\tilde]_{\tilde k_a}(x)
\eq 2\pi \tilde n_{a}
\quad\mbox{for}\quad
x\in \tilde{\contour{C}}_a,
\nln
\sheetsl[\hat]_{\hat l_a}(x)
-\sheetsl[\hat]_{\hat k_a}(x)
\eq 2\pi \hat n_{a}
\quad\mbox{for}\quad
x\in \hat{\contour{C}}_a.
\>

\item
The period $\contour{B}^\ast_a$ for a fermionic pole
connects $x=\infty$ to $x=x^\ast_a$ on sheet $k^\ast_a$ of $\tilde p(x)$.
It then continues from
$x=x^\ast_a$ to $x=\infty$ on sheet $l^\ast_a$ of $\hat p(x)$.
As fermionic singularities arise for coinciding
eigenvalues, the regular parts of $\tilde p(x)$ and $\hat p(x)$
must be equal modulo a shift by $2\pi n^\ast_a$
\[\label{eq:Moduli.DiscontFerm}
\sheetsl[\hat]_{\hat l_a}(x^\ast_a)
-\sheetsl[\tilde]_{\tilde k_a}(x^\ast_a)
= 2\pi n^\ast_{a}.
\]
Expressed as a B-period this yields
\[\label{eq:Moduli.BCycleFerm}
\pint_{\contour{B}^\ast_a}dp=2\pi n^\ast_a.
\]

\item
In addition to $\Mono(\infty)=I$
we also have $\Mono(0)=I$.
This means that
a period connecting $x=0$ with $x=\infty$
must be a multiple of $2\pi$.
In fact, the symmetry \eqref{eq:Moduli.Sym} enforces
\[\label{eq:Moduli.BCycleSing}
\tilde p_{1,2}(0)=
-\tilde p_{3,4}(0)=
\int_{\infty}^0 d\tilde p_{1,2}=-\int_{\infty}^0 d\tilde p_{3,4}=2\pi m,\qquad
\hat p_k(0)=\int_{\infty}^0 d\hat p_k=0.
\]
The integral for the $AdS_5$-part must vanish, because there cannot
be windings on the time circle of $AdS_5$ \cite{Kazakov:2004nh}. In
fact, for physical applications one needs to consider the universal
covering of $AdS_5$ where time circle has been decompactified.

\item
When no confusion arises, we may use a unified notation
$\contour{A}_a$ and $\contour{B}_a$ with $a=1,\ldots,2A$ for cuts
and poles,
$\tilde{\contour{A}}_a,\hat{\contour{A}}_a,\contour{A}^\ast_a$ and
$\tilde{\contour{B}}_a,\hat{\contour{B}}_a,\contour{B}^\ast_a$. The
total number of cuts and poles is $A=\tilde A+\hat A+A^\ast$. In
this case we label the sheets $p_k$ by $k=1,\ldots,8$ according to
\[\label{eq:Moduli.Beauty}
p_{1,2}=\hat p_{1,2},\qquad
p_{3,4,5,6}=\tilde p_{1,2,3,4},\qquad
p_{7,8}=\hat p_{3,4}.
\]
%
This ordering leads to the configuration of sheets as depicted in
\figref{fig:sheetssigma}. 
Some details of this representation
are discussed in \appref{sec:Beauty}.
It makes physical excitations and the comparison to gauge theory more
transparent.
%
%

\end{bulletlist}
%

\subsection{Ansatz}
\label{sec:Moduli.Ansatz}

The characteristic function of our algebraic curve is
rational
\[\label{eq:Moduli.Function}
F(y,x)=
\frac{\tilde F(y,x)}{\hat F(y,x)}
=
\frac
{\tilde F_4(x)y^4 + \tilde F_3(x)y^3 + \tilde F_2(x)y^2 + \tilde F_1(x)y + \tilde F_0(x)}
{\hat F_4(x)y^4 + \hat F_3(x)y^3 +\hat F_2(x)y^2 +\hat F_1(x)y +\hat F_0(x)}\,,
\]
with $\tilde F_k(x),\hat F_k(x)$ polynomials in $x$.
The curve $y(x)=\sset{\tilde y(x)}{\hat y(x)}$ obeys the algebraic equation
\[\label{eq:Moduli.Curve}
\tilde F(\tilde y(x),x)=0,\qquad
\hat F(\hat y(x),x)=0.
\]
We define the curve $y(x)$ with a different prefactor as
compared to the previous section as
\[\label{eq:Moduli.Ydef}
y(x)=(x-1/x)^2x\,p'(x).
\]
This definition removes the poles at $x=\pm 1$ \cite{Beisert:2004ag}.

\paragraph{Branch Points and Fermionic Poles.}

Bosonic branch points $\tilde x^\pm_a$
of the $S^5$ part
manifest themselves
as inverse square roots in $\tilde y(x)$.
An asymptotic analysis shows that they are obtained when
\[\label{eq:Moduli.BranchS5}
\tilde F_4(\tilde x^\pm_a)=\tilde F_3(\tilde x^\pm_a)=0
\qquad\mbox{while}\qquad
\tilde F_4'(\tilde x^\pm_a)\neq 0\neq \tilde F_3'(\tilde x^\pm_a).
\]
Similarly for branch points $\hat x^\pm_a$ in
the $AdS_5$ part
\[\label{eq:Moduli.BranchAdS5}
\hat F_4(\hat x^\pm_a)=\hat F_3(\hat x^\pm_a)=0
\qquad\mbox{while}\qquad
\hat F_4'(\hat x^\pm_a)\neq 0\neq \hat F_3'(\hat x^\pm_a).
\]
Fermionic singularities $x^\ast_a$ manifest themselves
as double poles in $y(x)$. A double pole is achieved by
\[\label{eq:Moduli.FermSing}
\tilde F_4(x^*_a)=\tilde F_4'(x^*_a)=
\hat F_4(x^*_a)=\hat F_4'(x^*_a)=0
\qquad
\mbox{while}
\qquad
\tilde F_3(x^*_a)\neq 0\neq \hat F_3(x^*_a).
\]
The behavior of $F_{2,1,0}$ is generic at these points.
Here we see that a non-zero $F_3$,
unlike in \cite{Beisert:2004ag},
is required due to fermions.
All these singularities are encoded in $F_4(x)$ as
\<\label{eq:Moduli.F4}
\tilde F_4(x)\eq
x^4
\prod_{a=1}^{2\tilde A}(x-\tilde x^+_a)
\prod_{a=1}^{2\tilde A}(x-\tilde x^-_a)
\prod_{a=1}^{2A^\ast}(x-x^\ast_a)^2,
\nln
\hat F_4(x)\eq
x^4
\prod_{a=1}^{2\hat A}(x-\hat x^+_a)
\prod_{a=1}^{2\hat A}(x-\hat x^-_a)
\prod_{a=1}^{2A^\ast}(x-x^\ast_a)^2.
\>
The factor $x^4$ is introduced for convenience as we shall see below.
For $\tilde F_4(x),\hat F_4(x)$ there are in total
$4\tilde A+4\hat A+2A^\ast$ degrees of freedom.

\paragraph{Asymptotics.}

At $x=\infty$ the curve behaves as $y(x)\sim x$ and at $x=0$ as
$y(x)\sim 1/x$. This is achieved by the following range of exponents
in the polynomials
\<\label{eq:Moduli.FBounds}
\tilde F_k(x)\eq {\ast}x^{4\tilde A+4A^\ast+8-k}+\ldots+{\ast}x^{k} ,
\nln
\hat F_k(x)\eq {\ast}x^{4\hat A+4A^\ast+8-k}+\ldots+{\ast}x^{k}.
\>
We can now count the remaining number of free coefficients. In
$\tilde F_k(x),\hat F_k(x)$, $k<4$, there are $4\tilde
A+4A^\ast+9-2k$ and $4\hat A+4A^\ast+9-2k$ degrees of freedom,
respectively. This leaves $20\tilde A+20\hat A+34A^\ast+48$ relevant
coefficients in total.

\paragraph{Unimodularity.}

The unimodularity condition
$\hat y_1+\hat y_2+\hat y_3+\hat y_4
=\tilde y_1+\tilde y_2+\tilde y_3+\tilde y_4$
is imposed as a relation of the
two leading coefficients of the algebraic equation
\[\label{eq:Moduli.FUnimodular}
\frac{\tilde F_3(x)}{\tilde F_4(x)}
=\frac{\hat F_3(x)}{\hat F_4(x)}
\,.
\]
This requires
\< \label{eq:Moduli.F3}
\tilde F_3(x)\eq F_3^\ast(x)
\prod_{a=1}^{2\tilde A}(x-\tilde x^+_a)
\prod_{a=1}^{2\tilde A}(x-\tilde x^-_a)
,
\nln
\hat F_3(x)\eq F_3^\ast(x)
\prod_{a=1}^{2\hat A}(x-\hat x^+_a)
\prod_{a=1}^{2\hat A}(x-\hat x^-_a)
,
\>
with some polynomial
\[\label{eq:Moduli.F3Fermi}
F_3^\ast(x)= {\ast}x^{4A^\ast+5}+\ldots+{\ast}x^3.
\]
It reduces the number of degrees of freedom
by $4\tilde A+4\hat A+4A^\ast+3$
to $16\tilde A+16\hat A+30A^\ast+45$.

\paragraph{Symmetry.}

The symmetry $y(1/x)=y(x)$ is realized by the conditions
\<\label{eq:Moduli.FSym}
\tilde F_k(1/x)\eq x^{-4\tilde A-4A^\ast-8}\tilde F_k(x),
\nln
\hat F_k(1/x)\eq x^{-4\hat A-4A^\ast-8}\hat F_k(x),
\nln
F^\ast_3(1/x)\eq x^{-4A^\ast-8} F^\ast_3(1/x).
\>
This yields $8\tilde A+8\hat A+15A^\ast+19$
constraints and leaves
$8\tilde A+8\hat A+15A^\ast+26$
coefficients.

\paragraph{Singularities.}

We have to group up the residues at $x=\pm 1$
according to \secref{sec:Super.Local}:
Out of the $16$ residues, there should only be $4$ independent ones.
This gives $12$ constraints, but two of them have already
been imposed by the unimodularity condition.
As the singularities are at the fixed points $x=\pm 1$ of the
symmetry $x\mapsto 1/x$, all $10$ constraints can be imposed independently.
This leaves
$8\tilde A+8\hat A+15A^\ast+16$
degrees of freedom.

\paragraph{Unphysical Branch Points.}

In addition to the physical branch points at $\tilde x^\pm_a,\hat x^\pm_a$
the algebraic curve might have further ones.
Generically, these singularities are square roots in contrast to
the physical one which are inverse square roots.
We can remove them using a condition of the discriminants%
\footnote{We could also use the equivalent condition:
All solutions to $dF=0$ are on the
curve unless there is a physical singularity at this value of $x$.
However, it is not quite clear how to count the number of constraints
from this condition.}
\<\label{eq:Moduli.Discri}
\tilde R\eq
- 4\tilde F_1^2\tilde F_2^3\tilde F_4
+ 16\tilde F_0\tilde F_2^4\tilde F_4
- 27\tilde F_1^4\tilde F_4^2
+ 144\tilde F_0\tilde F_1^2\tilde F_2\tilde F_4^2
- 128\tilde F_0^2\tilde F_2^2\tilde F_4^2
+ 256\tilde F_0^3\tilde F_4^3
\nl
+ 18\tilde F_1^3\tilde F_2\tilde F_3\tilde F_4
- 80\tilde F_0\tilde F_1\tilde F_2^2\tilde F_3\tilde F_4
- 192\tilde F_0^2\tilde F_1\tilde F_3\tilde F_4^2
- 6\tilde F_0\tilde F_1^2\tilde F_3^2\tilde F_4
+ 144\tilde F_0^2\tilde F_2\tilde F_3^2\tilde F_4
\nl
+ \tilde F_1^2\tilde F_2^2\tilde F_3^2
- 4\tilde F_0\tilde F_2^3\tilde F_3^2
- 4\tilde F_1^3\tilde F_3^3
+ 18\tilde F_0\tilde F_1\tilde F_2\tilde F_3^3
- 27\tilde F_0^2\tilde F_3^4
\>
and similarly for $\hat R$.
The discriminants measure the product of squared
distances of solutions $\tilde y_k(x)$ or $\hat y_k(x)$.
A single root of $\tilde R(x)=0$ or $\hat R(x)=0$
thus implies a square root behavior which can only occur at
$x=\tilde x^\pm_a$ or $x=\hat x^\pm_a$.
The discriminants must therefore have the form
\<\label{eq:Moduli.DiscriSqr}
\tilde R(x)\eq
x^{12}(x^2-1)^4
\prod_{a=1}^{2\tilde A}(x-\tilde x^+_a)
\prod_{a=1}^{2\tilde A}(x-\tilde x^-_a)
\,\tilde Q(x)^2,
\nln
\hat R(x)\eq
x^{12}(x^2-1)^4
\prod_{a=1}^{2\hat A}(x-\hat x^+_a)
\prod_{a=1}^{2\hat A}(x-\hat x^-_a)
\,\hat Q(x)^2.
\>
It is clear that $\tilde x^\pm_a$ and $\hat x^\pm_a$ are roots,
because all terms in \eqref{eq:Moduli.Discri}
contain $\tilde F_4$ or $\tilde F_3$.
Noting the generic form of the discriminants
\< \label{eq:Moduli.DiscriBounds}
\hat R(x)\eq {\ast} x^{24\tilde A+24A^\ast+36}+\ldots+{\ast}x^{12},
\nln
\tilde R(x)\eq {\ast}x^{24\hat A+24A^\ast+36}+\ldots+{\ast}x^{12}.
\>
together with the inversion symmetry
we find
$5\tilde A+5\tilde A+12A^\ast+8$
constraints and
$3\tilde A+3\hat A+3A^\ast+8$
remaining degrees of freedom.

\paragraph{Single Poles and A-Cycles.}

We need to remove all the single poles and A-cycles from the curve
$y(x)$ which would otherwise give rise to undesired logarithmic
behavior in the quasi-momentum when restoring the quasi-momentum from
its derivative. The symmetry $x\mapsto 1/x$ allows for $8$
independent single poles in $y(x)$ at $x=\pm 1$. There are 
$\tilde A+\hat A$ independent A-cycles around bosonic cuts. Fermionic
singularities contribute $2A^\ast$ independent single poles: one for
$\tilde y$ and one for $\hat y$ at each $x=x^\ast_a$ modulo
inversion symmetry. Among all these single poles and A-cycles, there
are $4$ relations from the sum over all residues, one for each pair
of sheets related by the symmetry.
In total this yields $\tilde A+\hat A+2A^\ast+4$ constraints and
leaves $2\tilde A+2\hat A+A^\ast+4$ coefficients.

\paragraph{B-Periods.}

For each bosonic cut and for each fermionic singularity there
is a B-period which must be integral.
Furthermore, for each pair of sheets related by the symmetry,
the B-period connecting $0$ and $\infty$ must also be integral.
Due to the unimodularity condition, only three of these periods
are independent.
In total we obtain $\tilde A+\hat A+A^\ast+3$ constraints
and are left with $\tilde A+\hat A+1$ degrees of freedom.

\paragraph{Hypercharge.}

One degree of freedom corresponds
to an irrelevant shift of the Lagrange multiplier,
c.f.~\secref{sec:Super.Central}.
The final number of moduli for admissible curves
is $\tilde A+\hat A$.

\subsection{Mode Numbers and Fillings}
\label{sec:Moduli.Fillings}

We will now associate each of the $\tilde A+\hat A$ moduli of the curve
to one parameter per pair of bosonic cuts.
We define the \emph{filling} of an $S^5$-cut $\tilde{\contour{C}}_a$
connecting sheets $\tilde k_a$ and $\tilde l_a$ as
\[\label{eq:Moduli.FillingS5}
\tilde K_a=
-\frac{\sqrt{\lambda}}{8\pi^2 i}\oint_{\tilde{\contour{A}}_a}
dx \lrbrk{1-\frac{1}{x^2}}\tilde p_{\tilde k_a}(x)
=\frac{\sqrt{\lambda}}{8\pi^2 i}\oint_{\tilde{\contour{A}}_a}
\lrbrk{x+\frac{1}{x}} d\tilde p_{\tilde k_a}.
\]
Our definition uses the sheet $\tilde k_a$,
alternatively we might use $\tilde l_a$
and invert the sign.
Equivalently, we define the filling for an $AdS_5$-cut $\hat{\contour{C}}_a$,
but now using the sheet $\hat l_a$
\[\label{eq:Moduli.FillingAdS5}
\hat K_a=
-\frac{\sqrt{\lambda}}{8\pi^2 i}\oint_{\hat{\contour{A}}_a}
dx \lrbrk{1-\frac{1}{x^2}}\hat p_{\hat l_a}(x)
=\frac{\sqrt{\lambda}}{8\pi^2 i}\oint_{\hat{\contour{A}}_a}
\lrbrk{x+\frac{1}{x}} d\hat p_{\hat l_a}.
\]
The corresponding definition using the sheet $\hat k_a$
would require an opposite sign.
For completeness, we also define a filling
for fermionic singularities $x^\ast_a$
\[\label{eq:Moduli.FillingFermi}
K^\ast_a=
-\frac{\sqrt{\lambda}}{8\pi^2 i}\oint_{\contour{A}^\ast_a}
dx \lrbrk{1-\frac{1}{x^2}}\tilde p_{k^\ast_a}(x)
=\frac{\sqrt{\lambda}}{8\pi^2 i}\oint_{\contour{A}^\ast_a}
\lrbrk{x+\frac{1}{x}} d\tilde p_{k^\ast_a}.
\]
which we could also write using $\hat p_{l^\ast_a}$. It is not an
independent modulus and it measures the residue at $x^\ast_a$.

In addition to the fillings, a curve is specified by the mode numbers
\[\label{eq:Moduli.ModeNumbers}
\tilde n_a=\frac{1}{2\pi}\int_{\tilde{\contour{B}}_a}dp,\qquad
\hat n_a=\frac{1}{2\pi}\int_{\hat{\contour{B}}_a}dp,\qquad
n^\ast_a=\frac{1}{2\pi}\pint_{\contour{B}^\ast_a}dp.
\]
These are discrete parameters and therefore not count as moduli.
Note that the B-periods all start at $x=\infty$ on sheet
$\tilde k_a,\hat k_a,k^\ast_a$ and end at $x=\infty$
on sheet $\tilde l_a,\hat l_a,l^\ast_a$, respectively.
Furthermore, there is one overall winding number defined as
\[\label{eq:Moduli.Winding}
m=\frac{1}{2\pi}\int_\infty^0 \tilde p_{1,2}=-\frac{1}{2\pi}\int_\infty^0 \tilde p_{3,4}.
\]
It is defined through the $S^5$-part of the curve and there is no
corresponding quantity for the $AdS_5$-part, because there cannot
be windings in the non-compact time direction of the
universal covering of $AdS_5$
\cite{Kazakov:2004nh}.

In most cases, the fillings give the right number of moduli, but
for $m=0$ there is a constraint among the fillings as we shall see below.
Therefore, let us introduce one further modulus which we call the
\emph{length}%
\footnote{The term `length' is due to analogy with spin chains.
For an alternative approach to identifying this
conserved charge in the sigma model,
see \cite{Mikhailov:2004au,Mikhailov:2005wn}.}
\[\label{eq:Moduli.Length}
L=
\frac{\sqrt{\lambda}}{16\pi^2 i} \oint_{+1} dx\, \sum_{k=1}^4 \sheetsign_k\tilde p_k
+\frac{\sqrt{\lambda}}{16\pi^2 i}\oint_{-1} dx\, \sum_{k=1}^4 \sheetsign_k\tilde p_k
+\sum_{a=1}^{A}\frac{\sqrt{\lambda}}{8\pi^2 i}
\oint_{\contour{A}_a} \frac{dx}{x^2}\, \sum_{k=1}^4 \sheetsign_k \tilde p_k.
\]
Note that we use only half of the $2A$ cuts for the definition of length,
one from each pair related by inversion symmetry.
This definition depends on which of the two cuts we select from each pair
and is therefore ambiguous.
In a particular limit, however, this choice is
obvious as we shall see in \secref{sec:Moduli.Limit}. The length is
related to the fillings by the constraint%
\footnote{This constraint reveals the ambiguity of $L$: For some cuts
the mode numbers and fillings of the mirror cut
are related by $n_{A+a}=2m-n_a, K_{A+a}=-K_a$.
If we interchange the cut $\contour{C}_a$ with its mirror image $\contour{C}_{A+a}$,
$L$ changes by $2K_a$.}
\[\label{eq:Moduli.QuadConstr}
m L=\sum_{a=1}^{A} n_{a} K_{a}
\]
which means that
among $\set{L,K_a}$ there are only $\tilde A+\hat A$
independent continuous parameters:
$\tilde A+\hat A-1$ independent fillings $K_a$ and the length $L$.
To derive it, consider the integral
\<\label{eq:Moduli.QuadConstDer}
0
\eq
\frac{\sqrt{\lambda}}{32\pi^3 i} \oint_{\infty}dx
\sum_{k=1}^4 \bigbrk{\tilde p_k^2(x)-\hat p_k^2(x)}
\nln\eq
\frac{\sqrt{\lambda}}{32\pi^3 i} \oint_{+1}dx
\sum_{k=1}^4 \bigbrk{\tilde p_k^2(x)-\hat p_k^2(x)}
+\frac{\sqrt{\lambda}}{32\pi^3 i}
\oint_{-1}dx
\sum_{k=1}^4 \bigbrk{\tilde p_k^2(x)-\hat p_k^2(x)}
\nl
+\sum_{a=1}^{2A}
\frac{\sqrt{\lambda}}{32\pi^3 i}
\oint_{\contour{A}_a}dx
\sum_{k=1}^4 \bigbrk{\tilde p_k^2(x)-\hat p_k^2(x)}
\nln\eq
m L
-\sum_{a=1}^{A} n_{a} K_{a}.
\>
The first integral is zero due to $p(x)\sim 1/x$ at $x=\infty$.
We then split up the contour of integration around
the singularities and cuts.
To obtain the last line,
we split up the integrals around $x=\pm 1$ evenly in two
and also split up the sum $\sum_{a=1}^{2A}$ into
$\sum_{a=1}^{A}$ and $\sum_{a=A+1}^{2A}$.
Then we transform half of the integrals to $1/x$
\[\label{eq:Moduli.QuadConstrTrans}
\int_{\contour{A}_{A+a}} dx\,f(x)=
-\int_{\contour{A}_{a}} \frac{dx}{x^2}\,f(1/x)
\]
and use the inversion symmetry
\[\label{eq:Moduli.QuadConstrSym}
\sum_{k=1}^4 \bigbrk{\tilde p_k^2(1/x)-\hat p_k^2(1/x)}
=
\sum_{k=1}^4 \bigbrk{\tilde p_k^2(x)-\hat p_k^2(x)}
-4\pi m\sum_{k=1}^4 \sheetsign_k\tilde p_k(x)
+16\pi^2 m^2
\]
to transform them back.
The terms proportional to $m^2$ drop out from the integrals,
they contain no residue, while the terms multiplying $m$ sum up to $L$.
The remaining integrals around $x=\pm 1$
\[\label{eq:Moduli.QuadConstrOne}
\frac{\sqrt{\lambda}}{64\pi^3 i}
\oint_{\pm 1}dx
\lrbrk{1-\frac{1}{x^2}}
\sum_{k=1}^4 \bigbrk{\tilde p_k^2(x)-\hat p_k^2(x)}
=
0.
\]
sum up to zero as discussed in \secref{sec:Super.Sing}.
In the final step we have
employed the identity
\[\label{eq:Moduli.QuadConstrKn}
\frac{\sqrt{\lambda}}{32\pi^3 i}
\oint_{\contour{A}_{a}}
dx \lrbrk{1-\frac{1}{x^2}}
\sum_{k=1}^4 \bigbrk{\tilde p_k^2(x)-\hat p_k^2(x)}
=-n_{a} K_{a}
\]
which one obtains after pulling the contour $\contour{A}_a$ tightly
around the cut $\contour{C}_a$.
Then the integrand
$p^2(x+\epsilon)-p^2(x-\epsilon)$ can be split into
symmetric and antisymmetric parts. The antisymmetric part
is equal on two sheets up to a sign. The symmetric parts
then combine using \eqref{eq:Moduli.DiscontBos,eq:Moduli.DiscontFerm}
and yield $2\pi n_a$. The remaining integral is the filling.

A more direct way to derive the constraint uses the 
Riemann bilinear identity
\[\label{eq:Moduli.Bilinear}
\frac{1}{2\pi i}\sum_{a}
\lrbrk{\oint_{\contour{A}_a}dp\int_{\contour{B}_a}dq-
\int_{\contour{B}_a}dp \oint_{\contour{A}_a}dq}=
\frac{1}{2\pi i}\sum_a \mathrm{Res}_a (p\,dq)
\]
valid for any curve with 
a set of independent cycles $\contour{A}_a,\contour{B}_a$
and two arbitrary holomorphic differentials $dp,dq$.
Let us briefly sketch the proof:
We take as $p$ the quasi-momentum and $dq=p\,dx$
and count the $S^5$-part and $AdS_5$-parts with 
opposite signs. 
The first product of integrals will be zero due to \eqref{eq:Moduli.ACycleBos}.
According to \eqref{eq:Moduli.BCycleBos,eq:Moduli.FillingS5,eq:Moduli.FillingAdS5}
the second product of integrals leads to the sum $\sum_a n_a K_a$ over
the bosonic cuts when the symmetry is taken into account as explained above.
The sum of residues of $p^2$ yields 
the contributions from the fermions
using \eqref{eq:Moduli.BCycleFerm,eq:Moduli.FillingFermi}.
The residues from $x=\pm 1$ cancel and the term $mL$ appears
during symmetrization as above.

\subsection{Moduli of String Solutions}
\label{sec:Moduli.Discussion}

At this point we briefly summarize our results on the number of
moduli and compare it to the general solution of strings in flat space
or on plane waves. We have found one continuous modulus, the filling
$K_a$, and one discrete parameter, $n_a$, per pair of cuts (related
by inversion symmetry). Furthermore we need to specify which of the
$4|4$ sheets are connected by the cut through $k_a,l_a$. The
situation for fermionic poles is similar, only that their filling
is not an independent parameter. In addition, there is one continuous global
modulus, the length $L$, and one discrete global parameter, $m$, but
also one global constraint which relates $K_a,n_a,L,m$. Note that we
have discarded $\lambda$ which can be considered as an external
parameter.

The classification for (classical) strings in flat space or on plane
waves is similar: Consider a solution with only a finite number of
active string modes. Let us furthermore assume a light-cone gauge to
focus on the physical excitations. Then each mode is described by
its mode number ($n_a$), amplitude ($K_a$) and orientation ($k_a,l_a$)
where we have indicated in brackets the corresponding quantities in
our sigma model. The amplitudes of fermions cannot be specified by
regular numbers and thus should not be counted as continuous moduli.
One overall level matching constraint relates the amplitudes and
mode numbers ($K_a,n_a$). The string tension ($\lambda$)
will again be considered external. The only difference between
strings in flat space and out model is the lack of a modulus
describing the effective curvature ($L$) and a parameter describing
winding ($m$).

While the relation between amplitudes and fillings
as well as integers $n$ and mode numbers is
obvious, the relation between sheets and orientation of the string
needs further explanations. For cuts related to $S^5$ we see that
there are $6$ pairs of sheets and thus $6$ choices for $(\tilde
k_a,\tilde l_a)$. Similarly for $AdS^5$. Fermions have to connect
one sheet of each type and thus there are $16$ choices. It thus
seems that there are $(6+6)|16$ orientations. There is however a
further criterion which we use to distinguish cuts and poles. We
denote the cuts/poles with $\sheetsign_k\neq \sheetsign_l$ as
\emph{physical}. The cuts/poles with $\sheetsign_k=\sheetsign_l$ are
considered \emph{auxiliary}. The explanation for this classification is
that precisely the physical cuts/poles appear within the combination
$q(x)$ in \eqref{eq:Super.LocalGenerator} which is used to define
the local charges (and also the energy shift, c.f.~the following
subsection). Among the $6$ types of bosonic cuts each, there are $4$
physical and $2$ auxiliary ones. The $16$ types of fermionic poles
split up evenly into $8$ physical and $8$ auxiliary ones. Thus the
counting of orientations for physical modes, $(4+4)|8$, is as expected
for a superstring.

In conclusion we see that the moduli of admissible curves are
in one to one correspondence to the moduli describing closed
superstrings in flat space. We expect that the
number of moduli and their types should be mostly independent of
the background. The only relevant properties
for the enumeration of moduli
(open/closed, bosonic/supersymmetric, number of spacetime dimensions,
smoothness of the target space, \ldots) are the same in both theories.
We take this as compelling evidence that all admissible curves,
as discussed in this section, indeed correspond to at least one
string solution.
We thus believe that we have not missed a relevant characteristic
feature in \secref{sec:Super} for the construction of admissible curves
and that our classification is complete.%
\footnote{We only refer to the action variables of string solutions.
Of course, the (time-dependent) angle variables are not described by
the algebraic curve. According to standard lore, they correspond
to a set of marked point on the Jacobian of the curve.}

\subsection{Global Charges}
\label{sec:Moduli.Global}

Here we shall relate the global charges of $\grp{PSU}(2,2|4)$ to the
fillings. Let us concentrate on $S^5$ at first and define global
fillings
\<\label{eq:Moduli.GlobFillS5}
\tilde K_1\eq
-\sum_{a=1}^{A}
\frac{\sqrt{\lambda}}{8\pi^2 i}\oint_{\contour{A}_a} dx\lrbrk{1-\frac{1}{x^2}}
\bigbrk{\sfrac{3}{4}\tilde p_1-\sfrac{1}{4}\tilde p_2-\sfrac{1}{4}\tilde p_3-\sfrac{1}{4}\tilde p_4},
\nln
\tilde K_2\eq
-\sum_{a=1}^{A}
\frac{\sqrt{\lambda}}{8\pi^2 i}\oint_{\contour{A}_a} dx\lrbrk{1-\frac{1}{x^2}}
\bigbrk{\half\tilde p_1+\half\tilde p_2-\half\tilde p_3-\half\tilde p_4},
\nln
\tilde K_3\eq
-\sum_{a=1}^{A}
\frac{\sqrt{\lambda}}{8\pi^2 i}\oint_{\contour{A}_a} dx\lrbrk{1-\frac{1}{x^2}}
\bigbrk{\sfrac{1}{4}\tilde p_1+\sfrac{1}{4}\tilde p_2+\sfrac{1}{4}\tilde p_3-\sfrac{3}{4}\tilde p_4}.
\>
These can also be represented as a sum of fillings $\tilde K_a$
of the individual cuts and residues $K^\ast_a$ of fermionic poles.
We will not do this explicitly, as there are too many pairs of sheets
and thus too many types of cuts.
The Dynkin labels $[\tilde r_1,\tilde r_2,\tilde r_3]$ of $\grp{SU}(4)$
are given by the following combinations
\[\label{eq:Moduli.DynkinS5Def}
\tilde r_j=
\frac{\sqrt{\lambda}}{8\pi^2 i}\oint_\infty dx\, \bigbrk{\tilde p_j(x)-\tilde p_{j+1}(x)}.
\]
Their relation to the global fillings is as follows
%
\[\label{eq:Moduli.DynkinS5}
\begin{array}{rclcrcl}
\tilde r_1\eq \tilde K_2-2\tilde K_1,&&
\tilde K_1\eq \half L-\sfrac{3}{4}\tilde r_1-\half\tilde r_2-\sfrac{1}{4}\tilde r_3,
\\[0.7ex]
\tilde r_2\eq L-2\tilde K_2+\tilde K_1+\tilde K_3,&&
\tilde K_2\eq \phantom{\half}L-\sfrac{1}{2}\tilde r_1-\phantom{\half}\tilde r_2-\sfrac{1}{2}\tilde r_3,
\\[0.7ex]
\tilde r_3\eq \tilde K_2-2\tilde K_3,&&
\tilde K_3\eq \half L-\sfrac{1}{4}\tilde r_1-\half\tilde r_2-\sfrac{3}{4}\tilde r_3.
\end{array}
\]
To derive these, it is convenient to make use of the inversion symmetry,
c.f.~the previous subsection.

For $AdS_5$ the results are very similar. Again we define the global fillings
\<\label{eq:Moduli.GlobFillAdS5}
\hat K_1\eq
\sum_{a=1}^{A}
\frac{\sqrt{\lambda}}{8\pi^2 i}\oint_{\contour{A}_a} dx\lrbrk{1-\frac{1}{x^2}}
\bigbrk{\sfrac{3}{4}\hat p_1-\sfrac{1}{4}\hat p_2-\sfrac{1}{4}\hat p_3-\sfrac{1}{4}\hat p_4},
\nln
\hat K_2\eq
\sum_{a=1}^{A}
\frac{\sqrt{\lambda}}{8\pi^2 i}\oint_{\contour{A}_a} dx\lrbrk{1-\frac{1}{x^2}}
\bigbrk{\half\hat p_1+\half\hat p_2-\half\hat p_3-\half\hat p_4},
\nln
\hat K_3\eq
\sum_{a=1}^{A}
\frac{\sqrt{\lambda}}{8\pi^2 i}\oint_{\contour{A}_a} dx\lrbrk{1-\frac{1}{x^2}}
\bigbrk{\sfrac{1}{4}\hat p_1+\sfrac{1}{4}\hat p_2+\sfrac{1}{4}\hat p_3-\sfrac{3}{4}\hat p_4},
\>
which we might write as sums of the individual fillings.
Then the Dynkin labels are given by
\[\label{eq:Moduli.DynkinAdS5Def}
\hat r_j=
\frac{\sqrt{\lambda}}{8\pi^2 i}\oint_\infty dx\, \bigbrk{\hat p_{j+1}(x)-\hat p_j(x)}.
\]
and related to the global fillings by
%
\[\label{eq:Moduli.DynkinAdS5}
\begin{array}{rclcrcl}
\hat r_1\eq \hat K_2-2\hat K_1,&&
\hat K_1\eq -\half L-\half\delta E-\sfrac{3}{4}\hat r_1-\half\hat r_2-\sfrac{1}{4}\hat r_3,
\\[0.7ex]
\hat r_2\eq -L-\delta E-2\hat K_2+\hat K_1+\hat K_3,&&
\hat K_2\eq -\phantom{\half}L-\phantom{\half}\delta E-\sfrac{1}{2}\hat r_1-\phantom{\half}\hat r_2-\sfrac{1}{2}\hat r_3,
\\[0.7ex]
\hat r_3\eq \hat K_2-2\hat K_3,&&
\hat K_3\eq -\half L-\half\delta E-\sfrac{1}{4}\hat r_1-\half\hat r_2-\sfrac{3}{4}\hat r_3.
\end{array}
\]
Here we have introduced a new quantity $\delta E$, the \emph{energy shift}
\[\label{eq:Moduli.EnergyShift}
\delta E=
\sum_{a=1}^{A}
\frac{\sqrt{\lambda}}{8\pi^2 i}
\oint_{\contour{A}_a} \frac{dx}{x^2}
\sum_{k=1}^4
\bigbrk{-\sheetsign_k\tilde p_k+\sheetsign_k\hat p_k}
=
-\sum_{a=1}^{A}
\frac{\sqrt{\lambda}}{4\pi^2 i}
\oint_{\contour{A}_a} \frac{dx}{x^2}\,q(x)
\]
with $q(x)$ defined in \eqref{eq:Super.LocalGenerator}.
When we write $\hat r_2$ in terms of the $AdS_5$ energy $E$
\[\label{eq:Moduli.Energy}
E=-r_2-\half r_1-\half r_3=L+\hat K_2+\delta E
\]
we see that $\delta E$ is indeed the energy shift
when $L+\hat K_2$ is interpreted as the bare energy. Finally, we
introduce the global fermionic filling
\[\label{eq:Moduli.GlobFillFermi}
K^\ast=-
\sum_{a=1}^{A}
\frac{\sqrt{\lambda}}{8\pi^2 i}
\oint_{\contour{A}_a} dx\lrbrk{1-\frac{1}{x^2}}
\sum_{k=1}^4
\bigbrk{\half\tilde p_k+\half\hat p_k}.
\]
It is related to the hypercharge eigenvalue $r^\ast$
\[\label{eq:Moduli.Hypercharge}
r^\ast=\frac{\sqrt{\lambda}}{8\pi^2 i}\oint_{\infty}dx
\sum_{k=1}^4 \bigbrk{\half\tilde p_k(x)+\half\hat p_k(x)}
=
2B-K^\ast.
\]
We have introduced a charge $B$
which is related to the Lagrange multiplier,
see \secref{sec:Super.Central}.
Under the symmetry it transforms as
$B\mapsto B+\half\upsilon\sqrt{\lambda}$.

\subsection{Superstrings on $AdS_3\times S^3$}
\label{sec:Moduli.AdS3S3}

Let us consider solutions of the supersymmetric $AdS_5\times S^5$
sigma model which extend only over a supersymmetric $AdS_3\times S^3$ subspace, 
which in fact is given by the group manifold $\grp{PSU}(1,1|2)$.
For this class of solutions, the algebraic curve
will split into two disconnected parts. 
The first component consists of $\tilde p_2,\tilde p_3$ 
and $\hat p_1,\hat p_4$ and the other
component consists of the remaining four sheets. There are no branch cuts or
fermionic poles connecting the two parts. Both components are
isomorphic to algebraic curves obtained from the $\grp{PSU}(1,1|2)$
sigma model \cite{Berkovits:1999im,Dolan:1999dc,Metsaev:2000mv}. 
One of them corresponds to the monodromy in the
fundamental representation, the other one to the monodromy in the
antifundamental representation. These two curves are not unrelated,
for a sigma model on a group manifold they should map into each
other under inversion $x\mapsto 1/x$. Indeed, this is precisely what
the $AdS_5\times S^5$ sigma model implies, see
\secref{sec:Super.Sym}. There are several conceptual differences
which make it interesting to consider the $AdS_3\times S^3$ model
separately.

First of all, the $AdS_3\times S^3$ model
leads to one algebraic curve without inversion symmetry
(or, equivalently, two related algebraic curves) whereas the
full $AdS_5\times S^5$ model has only one self-symmetric curve.
This also means that we can distinguish between a cut
and its image under inversion: They reside on different
components of the algebraic curve and we shall consider only
one component. The definitions of length \eqref{eq:Moduli.Length}
and energy shift \eqref{eq:Moduli.EnergyShift}
thus become natural and unambiguous. In fact they become two
of the global charges. Together with the spin $S$ on
$AdS_3$ and spin $J_2$ on $S^3$ they are the four
charge eigenvalues of the isometry group
$\grp{PSU}(1,1|2)\times\grp{PSU}(1,1|2)$.
Again we can express the global charges through the fillings
\<\label{eq:Moduli.AdS3S3Filling}
J_2=\tilde K\eq
-\sum_{a=1}^{A}
\frac{\sqrt{\lambda}}{8\pi^2 i}\oint_{\contour{A}_a} dx\lrbrk{1-\frac{1}{x^2}}
\bigbrk{\half\tilde p_2-\half\tilde p_3},
\nln
S=\hat K\eq
\sum_{a=1}^{A}
\frac{\sqrt{\lambda}}{8\pi^2 i}\oint_{\contour{A}_a} dx\lrbrk{1-\frac{1}{x^2}}
\bigbrk{\half\hat p_1-\half\hat p_4},
\>
Here we have defined $J_2$ and $S$ to match the fillings.
The expansion of the quasi-momentum at $x=\infty$
is related to the charges of one of the global $\grp{PSU}(1,1|2)$ factors
\<\label{eq:Moduli.AdS3S3Infty}
\frac{\sqrt{\lambda}}{8\pi^2 i}\oint_\infty dx\, \bigbrk{\tilde p_2-\tilde p_3}
\eq L-2\tilde K=J_1-J_2,
\nln
\frac{\sqrt{\lambda}}{8\pi^2 i}\oint_\infty dx\, \bigbrk{\hat p_4-\hat p_1}
\eq -L-\delta E-2\hat K=-E-S.
\>
For convenience,
we have defined the spin $J_1$ and energy $E$ to replace
the length and energy shift as follow
\[\label{eq:Moduli.AdS3S3ChargeRel}
L=J_1+J_2,\qquad \delta E=E-L-S.
\]
The expansion at $x=0$ relates to the charges of the other global $\grp{PSU}(1,1|2)$ factor
\<\label{eq:Moduli.AdS3S3Zero}
\frac{\sqrt{\lambda}}{8\pi^2 i}\oint_0 \frac{dx}{x^2}\, \bigbrk{\tilde p_2-\tilde p_3}
\eq L=J_1+J_2,
\nln
\frac{\sqrt{\lambda}}{8\pi^2 i}\oint_0 \frac{dx}{x^2}\, \bigbrk{\hat p_4-\hat p_1}
\eq -L-\delta E=-E+S.
\>
These expressions agree precisely with the rank-one subsectors
of $\Real\times S^3$ and $AdS_3\times S^1$ considered in
\cite{Kazakov:2004qf,Kazakov:2004nh}.

\subsection{The Frolov-Tseytlin Limit}
\label{sec:Moduli.Limit}

In this section we shall discuss the
Frolov-Tseytlin limit $L/\sqrt{\lambda}\to\infty$ of the curve.%
\footnote{At the level of the action and the Hamiltonian,
this limit was studied in \cite{Kruczenski:2003gt,Kruczenski:2004kw}.}
In this limit, half of the cuts and poles approach $x=\infty$ and
half of them approach $x=0$.%
\footnote{Solutions with self-symmetric cuts
do not have a proper Frolov-Tseytlin limit.}
Let us label those cuts and poles which escape to
$x=\infty$ by $a=1,\ldots,A$,
those which approach $x=0$ will be labelled by
$a=A+1,\ldots,2A$.
Therefore, there is a natural choice for those
cuts which contribute to the definition of
the length in \eqref{eq:Moduli.Length}.

\begin{figure}\centering
\includegraphics{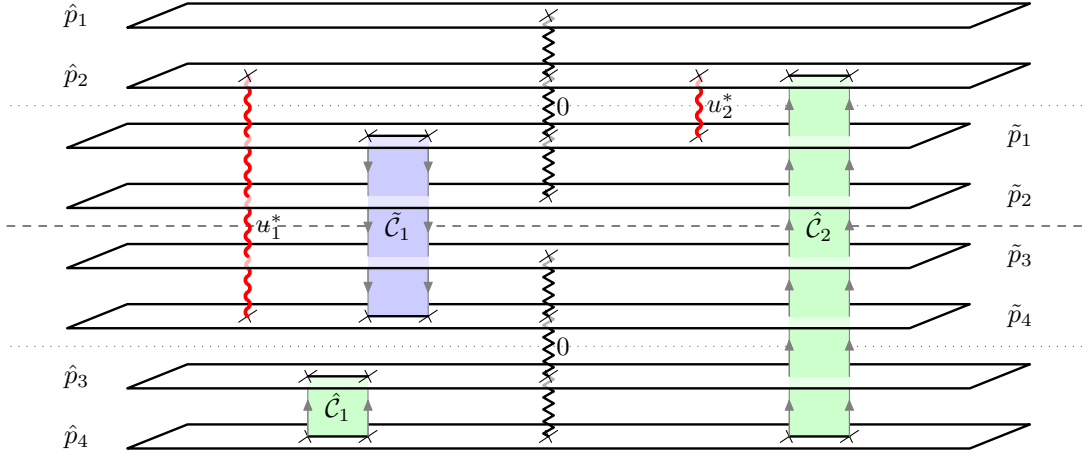}
\caption{The Frolov-Tseytlin limit of the configuration of cuts and poles in
\protect\figref{fig:sheetssigma}.
All inverse cuts and poles
as well as the poles at $x=\pm 1$ have
been scaled to $u=0$ and absorbed into an effective pole.}
\label{fig:sheetslimit}
\end{figure}

We will define for convenience a rescaled variable%
\footnote{This relationship needs to be refined at higher
orders in $L/\sqrt{\lambda}$ \cite{Kazakov:2004qf,Beisert:2004hm}.}
\[\label{sec:Moduli.FTRescale}
u=\frac{\sqrt{\lambda}}{4\pi}\,x.
\]
In other words, in the $u$-plane all cuts and poles
with $a=1,\ldots,A$ approach finite values of $u$
while their images approach $u=0$.
Similarly, the poles at $x=\pm 1$ approach $u=0$,
see \figref{fig:sheetslimit}.
This means that the point $u=0$ is special and
$p(u)$ near $u=0$ is not directly
related to $p(x)$ near $x=0$, but there
are contributions from the cuts and poles.
To understand the expansion of $p(u)$ at $u=0$
we shall define a contour $\contour{C}$ in the $x$-plane which
encircles the poles at $x=\pm 1$ and all the cuts and poles
with $a=A+1,\ldots,2A$.
Equivalently, this may be considered a contour which excludes
$x=\infty$ and all the cuts and poles with $a=1,\ldots,A$.
After rescaling $\contour{C}$ merely encircles the point $u=0$
in the $u$-plane which can be used to obtain the expansion
of $p(u)$ according to the formula%
\footnote{This formula explains why there are two different
transfer matrices $T(x),\bar T(u)$ in \cite{Beisert:2004hm}.
The transfer matrix $\bar T(u)$ is the suitable
one for finite $g$, while $T(x)$ is the effective one according
to this formula.
It should also be useful to understand
the relationship between conserved local charges in
string theory and gauge theory in \cite{Arutyunov:2003rg,Arutyunov:2004xy}.}
\[\label{eq:Moduli.FTExtraction}
\frac{\partial^{r-1} p_k}{\partial u^{r-1}}(0)
=\lrbrk{\frac{4\pi L}{\sqrt{\lambda}}}^{r-1}\frac{1}{2\pi i}
\oint_{\contour{C}} \frac{dx}{x^{r}}\,p_k(x).
\]

Using the identities and definitions in \secref{sec:Moduli.Fillings,sec:Moduli.Global}
we find the singular behavior of all sheets at $u=0$
\<\label{eq:Moduli.FTExtractSing}
\frac{1}{2\pi i}\oint_{\contour{C}} \tilde p_k(x)\,dx
\eq
\frac{2\pi}{\sqrt{\lambda}}
\bigbrk{B+\sheetsign_k L}
-\sum_{a=1}^{A}
\frac{1}{2\pi i}\oint_{\contour{A}_a}
\frac{dx}{x^{2}}\,\tilde p_k(x),
\nln
\frac{1}{2\pi i}\oint_{\contour{C}}\hat p_k(x)\,dx
\eq
\frac{2\pi}{\sqrt{\lambda}}\bigbrk{B+\sheetsign_k L+\sheetsign_k\delta E}
-\sum_{a=1}^{A}
\frac{1}{2\pi i}\oint_{\contour{A}_a}
\frac{dx}{x^{2}}\,\hat p_k(x)
\>
and the first few moments of the generator of local charges
\eqref{eq:Super.LocalGenerator}
\<\label{eq:Moduli.FTExtractLocal}
\frac{1}{2\pi i}\oint_{\contour{C}} dx\,q(x)\eq
-\frac{2\pi\,\delta E}{\sqrt{\lambda}}\,,
\nln
\frac{1}{2\pi i}\oint_{\contour{C}} \frac{dx}{x}\,q(x)\eq
2\pi m,
\nln
\frac{1}{2\pi i}\oint_{\contour{C}} \frac{dx}{x^2}\,q(x)\eq
\frac{2\pi\,\delta E}{\sqrt{\lambda}}\,.
\>
Now we note that the energy shift \eqref{eq:Moduli.EnergyShift}
\[\label{eq:Moduli.FTEnergyShift}
\delta E=
-\sum_{a=1}^{A}
\frac{\sqrt{\lambda}}{4\pi^2 i}
\oint_{\contour{A}_a} \frac{dx}{x^2}\,q(x)
=
-\frac{\lambda}{8\pi^2 L}\sum_{a=1}^{A}
\frac{1}{2\pi i}
\oint_{\contour{A}_a} \frac{du}{u^2}\,q(u).
\]
is of order $\order{\lambda/L}$.
After rescaling we obtain the singular behavior of $p(u)$ at $u=0$
from \eqref{eq:Moduli.FTExtraction,eq:Moduli.FTExtractSing}
\[\label{eq:Moduli.FTSingular}
\tilde p_k(u)\sim\hat p_k(u)\sim
\frac{1}{u}\lrbrk{\frac{\sheetsign_k}{2}+\frac{B}{2L}}
\]
and the first few local charges from \eqref{eq:Moduli.FTExtraction,eq:Moduli.FTExtractLocal}
\[\label{eq:Moduli.FTCharges}
q(u)=
2\pi m+
\frac{8\pi^2 L}{\lambda}\,\delta E\, u+\order{u^2}.
\]
In particular, the momentum constraint is
$q(0)=2\pi\Integers$
while the individual sheets $p_k(u)$ no longer have a fixed
finite value at $u=0$.

The above curve apparently is the spectral curve
of the supersymmetric Landau-Lifshitz 
model in \cite{Mikhailov:2004xw}.%
\footnote{We thank A.~Mikhailov for discussions on this point.}
This model is related to 
the coherent state approach to gauge theory 
\cite{Kruczenski:2003gt,Kruczenski:2004kw}.
Unlike the curve of the full superstring, 
this curve has only one singular point at $u=0$
and thus seems to be similar to the one of the
classical Heisenberg magnet discussed in \cite{Kazakov:2004qf}.
We expect the expansion of the function $q(u)$ around $u=0$
to yield the local charges of the model, while the point
$u=\infty$ should be related to Noether and multi-local
charges.



\section{Integral Representation of the Sigma Model}
\label{sec:Integral}

We can reformulate the algebraic curve in terms
of a Riemann-Hilbert problem,
i.e.~as integral equations on some density functions.
This formulation is similar to the thermodynamic limit
of the Bethe equations for the gauge theory counterpart.
It is thus suited well for a comparison of both theories,
especially at higher loops (once the gauge theory results
become available).
We start by representing
the various discontinuities of cuts of the algebraic curve by
integrals over densities. We then match this representation
to the properties derived earlier and thus fix several of
the parameters. The remaining properties lead to equations
which are of the same nature as the Bethe equations in
the thermodynamic limit.
In order to be more explicit,
we specify the equations for a number of subsectors,
while full equations are written out only
in \appref{sec:Beauty}.
Finally, we compare to one-loop gauge theory and find complete
agreement.

\subsection{Parametrization of the Quasi-Momentum}
\label{sec:Integral.Integral}

The quasi-momentum $p(x)$ is a function with two sets of four sheets
$\tilde p_k(x)$ and $\hat p_k(x)$, $k=1,2,3,4$.
These are analytic functions of $x$ except at
the singular points $x=\pm1$,
a set of branch cuts and additional fermionic poles.

\paragraph{Ansatz.}

Let us construct a generic ansatz for $p(x)$: It is straightforward
to incorporate the poles at $x=\pm 1$ with undetermined residues
$\tilde a_k^\pm$ and $\hat a_k^\pm$.
The branch cuts and fermionic poles will be contained in
several resolvents $G(x)$. All the resolvents $G(x)$ will be
defined to vanish at $x=\infty$, consequently we should add
an undetermined constant $\tilde b_k,\hat b_k$
for each sheet.

The branch cuts connect two sheets of either $\tilde p(x)$ or $\hat p(x)$.
The discontinuity of the branch cuts between sheets $k$ and $l$ is
contained in the resolvent $\tilde G_{kl}(x)$ or $\hat G_{kl}(x)$.
As the graded sum of all sheets $p_k$ should be zero,
the sum of discontinuities must cancel. In other words,
$\tilde G_{kl}$ and $\hat G_{kl}$ must be antisymmetric in $k,l$.
A fermionic pole appears on a sheet $\tilde p_k$ and a sheet $\hat p_l$
with the same residue for both sheets. They are contained in the
resolvent $G^\ast_{kl}(x)$.
We shall include only the cuts/poles with $a=1,\ldots,A$
in the resolvent $G(x)$. Their images with $a=A+1,\ldots,2A$ under the inversion symmetry
\eqref{eq:Moduli.Sym} will be incorporated by $G(1/x)$.
This leaves some discrete arbitrariness of which cuts/poles belong to
$G(x)$ and which to $G(1/x)$. Note that although the algebraic curve is
invariant under such permutations, our interpretation
of some quantities will have to change.

Taking the above constraints into account, we arrive at the following ansatz
\<\label{eq:Integral.Ansatz}
\tilde p_k(x)\eq
\sum_{l=1}^4
\lrbrk{\tilde G_{kl}(x)-\tilde G_{k'l}(1/x)
+G^\ast_{kl}(x)-G^\ast_{k'l}(1/x)}
+\frac{\tilde a^+_k}{x-1}
+\frac{\tilde a^-_k}{x+1}
+\tilde b_k
,
\nln
\hat p_k(x)\eq
\sum_{l=1}^4
\lrbrk{\hat G_{lk}(x)-\hat G_{lk'}(1/x)
+G^\ast_{lk}(x)-G^\ast_{lk'}(1/x)}
+\frac{\hat a^+_k}{x-1}
+\frac{\hat a^-_k}{x+1}
+\hat b_k.\qquad
\>
where the permutation $k'$ of sheets is defined in \eqref{eq:Super.Perm}.
We will now determine the constants
using the known properties of $p(x)$.

\paragraph{Resolvents.}

The bosonic resolvents $\tilde G(x),\hat G(x)$
are defined in terms of the densities $\tilde\rho(x),\hat\rho(x)$
as follows
\[\label{eq:Integral.Resolvent}
\tilde G_{kl}(x)=\int_{\tilde{\contour{C}}_{kl}}\frac{dy\,\tilde\rho_{kl}(y)}{1-1/y^2}\,\frac{1}{y-x}\,,\qquad
\hat G_{kl}(x)=\int_{\hat{\contour{C}}_{kl}}\frac{dy\,\hat\rho_{kl}(y)}{1-1/y^2}\,\frac{1}{y-x}\,.
\]
The fermionic resolvent $G^\ast_{kl}(x)$ is given by a set
of poles%
\footnote{Strictly speaking the residues must
be nilpotent numbers because they represent 
a product of two Grassmann odd numbers, 
but we can mostly ignore this fact.}
\[\label{eq:Integral.FermiRes}
G^\ast_{kl}(x)=\sum_{a=1}^{A^\ast_{kl}}
\frac{\alpha^\ast_{kl,a}}{1-1/x^{\ast\,2}_{kl,a}}\,\frac{1}{x^\ast_{kl,a}-x}\,.
\]
All the resolvents vanish at $x=\infty$ and are analytic functions
except on the curves $\contour{C}_a$ or at the poles $x^\ast_a$.
They are obviously single-valued, i.e.~the cycles of $dG$ around
cuts/poles vanish.
The filling of a cut \eqref{eq:Moduli.FillingS5,eq:Moduli.FillingAdS5,eq:Moduli.FillingFermi}
is given by
\[\label{eq:Integral.ResFilling}
\tilde K_{kl,a}=
\frac{\sqrt{\lambda}}{4\pi}\int_{\tilde{\contour{C}}_{kl,a}}dy\,\tilde\rho_{kl}(y),
\qquad
\hat K_{kl,a}=
\frac{\sqrt{\lambda}}{4\pi}\int_{\hat{\contour{C}}_{kl,a}}dy\,\hat\rho_{kl}(y),
\qquad
K^\ast_{kl,a}=\frac{\sqrt{\lambda}}{4\pi}\,\alpha^\ast_{kl,a}.
\]
%

\paragraph{Singularities.}

{}From \secref{sec:Super.Sing} we know
the general structure of the
residues at $x=\pm 1$. The residues $\tilde a^\pm_k$
for $S^5$ are linked to the residues $\hat a^\pm_k$
for $AdS_5$. Furthermore, all residues are paired.
We can thus write them in terms
of four independent constants
$c_{1,2},d_{1,2}$ using
\eqref{eq:Super.Epsi}
\[\label{eq:Integral.Singularity}
\tilde a^\pm_k=\hat a^\pm_k=:
\half c_1\pm \half c_2+\half d_1\sheetsign_k\pm \half d_2\sheetsign_k.
\]

\paragraph{Asymptotics.}

The asymptotics $p(x)\sim 1/x$ at $x=\infty$ fixes all the
constants $b$
\[\label{eq:Integral.Asympotics}
\tilde b_k=
\sum_{l=1}^4
\lrbrk{\tilde G_{k'l}(0)+G^\ast_{k'l}(0)},
\qquad
\hat b_k=
\sum_{l=1}^4
\lrbrk{\hat G_{lk'}(0)+G^\ast_{lk'}(0)}.
\]

\paragraph{Unimodularity.}

The graded sum
\[\label{eq:Integral.Unimodular}
\sum_{k=1}^4 \bigbrk{\tilde p_k(x)-\hat p_k(x)}=0
\]
of all sheets indeed vanishes trivially
by antisymmetry of $\tilde G_{kl},\hat G_{kl}$ in $k,l$.

\paragraph{Symmetry.}

The inversion symmetry leads to the following expressions
\<\label{eq:Integral.SymConstraint}
\tilde p_{k}(1/x)\eq
-\tilde p_{k'}(x)
-\tilde a^+_k
+\tilde a^-_k
+\tilde b_{k}
+\tilde b_{k'}
\stackrel{!}{=}
-\tilde p_{k'}(x)
+2\pi \sheetsign_k m
,
\nln
\tilde p_{k}(1/x)\eq
-\hat p_{k'}(x)
-\hat a^+_k
+\hat a^-_k
+\hat b_{k}
+\hat b_{k'}
\stackrel{!}{=}
-\hat p_{k'}(x)
\>
with the permutation $k'$ of sheets is defined in \eqref{eq:Super.Perm}.
When we substitute the above expressions for $\tilde a^\pm_k,\hat a^\pm_k$, we find
\[\label{eq:Integral.SymConstant}
c_2=
\half \sum_{k,l=1}^4 G^\ast_{kl}(0),\qquad
d_2=\half \sum_{k,l=1}^4 \sheetsign_k
\bigbrk{\hat G_{lk}(0)+G^\ast_{lk}(0)}
\]
as well as the (momentum) constraint
\[\label{eq:Integral.MomConstraint}
\half\sum_{k,l=1}^4 \sheetsign_k
\bigbrk{\tilde G_{kl}(0)+\hat G_{kl}(0)+G^\ast_{kl}(0)-G^\ast_{lk}(0)}
=2\pi m.
\]

\paragraph{Length.}
We substitute $\tilde p_k$ in the definition of length and obtain
\[\label{eq:Integral.Length}
L=
\frac{\sqrt{\lambda}}{8\pi}\sum_{k=1}^4 \sheetsign_k (\tilde a^+_k +\tilde a^-_k)
-\frac{\sqrt{\lambda}}{4\pi}
\sum_{k,l=1}^4 \sheetsign_k
\bigbrk{\tilde G'_{kl}(0)+G^{\ast\prime}_{kl}(0)}
\]
This leads to
\[\label{eq:Integral.LengthConstant}
d_1=
\frac{2\pi\, L}{\sqrt{\lambda}}
+\half \sum_{k,l=1}^4 \sheetsign_k \bigbrk{\tilde G'_{kl}(0)+G^{\ast\prime}_{kl}(0)}.
\]
Note that we have assumed that all cuts $a=1,\ldots,\tilde A$
are captured by $G(x)$
while the cuts $a=\tilde A+1,\ldots,2\tilde A$
are captured by $G(1/x)$.

\paragraph{Hypercharge.}

The remaining constant $c_1$ corresponds to a
shift of the Lagrange multiplier and is irrelevant.
We shall express it by means of the
hypercharge $B$ defined in \eqref{eq:Moduli.Hypercharge}
\[\label{eq:Integral.Hypercharge}
c_1=\half \sum_{k,l=1}^4 G^{\ast\prime}_{kl}(0)+\frac{2\pi\,B}{\sqrt{\lambda}}\,.
\]

\paragraph{Energy Shift.}
We substitute $p_k$ in the definition of energy shift and obtain
\[\label{eq:Integral.EnergyShift}
\delta E=
\frac{\sqrt{\lambda}}{4\pi}
\sum_{k,l=1}^4\sheetsign_k \bigbrk{\tilde G'_{kl}(0)+G^{\ast\prime}_{kl}(0)-\hat G'_{lk}(0)-G^{\ast\prime}_{lk}(0)}.
\]

\subsection{Integral Equations}
\label{sec:Integral.Bethe}

Let us now assemble and simplify the various findings
of the previous section. As a first step, we write
$G(x)$, but not $G(1/x)$, $G(0)$ or $G'(0)$,
in terms of the inversion-symmetric function
\[\label{eq:Integral.SymResolv}
H_{kl}(x):=G_{kl}(x)+G_{kl}(1/x)-G_{kl}(0).
\]
The terms in the integrand of $H$ combine as follows
\[\label{eq:Integral.PoleCombine}
\frac{1}{1-1/y^2}\lrbrk{\frac{1}{y-x}+\frac{1}{y-1/x}-\frac{1}{y}}
=
\frac{1}{(y+1/y)-(x+1/x)}
\]
which means that we can write $H$ as
\[\label{eq:Integral.SymResolvDens}
H(x):=\int_{\contour{C}}\frac{du\,\rho(u)}{u-(x+1/x)}\,,
\]
where $\rho$ transforms as a density, $dx\,\rho(x)=du\,\rho(u)$,
under the map
\[\label{eq:Integral.xuMap}
u(x)=x+1/x,\qquad
x(u)=\half u+\half u\sqrt{1-4/u^2}\,.
\]
The conversion of $G(x)$ to $H(x)$
creates a few more instances of $G_{kl}(1/x)$
which turn out to pair up with
existing instances of $G_{k'l}(1/x)$ in all cases.
These can be rewritten by applying
\[\label{eq:Integral.SumSheetSym}
f_{k}+f_{k'}
=
\half\sum_{l=1}^4 f_l
+\half \sheetsign_k\sum_{l=1}^4 \sheetsign_l f_l
\]
This identity for any $f_k$ can easily be verified by evaluating it
for all possible values $k=1,2,3,4$.
It is convenient to introduce the following combinations
of resolvents
\<\label{eq:Integral.ResolvSum}
\tilde G\indup{sum}(x)\eq
\half \sum_{k,l=1}^4 \sheetsign_k \bigbrk{\tilde G_{kl}(x)+G^\ast_{kl}(x)},
\nln
\hat G\indup{sum}(x)\eq
\half \sum_{k,l=1}^4 \sheetsign_k \bigbrk{\hat G_{lk}(x)+G^\ast_{lk}(x)},
\nln
G^\ast\indup{sum}(x)\eq
\half \sum_{k,l=1}^4 G^\ast_{kl}(x),
\nln
G\indup{mom}(x)\eq
\tilde G\indup{sum}(x)-
\hat G\indup{sum}(x)
\>
We can now write down the simplified quasi-momentum
\<\label{eq:Integral.SheetsSimp}
\tilde p_k(x)\eq
\sum_{l=1}^4
\bigbrk{\tilde H_{kl}(x)+H^\ast_{kl}(x)}
+\sheetsign_k \tilde F(x)+F^\ast(x),
\nln
\hat p_k(x)\eq
\sum_{l=1}^4
\bigbrk{\hat H_{lk}(x)+H^\ast_{lk}(x)}
+\sheetsign_k \hat F(x)+F^\ast(x).
\>
All the terms which do not follow the regular
pattern of resolvents $H$ could be absorbed into
three potentials $F$
\<\label{eq:Integral.Potentials}
\tilde F(x)\eq
\lrbrk{\frac{2\pi\, L}{\sqrt{\lambda}}+\tilde G'\indup{sum}(0)}
\frac{1/x}{1-1/x^2}
+\frac{\hat G\indup{sum}(0)}{1-1/x^2}
-\tilde G\indup{sum}(1/x)
+G\indup{mom}(0)
,
\nln
\hat F(x)\eq
\lrbrk{\frac{2\pi\, L}{\sqrt{\lambda}}+\tilde G'\indup{sum}(0)}
\frac{1/x}{1-1/x^2}
+\frac{\hat G\indup{sum}(0)}{1-1/x^2}
-\hat G\indup{sum}(1/x),
\nln
F^\ast(x)\eq
\lrbrk{\frac{2\pi\,B}{\sqrt{\lambda}}+G^{\ast\prime}\indup{sum}(0)}\frac{1/x}{1-1/x^2}
+\frac{G^\ast\indup{sum}(0)}{1-1/x^2}
-G^\ast\indup{sum}(1/x)
\,.
\>
It might be useful to note the transformation of the potentials $F$
under the symmetry%
\footnote{The summed resolvents $H\indup{sum}$ are defined
in analogy to \eqref{eq:Integral.ResolvSum}.}
\<\label{eq:Integral.PotentialSym}
\tilde F(1/x)\eq
-\tilde F(x)
-\tilde H\indup{sum}(x)
+G\indup{mom}(0),
\nln
\hat F(1/x)\eq
-\hat F(x)
-\hat H\indup{sum}(x),
\nln
F^\ast(1/x)\eq
-F^\ast(x)
-H^\ast\indup{sum}(x).
\>

The integral equations
\eqref{eq:Moduli.DiscontBos,eq:Moduli.DiscontFerm}
enforcing integrality of the B-periods
\eqref{eq:Moduli.BCycleBos,eq:Moduli.BCycleFerm}
read
\<\label{eq:Integral.Bethe}
\sheetsl[\tilde]_l(x)-\sheetsl[\tilde]_k(x)\eq
2\pi \tilde n_{kl,a}
\quad\mbox{for }
x\in \tilde{\contour{C}}_{kl,a},
\nln
\sheetsl[\hat]_l(x)-\sheetsl[\hat]_k(x)\eq
2\pi \hat n_{kl,a}
\quad\mbox{for }
x\in \hat{\contour{C}}_{kl,a},
\nln
\sheetsl[\hat]_l(x)-\sheetsl[\tilde]_k(x)\eq
2\pi n^\ast_{kl,a}
\quad\mbox{for }
x= x^\ast_{kl,a}.
\>
These equations must be supplemented by the momentum constraint
\eqref{eq:Integral.MomConstraint}
\[\label{eq:Integral.Momentum}
G\indup{mom}(0)
=2\pi m.
\]
Note that the potential can appear in various combinations depending
on the type of cut/pole. Let us denote those cuts/poles
with $\sheetsign_k\neq \sheetsign_l$ as \emph{physical},
the others are considered as \emph{auxiliary}.
The physical cuts are precisely the ones
that contribute to $G\indup{mom}$
which in turn contains the total momentum for
the momentum constraint \eqref{eq:Integral.MomConstraint}
and the energy shift \eqref{eq:Integral.EnergyShift}.
They connect sheets $1,2$ to sheet $3,4$ of either type.
A physical cut is subject to the potential $2\tilde F$ or $2\hat F$
depending on whether it is of $S^5$-type or $AdS_5$-type.
For an auxiliary bosonic cut there is no effective potential.
For physical fermionic poles we get the potential $\tilde F+\hat F$
and $\tilde F-\hat F$ for auxiliary fermions.

The global charges are found at $x=\infty$,
they are determined through the fillings $K$,
the length $L$ and the energy shift $\delta E$:
The expansion of $H$ gives the total filling $K$ of the
cuts in $H$
\[\label{eq:Integral.GlobalFilling}
H(x)=-\frac{1}{x}\sum_{a=1}^{A} \frac{4\pi\, K_a}{\sqrt{\lambda}}
+\order{1/x^2}
=-\frac{1}{x}\frac{4\pi\, K}{\sqrt{\lambda}}+\order{1/x^2}.
\]
The expansion of $F$ provides the length
and the energy shift \eqref{eq:Integral.EnergyShift}
\[\label{eq:Integral.GlobalPotential}
\tilde F(x)=
\frac{1}{x}\,\frac{2\pi\, L}{\sqrt{\lambda}}
+\order{1/x^2},\qquad
\hat F(x)=
\frac{1}{x}\,\frac{2\pi\, (L+\delta E)}{\sqrt{\lambda}}
+\order{1/x^2}.
\]
The energy shift is given by
\[\label{eq:Integral.EnergySimp}
\delta E=
\frac{\sqrt{\lambda}}{2\pi}\,G'\indup{mom}(0).
\]
%

\subsection{Rank-One Sectors}
\label{sec:Integral.Rank1}

We will now investigate the cases when only one
of the physical resolvents is non-zero.
The final result depends on the type of resolvent,
$\tilde G$, $\hat G$ or $G^\ast$.

\paragraph{Bosonic, Compact.}

We turn on only $G=\tilde G_{23}$ and consider
one quasi-momentum $p=\tilde p_2=-\tilde p_3$.
This reduces to the case of strings on $\Real\times S^3$
investigated in \cite{Kazakov:2004qf}
\[\label{eq:Integral.Rank1RS3}
2\sheetsl(x)=+2\resolvHsl(x)+2\tilde F(x)
=
2\resolvsl(x)
+\frac{2G'(0)/x}{1-1/x^2}
+\frac{4\pi L}{\sqrt{\lambda}}\,\frac{1/x}{1-1/x^2}
=-2\pi n_a
\quad\mbox{for }x\in \contour{C}_a.
\]
Note that the term $G(1/x)$ has precisely cancelled out
and $p(x)$ is no longer symmetric under $x\mapsto 1/x$.
This is related to the fact that spacetime is now
a group manifold. In this case, the image under inversion
is given by a different quasi-momentum, here $\tilde p_1$.
Also the length $L$ now becomes a true global charge
next to $J$. This is related to the left and right symmetry of
group manifolds.

\paragraph{Bosonic, Non-Compact.}

We turn on only $G=\hat G_{14}$ and consider
the single quasi-momentum $p=\hat p_1=-\hat p_4$.
This reduces to the case of strings on $AdS_3\times S^1$
investigated in \cite{Kazakov:2004nh}
\[\label{eq:Integral.Rank1AdS3S1}
2\sheetsl(x)=
-2\resolvHsl(x)
+2\hat F(x)
=
-2\resolvsl(x)
-\frac{2G(0)/x^2}{1-1/x^2}
+\frac{4\pi L}{\sqrt{\lambda}}\,\frac{1/x}{1-1/x^2}
=
-2\pi n_a\quad\mbox{for }x\in \contour{C}_a.
\]
%

\paragraph{Fermionic.}

We turn on only $G=G^\ast_{24}$ and consider
the quasi-momenta $\tilde p=\tilde p_2$,
$\hat p=\hat p_4$. The two quasi-momenta are given by
$\tilde p(x)=H(x)+\tilde F(x)+F^\ast(x)$ and
$\hat p(x)=H(x)-\hat F(x)+F^\ast(x)$
\<\label{eq:Integral.Rank1FermiSheets}
\tilde p(x)\eq
G(x)+\lrbrk{\frac{2\pi\,(B+L)}{\sqrt{\lambda}}+G'(0)}
\frac{1/x}{1-1/x^2}\,,
\\\nn
\hat p(x)\eq
G(x)
+\frac{2\pi\,(B-L)}{\sqrt{\lambda}}\,
\frac{1/x}{1-1/x^2}
+\frac{G(0)/x^2}{1-1/x^2}\,.
\>
The relevant combination for the integral equation
is the difference of sheets
$\tilde p(x)-\hat p(x)=\tilde F(x)+\hat F(x)$
\[\label{eq:Integral.Rank1FermiDifference}
\sheetsl[\tilde](x)-\sheetsl[\hat](x)=
-\frac{G(0)/x^2}{1-1/x^2}
+\frac{G'(0)/x}{1-1/x^2}
+\frac{4\pi L}{\sqrt{\lambda}}\,\frac{1/x}{1-1/x^2}
=-2\pi n_a\quad\mbox{for }x=x^\ast_a.
\]
This agrees precisely with the expression derived
from the near-plane-wave limit in \cite{Staudacher:2004tk}.

\subsection{Superstrings on $AdS_3\times S^3$}
\label{sec:Integral.AdS3S3}

The above three subsectors can be combined into one larger sector.
Let us consider only the following four sheets
$p_1=\hat p_1$,
$p_2=\tilde p_2$,
$p_3=\tilde p_3$,
$p_4=\hat p_4$ and corresponding resolvents
$\hat G_{41},\tilde G_{23},G^\ast_{21},G^\ast_{31},G^\ast_{24},G^\ast_{34}$
so that again there is no apparent inversion symmetry.
By inspection of the quasi-momenta there are only three
independent combinations of resolvents appearing:
\<\label{eq:Integral.AdS3S3SimpleRes}
G_1\eq -G^\ast_{21}-G^\ast_{31}\phantom{\mathord{}+G^\ast_{24}}-\hat G_{41},
\nln
G\indup{mom}=G_2\eq +\tilde G_{23}-G^\ast_{31}+G^\ast_{24}-\hat G_{41},
\nln
G_3\eq +G^\ast_{34}\phantom{\mathord{}-G^\ast_{31}}+G^\ast_{24}-\hat G_{41}.
\>
The quasi-momenta then read
\<\label{eq:Integral.AdS3S3Sheets}
p_1(x)\eq
\phantom{G_2(x)}
-G_{1}(x)
+\frac{1/x}{1-1/x^2}
\lrbrk{\frac{2\pi (B+L)}{\sqrt{\lambda}}+G'_2(0)-G'_1(0)}
-\frac{G_{1}(0)/x^2}{1-1/x^2}\,,
\nln
p_2(x)\eq
G_{2}(x)
-G_{1}(x)
+\frac{1/x}{1-1/x^2}
\lrbrk{\frac{2\pi (B+L)}{\sqrt{\lambda}}+G'_2(0)-G'_1(0)}
-\frac{G_{1}(0)/x^2}{1-1/x^2}\,,
\nln
p_3(x)\eq
G_{3}(x)
-G_{2}(x)
+\frac{1/x}{1-1/x^2}
\lrbrk{\frac{2\pi (B-L)}{\sqrt{\lambda}}+G'_{3}(0)-G'_{2}(0)}
+\frac{G_{3}(0)/x^2}{1-1/x^2}
\,,
\nln
p_4(x)\eq
G_{3}(x)
\phantom{\mathord{}-\mathord{G_2(x)}}
+\frac{1/x}{1-1/x^2}
\lrbrk{\frac{2\pi (B-L)}{\sqrt{\lambda}}+G'_{3}(0)-G'_{2}(0)}
+\frac{G_{3}(0)/x^2}{1-1/x^2}\,.
\nln\earel{}
\>
The differences of adjacent sheets which appear in the
integral equations are given by
\<\label{eq:Integral.AdS3S3Differences}
p_1(x)-p_2(x)\eq
-G_2(x)
,
\nln
p_2(x)-p_3(x)\eq
+2G_2(x)
+\frac{2G'_2(0)/x}{1-1/x^2}
+\frac{4\pi L}{\sqrt{\lambda}}\,\frac{1/x}{1-1/x^2}
\nl
-\tilde G_1(x)
-\frac{G'_1(0)/x}{1-1/x^2}
-\frac{G_1(0)/x^2}{1-1/x^2}
\nl
-G_3(x)
-\frac{G'_3(0)/x}{1-1/x^2}
-\frac{G_3(0)/x^2}{1-1/x^2}\,,
\nln
p_3(x)-p_4(x)\eq
-G_2(x)
.
\>
Differences of non-adjacent sheets are obtained by summing up the equations.

For purely bosonic solutions on $AdS_3\times S^3$ we set
$G_1(x)=G_3(x)$ and
\<\label{eq:Integral.AdS3S3BosRes}
\tilde{G}(x)\eq
\tilde{G}_{23}(x)=
\int_{\tilde{\contour{C}}}
\frac{dy\,\tilde{\rho} (y)}{1-1/y^2}\,\frac{1}{y-x}
=G_2(x)-G_1(x),
\nln
\hat{G}(x)\eq
\hat{G}_{14}(x)=
\int_{\hat{\contour{C}}}
\frac{dy\,\hat{\rho} (y)}{1-1/y^2}\,\frac{1}{y-x}
=G_1(x)=G_3(x).
\>
The densities satisfy the following set of equations on the
respective cuts:
\<\label{eq:Integral.AdS3S3BosBethe}
2\resolvsl[\tilde](x)+F(x)
\eq -2\pi \tilde n_{a}
\quad\mbox{for }
x\in \tilde{\contour{C}}_{a},
\nln
2\resolvsl[\hat](x)-F(x)
\eq
-2\pi \hat n_{a}
\quad\mbox{for }
x\in \hat{\contour{C}}_{a},
\>
where the potential $F(x)$ is given by
\[\label{eq:Integral.AdS3S3BosPotential}
F(x)=\left(\frac{4\pi L}{\sqrt{\lambda}}+2\tilde{G}'(0)\right)\frac{1/x}{1-1/x^2}
-\frac{2\hat{G}(0)/x^2}{1-1/x^2}\,.
\]
The momentum constraint
\eqref{eq:Integral.Momentum}
and energy shift
\eqref{eq:Integral.EnergySimp}
are contained in the combination
\[\label{eq:Integral.AdS3S3BosMom}
G\indup{mom}(x)=\tilde G(x)+\hat G(x).
\]

\subsection{Comparison to Gauge Theory}
\label{eq:Integral.Compare}

Let us briefly comment on the comparison to $\superN=4$ gauge theory.
An in-depth comparison of the spectral curves 
can be found in \cite{Beisert:2005di}.
The complete one-loop dilatation generator
has been derived in \cite{Beisert:2003jj,Beisert:2004ry}.
It is integrable and one can use a Bethe ansatz to
find its energy (scaling dimension) eigenvalues \cite{Beisert:2003jj}.
In the thermodynamic limit \cite{Sutherland:1995aa,Beisert:2003xu},
which should be related to string theory \cite{Frolov:2003qc},
the Bethe equations have been written in \cite{Beisert:2004ry}.
Their form does not resemble the equations
\eqref{eq:Integral.SheetsSimp,eq:Integral.Potentials,eq:Integral.Bethe}
very much, but rather the one
in the previous \secref{sec:Integral.AdS3S3}.
We shall refrain from transforming the equations here and refer
the reader to~\appref{sec:Beauty}.
The resulting equations \eqref{eq:Beauty.IntegralBethe}
\[\label{eq:Integral.BeautyBethe}
\sum_{j'=1}^7 M_{j,j'}\resolvHsl_{j'}(x)+F_j(x)
=-2\pi n_{j,a}\,
\quad\mbox{for }x\in \contour{C}_{j,a}
\]
can be seen to agree with the equations in
\cite{Beisert:2004ry}.
Also the expressions for the momentum constraint and energy shift
as well as the local and global charges agree.
Note that in the Frolov-Tseytlin limit,
see \secref{sec:Moduli.Limit},
the potential $F_j$ reduces to a term proportional to $V_j L/u$,
c.f.~\cite{Kazakov:2004qf,Beisert:2004ag} for similar results.
We have thus proven the agreement of the spectra of one-loop
planar gauge theory and classical string theory.

\section{Conclusions and Outlook}
\label{sec:Concl}

We solve the problem of describing all classical solutions of the
superstring sigma-model in $AdS_5\times S^5$ in terms of their spectral curves. 
Let us underline the importance of dealing with the whole supersymmetric 
string theory on the $AdS_5\times S^5$ space, 
including the fermionic degrees of freedom, 
for its quantization. 
For the classical string we can drop the fermions and the 
two bosonic sectors, $AdS_5$ and $S^5$, appear to be completely 
factorized (up to the constraints on the fixed poles and total momentum).
Conversely, the quantum corrections at higher powers of 
$\hbar\sim 1/\sqrt{\lambda}$\,, make the two sectors interact nontrivially,
an effect which is already seen in the super spin chain for the 
one-loop approximation to gauge theory. 
It is also clear that the direct quantization of
sigma models in closed subsectors, like $\Real\times S^5$, does not make much
sense since those models are even not conformal. It seems that it is better to
attack directly the full supersymmetric quantum theory 
on $AdS_5\times S^5$. 
Our paper shows that, at least at the classical level, 
the full string theory has no principal
difficulties comparing to the simpler subset sigma models.

The curves are solutions of a Riemann-Hilbert problem and as
such can be encoded in the set of singular integral equations
which we have derived. We hope that this classical result
will be a useful starting point for a quantization.
Some indications that the integrable
structures persist in the quantum regime are found in
\cite{Arutyunov:2004vx,Beisert:2004jw,Swanson:2004qa,Arutyunov:2004yx,Berkovits:2004xu,Beisert:2005mq,Hernandez:2005nf}.
There are several benefits of the formulation
in terms of algebraic curves which might facilitate quantization:
For one, the formulation is completely gauge independent,
at no point one is required to fix a gauge;
especially we can preserve full kappa symmetry \cite{Bena:2003wd}.
Moreover, the curve consists only of action variables.
Due to the Heisenberg principle this is all we can
ask for to know in the quantum theory.
Finally, there are no unphysical degrees of freedom associated to
the curve. These would usually contain spurious infinities and
their absence should make the curve completely finite.
Our integral equations can be interpreted as the classical limit of the yet
to be found discrete Bethe equations, which describe the exact quantum
spectrum of the string, c.f.~the ansatz by Arutyunov, Frolov 
and Staudacher \cite{Arutyunov:2004vx} and
a corresponding `string chain' \cite{Beisert:2004jw} for some initial steps 
in this direction.
The existence of such equations is an assumption, 
but since many quantum integrable systems are solvable
by a Bethe ansatz, in particular some sigma models
\cite{Polyakov:1983tt,Polyakov:1984et,Faddeev:1985qu,Ogievetsky:1987vv},
this assumption does not look inconceivable. 
We believe that in any
event integrability will be an important ingredient in solving
string theory in $AdS_5\times S^5$, be it a Bethe ansatz or some other
method, and hope that our findings will be helpful in attacking this
challenging problem.

We should mention that many classical string solutions 
(those without self-symmetric cuts)
admit a regular expansion in the 't~Hooft coupling.%
\footnote{Typically, this is an expansion in inverse powers 
of some large conserved charge. In our framework a useful 
conserved charge is the length $L$, 
see also \cite{Mikhailov:2004au,Mikhailov:2005wn}.
The length is related to angular momentum $J$ on $S^5$ 
in many cases and thus gives a natural generalization of 
the Frolov-Tseytlin proposal \cite{Frolov:2003qc,Tseytlin:2003ii}.} 
We have compared the energy spectrum of the classical string with
the spectrum of anomalous dimensions in $\superN=4$ SYM which
at one loop is given by a Bethe ansatz. 
Our comparison is based on Bethe equations
but it can also be performed at at the level of spectral curves.
Although we should not expect agreement beyond third order
in the effective coupling \cite{Serban:2004jf}, 
we might use the present result as a source of inspiration 
for higher-loop gauge theory.

The method of solving the spectral problem in terms of algebraic curves 
which we employ for this sigma model is usually called the
finite gap method (see the book \cite{Novikov:1984id} for a good
introduction). This means that we look for a finite genus algebraic
curve characterizing some solution. It was first proposed in
\cite{Its:1975aa,Dubrovin:1976xx} for the KdV system and later generalized to 
KP equations in \cite{Krichever:1980aa}. 
In principle, for any given algebraic curve one can
construct explicitly the corresponding solution of the
equations of motion in terms of Riemann theta functions.
The dependence on time
enters linearly in the argument of the Riemann theta function 
and the frequencies are given by 
the periods of certain Abelian differentials.
Note that in our investigation 
we have the angle variables, these
enter as the initial phase for the arguments of the theta function.
Moreover, to construct the solution 
for the $AdS_5\times S^5$ coset model, 
one would first have to fix a gauge for the local symmetries.
Finally, only up to genus one the theta functions can be 
expressed in terms of conventional algebraic and elliptic functions. 
Beyond that they are known only as integrals or series and
therefore less efficient.

\subsection*{Acknowledgements}

It is a pleasure to thank A.~Agarwal, G.~Arutyunov, C.~Callan, L.~Dolan, A.~Gorsky, 
A.~Mikhailov, J.~Minahan, R.~Plesser, S.~G.~Rajeev, L.~Rastelli, R.~Roiban, 
A.~Ryzhov, F.~Smirnov, D.~Serban, A.~Sorin, M.~Staudacher A.~Tseytlin, H.~Verlinde 
and M.~Wijnholt for discussions.
N.~B.~would like to thank the Kavli Institute for Theoretical Physics, 
the Ecole Normale Sup\'erieure
and the Albert-Einstein-Institut for the kind hospitality during
parts of this work.  K.~Z. would like to thank the Kavli Institute
for Theoretical Physics for hospitality. The work of N.~B.~is
supported in part by the U.S.~National Science Foundation Grants
No.~PHY99-07949 and PHY02-43680. Any opinions, findings and
conclusions or recommendations expressed in this material are those
of the authors and do not necessarily reflect the views of the
National Science Foundation. V.~K.~would like to thank the Princeton
Institute for Advanced Study for the kind hospitality during a part
of this work. The work of V.~K.~was partially supported by European
Union under the RTN contracts HPRN-CT-2000-00122 and 00131 and by
NATO grant PST.CLG.978817. The work of K.~S.~is supported by the
Nishina Memorial Foundation. The work of K.Z. was supported in part
by the Swedish Research Council under contracts 621-2002-3920 and
621-2004-3178, by the G\"oran Gustafsson Foundation, and by the RFBR
grant 02-02-17260.

\appendix






















\section{Supermatrices}
\label{sec:SuperAlgebra}

We shall write supermatrices as matrices
where horizontal/vertical bars
separate between rows/columns with
even and odd grading. We shall only consider
bosonic supermatrices.
The grading matrix $\sgrad$ consequently is given by
\[\label{eq:SuperAlgebra.Grading}
\sgrad=\smatr{+I}{0}{0}{-I}.
\]
For example, it can be used to define the supertrace
of a supermatrix
as a regular trace of a supermatrix
\[\label{eq:SuperAlgebra.STr}
\str A=\tr \sgrad\,A
\]
The supertrace is cyclic
\[\label{eq:SuperAlgebra.STrCyclicity}
\str A \,B=\str B \,A.
\]
The superdeterminant is defined as
\[\label{eq:SuperAlgebra.SDet}
\sdet \smatr{A}{B}{C}{D}=\frac{\det(A-BD^{-1}C)}{\det D}
=\frac{\det A}{\det (D-CA^{-1}B)},
\]
it obeys
\[\label{eq:SuperAlgebra.SDetProd}
\sdet(AB)=\sdet A\sdet B
\]
and is compatible with the identity
\[\label{eq:SuperAlgebra.SDetExpSTr}
\sdet\exp A=\exp\str A.
\]
%

\paragraph{Supertranspose.}

The supertranspose is defined as
\[\label{eq:SuperAlgebra.ST}
\smatr{A}{B}{C}{D}^\strans=
\smatr{A^\trans}{C^\trans}{-B^\trans}{D^\trans}.
\]
Like the common transpose, it inverts the order
within a product of matrices
\[\label{eq:SuperAlgebra.STProd}
(A\,B)^\strans= B^\strans\,A^\strans.
\]
and does not change supertraces and superdeterminants
\[\label{eq:SuperAlgebra.STSTrSDet}
\str A^\strans = \str A,
\qquad
\sdet A^\strans = \sdet A.
\]
Unlike the common transpose, it
is not an involution,
but a $\Integers_4$-operation due to the identity
\[\label{eq:SuperAlgebra.STST}
(A^\strans)^\strans=\sgrad\, A\, \sgrad
\]
Furthermore, a supersymmetric or a superantisymmetric matrix
require a slightly modified definition
\[\label{eq:SuperAlgebra.STSym}
A=+\sgrad \,A^\strans,\qquad
A=-\sgrad \,A^\strans.
\]
%

\paragraph{$(1|1)\times(1|1)$ Supermatrices.}

Let us collect some formulas for $(1|1)\times (1|1)$ supermatrices
\[\label{eq:SuperAlgebra.2x2}
A=\smatr{a}{b}{c}{d}
\]
The inverse is given by
\[\label{eq:SuperAlgebra.2x2Inv}
A^{-1}=\smatr{\frac{1}{a}+\frac{bc}{a^2d}}{-\frac{b}{ad}}{-\frac{c}{ad}}{\frac{1}{d}-\frac{bc}{ad^2}}.
\]
The supertrace and superdeterminant read
\[\label{eq:SuperAlgebra.2x2STrSDet}
\str A=a-d,\qquad
\sdet A=\frac{a}{d}-\frac{bc}{d^2}\,.
\]
It can be diagonalized by the matrices
\[\label{eq:SuperAlgebra.2x2Diag}
T=\smatr{1-\frac{bc}{2(a-d)^2}}{+\frac{b}{a-d}}
        {-\frac{c}{a-d}}{1+\frac{bc}{2(a-d)^2}},\qquad
T^{-1}=\smatr{1-\frac{bc}{2(a-d)^2}}{-\frac{b}{a-d}}
             {+\frac{c}{a-d}}{1+\frac{bc}{2(a-d)^2}},
\]
such that
\[\label{eq:SuperAlgebra.2x2Eigen}
TAT^{-1}
=\smatr{\alpha_1}{0}{0}{\alpha_2},
\qquad \alpha_1=a+\frac{bc}{a-d}\,,\quad
\alpha_2=d+\frac{bc}{a-d}\,.
\]
The eigenvalues satisfy the sum and product rules
\[\label{eq:SuperAlgebra.2x2EigenSTrSDet}
\alpha_1-\alpha_2=\str A,\qquad \frac{\alpha_1}{\alpha_2}=\sdet A
\]
as well as the characteristic equation
\[\label{eq:SuperAlgebra.2x2EigenChar}
\sdet(A-\alpha_1)=0,\qquad \sdet(A-\alpha_2)=\infty.
\]
Clearly $\alpha_1$ and $\alpha_2$ are associated to different gradings.
For $\str A=0$, i.e.~$a=d$, the eigenvalues degenerate
and some problems arise as in the case of bosonic matrices.

\section{Cosets and Vectors}
\label{sec:Vector}

In this appendix we shall explain the relationship
between the vector model used in \cite{Beisert:2004ag}
and the coset model used in this paper.

We will start with the coset model.
The physically relevant cosets $S^5=\grp{SU}(4)/\grp{Sp}(2)$
and $AdS^5=\grp{SU}(2,2)/\grp{Sp}(1,1)$
require several $i$'s at various places.
These can be avoided by considering
the coset $\grp{SL}(4,\Real)/\grp{Sp}(4,\Real)$.
Its algebraic structure is precisely the same but
the formulas are slightly easier to handle.
For the breaking of $\grp{SL(4,\Real)}$
we can use a fixed $4\times 4$ antisymmetric matrix $E$, say
\[\label{eq:Vector.Metric}
E=\matr{cc}{0&+I\\-I&0},
\]
where each block corresponds to a $2\times 2$ matrix
and $I$ is the identity.
The currents of the coset model in the moving frame
are
\< \label{eq:Vector.MovingCurrents}
J\eq -g^{-1}dg,\nln H\eq\half J-\half EJ^\trans E^{-1},\nln
K\eq\half J+\half EJ^\trans E^{-1}. \>
In the fixed frame, which is related
to the moving frame by $j=gJg^{-1}$, etc.,
they are given by
\< \label{eq:Vector.FixedCurrents}
j\eq -dg\,g^{-1},\nln
h\eq-\half dg\,g^{-1}+\half gEdg^\trans g^{-\trans} E^{-1}g^{-1},\nln
k\eq-\half dg\,g^{-1}-\half gEdg^\trans g^{-\trans} E^{-1}g^{-1}.
\>
We can now rewrite $k$ as
\[\label{eq:Vector.CurrentRewrite}
k=-\half dg\,g^{-1}-\half gEdg^\trans (g E g^\trans)^{-1} =-\half
d(g E g^\trans)\,(g E g^\trans)^{-1} .
\]
and see that it can be rewritten as
\[\label{eq:Vector.CurrentX}
k=-\half dX\,X^{-1}
\qquad \mbox{with}\qquad
X=g E g^\trans.
\]

We would now like to interpret $X$ as the fundamental field of the theory.
For all $g\in \grp{Sp}(4,\Real)$ we find $X=E$,
thus $X$ parametrizes the coset $\grp{SL}(4,\Real)/\grp{Sp}(4,\Real)$.
Note that we can define a norm for $X$ by
\[\varepsilon_{\alpha\beta\gamma\delta}X^{\alpha\beta}X^{\gamma\delta}
=\varepsilon_{\alpha\beta\gamma\delta}E^{\alpha\beta}E^{\gamma\delta}\,
\det g =-8.
\]
Starting from a generic matrix $X$, the conditions $X=-X^\trans$ and
$\varepsilon_{\alpha\beta\gamma\delta}X^{\alpha\beta}X^{\gamma\delta}=-8$
leave $5$ degrees of freedom for $X$ and therefore such an $X$ indeed
parametrizes the coset $\grp{SL}(4,\Real)/\grp{Sp}(4,\Real)$ which
has $15-10=5$ dimensions.
We now parametrize $X$ as
\[\label{eq:Vector.Xvector}
X=\vec\sigma\cdot \vec X,\qquad
\vec X^2=-1.
\]
where $\vec X$ is a vector of $\grp{SO}(3,3)$
and $\vec\sigma$ is a chiral component of the
Clifford algebra.
This reveals the connection between the sigma model
\eqref{eq:Vector.CurrentX} and the vector model \eqref{eq:Vector.Xvector},
they are merely reparametrizations of the same model.
More explicitly the matrix $X$ is given through the
components of the vector $\vec X$ by
\[\label{eq:Vector.Xmatrix}
X= \left(\begin{array}{cccc}
       0&+X_1+X_4&+X_2+X_5&+X_3+X_6\\
-X_1-X_4&       0&+X_3-X_6&-X_2+X_5\\
-X_2-X_5&-X_3+X_6&       0&+X_1-X_4\\
-X_3-X_6&+X_2-X_5&-X_1+X_4&       0
\end{array}\right)
\]
such that
\[\label{eq:Vector.Xconstraint}
\sfrac{1}{8}\varepsilon_{\alpha\beta\gamma\delta}X^{\alpha\beta}X^{\gamma\delta}=
\vec X^2=
{X_1}^2+{X_2}^2+{X_3}^2-{X_4}^2-{X_5}^2-{X_6}^2.
\]
The corresponding expressions for $\grp{SO}(6)$ are
%
\[\label{eq:Vector.XmatrixSU4}
X= \left(\begin{array}{cccc}
        0&+X_1+iX_2&+X_3+iX_4&+X_5+iX_6\\
-X_1-iX_2&        0&+X_5-iX_6&-X_3+iX_4\\
-X_3-iX_4&-X_5+iX_6&        0&+X_1-iX_2\\
-X_5-iX_6&+X_3-iX_4&-X_1+iX_2&        0
\end{array}\right)
\]
where again
$\sfrac{1}{8}\varepsilon_{\alpha\beta\gamma\delta}X^{\alpha\beta}X^{\gamma\delta}=\vec X^2$,
but with a positive signature of the norm.
For $\grp{SO}(2,4)$ we find
\[\label{eq:Vector.XmatrixSU22}
X= \left(\begin{array}{cccc}
        0&+X_5+iX_0&+X_1+iX_2& X_3+iX_4\\
-X_5-iX_0&        0&+X_3-iX_4&-X_1+iX_2\\
-X_1-iX_2&-X_3+iX_4&        0&-X_5+iX_0\\
-X_3-iX_4&+X_1-iX_2&+X_5-iX_0&        0
\end{array}\right)
\]
where $\vec{X}^2$ has signature $+----+$.
These expressions only differ from the
expressions for $\grp{SO}(3,3)$ by
relabelling the $X_k$ and multiplying some of them by $i$.

\section{A Local Charge}
\label{sec:Local}

Here we compute the first local charge
as outlined in \secref{sec:Super.Local}. 
Our starting point is 
\eqref{eq:Super.LocalLaxLeadFinal}
where we have already diagonalized the
leading order $\bar A_{-2}$ of the Lax connection 
using some matrix $T_0$,
\[
\bar A_{-2}=\half T_0(P_+-\Lambda_\sigma)T_0^{-1}=\matr{cc}{\alpha I&0\\0&\beta I}.
\]
Let us see how this matrix can be used to block-diagonalize 
\[
\bar A_{-1}=\bar X+\bigcomm{T_1T_0^{-1}}{\bar A_{-2}}
\qquad\mbox{with}\qquad
\bar X=T_0^{-1}Q_{1,\sigma}T_0=\matr{cc}{u&v\\x&y}.
\]
The key insight is that the double commutator 
\[
\bigcomm{\bar A_{-2}}{\comm{\bar A_{-2}}{\bar X}}
=
(\alpha-\beta)^2\matr{cc}{0&v\\x&0}
\]
can be used to extract the off-diagonal elements. 
We can thus cancel them in $\bar A_{-1}$ by setting
\[
T_1=\frac{1}{(\alpha-\beta)^2}\,\comm{\bar A_{-2}}{\bar X}T_0
\]
and obtain 
\[
\bar A_{-1}=\bar X-\frac{1}{(\alpha-\beta)^2}\bigcomm{\bar A_{-2}}{\comm{\bar A_{-2}}{\bar X}}
=\matr{cc}{u&0\\0&y}.
\]
We can continue and block-diagonalize $\bar A_r$ order by order in
this fashion. 
Note that $A_r$ is block-diagonal if and only if 
\[
\bigcomm{\bar A_{-2}}{\bar A_r}=0.
\]
Together with the identity for any matrix $\bar Y$
\[
\bigcomm{\bar A_{-2}}{\bigcomm{\bar A_{-2}}{\comm{\bar A_{-2}}{\bar Y}}}
=(\alpha-\beta)^2\comm{\bar A_{-2}}{\bar Y}
\]
one can construct the higher order transformation matrices quite conveniently. 
The local charges are defined via the trace of only 
one block $a_r$ of $\bar A_r$. 
Again this can be achieved using the matrix $\bar A_{-2}$ as follows
\[
\frac{1}{\alpha-\beta}\,\str \bar A_{-2} \bar A_r
=
\frac{1}{\alpha-\beta}\,\bigbrk{\alpha\str a_r+\beta \str b_r}
=
\str a_r.
\]
Here it is important that $\str \bar A_r=\str a_r-\str b_r=0$.

Finally, we would like to express the local charges in terms of 
the physical currents $P,Q_{1,2}$.
Note that all the expressions occurring in 
the conjugated 
$T_0^{-1}\bar A_r T_0$ are commutators of the currents,
e.g.
\[
T_0^{-1}\bar A_{-1}T_0=Q_{1,\sigma}
-\Delta_+^{-2}\bigcomm{P_+}{\comm{P_+}{Q_{1,\sigma}}}.
\]
where $\Delta_+=2(\alpha-\beta)$ is the difference of eigenvalues of $P_+$.
The only exception is a term related to the diagonalization 
using $T_0$, i.e.~$T^{-1}_0\partial_\sigma T_0$.
Within the trace they can be eliminated by making use of
\[
\bigcomm{\bar A_{-2}}{\partial_\sigma \bar A_{-2}}=0
\]
which is equivalent to the statement that 
$\partial_\sigma \bar A_{-2}$ is block-diagonal.
This leads to 
\[
\bigcomm{P_+ }{\comm{P_+}{T_0^{-1}\partial_\sigma T_0}}
=\bigcomm{P_+}{\partial_\sigma P_+}
\]
and is sufficient to 
write every instance of $T_0^{-1}\partial_\sigma T_0$
within $\str a_r$ in terms of 
$\partial_\sigma P_+$.
Putting everything together 
we find 
\<
\str a_2\eq
\half \Delta_+^{-1}\str P_+P_-
+\Delta_+^{-5}\str \comm{P_+}{D_\sigma P_+}\comm{P_+}{D_\sigma P_+}
\nl
-6\Delta_+^{-5}\str \bigcomm{\comm{P_+}{Q_{1,\sigma}}}{Q_{1,\sigma}}\comm{P_+}{D_\sigma P_+}
\nl
+2\Delta_+^{-3}\str \comm{P_+}{Q_{1,\sigma}}D_\sigma Q_{1,\sigma}
\nl
-2\Delta_+^{-3}\str \comm{P_+}{Q_{1,\sigma}}\comm{P_+}{Q_{2,\sigma}}
\nl
-\Delta_+^{-5}\str \bigcomm{\comm{P_+}{Q_{1,\sigma}}}{Q_{1,\sigma}}
\bigcomm{\comm{P_+}{Q_{1,\sigma}}}{Q_{1,\sigma}}
\nl
-5\Delta_+^{-7}\str \bigcomm{\bigcomm{\comm{P_+}{Q_{1,\sigma}}}{P_+}}{Q_{1,\sigma}}
\bigcomm{\bigcomm{\comm{P_+}{Q_{1,\sigma}}}{P_+}}{Q_{1,\sigma}}.
\>
Here $D_\sigma X=\partial_\sigma X-\comm{H_\sigma}{X}$
is the world-sheet covariant derivative.
The conserved charge corresponding to
$\str a_2$ is the integral
\[
q_2^+=-i\int_{0}^{2\pi}d\sigma\, \str a_{2}.
\]
Furthermore there exists a world-sheet parity conjugate charge
$q_2^-$ from the expansion around $z=\infty$ instead
of $z=0$. It is obtained from $q_2^+$ with the 
replacements $P_{\pm}\to -P_{\mp}$ and $Q_{1,2}\to Q_{2,1}$.




\section{Sleeping Beauty}
\label{sec:Beauty}

\begin{figure}\centering
\includegraphics{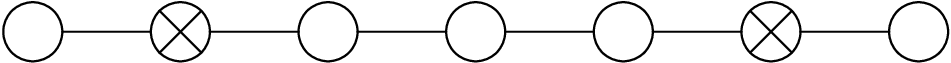}
\caption{`Beauty' Dynkin diagram of $\alg{su}(2,2|4)$ \protect\cite{Beisert:2003yb}.}
\label{fig:Beauty}
\end{figure}

This appendix contains lengthy expressions related
to the complete superalgebra
using the `Beauty' form of $\alg{psu}(2,2|4)$ \cite{Beisert:2003yb},
c.f.~\figref{fig:Beauty}.
In this form, the grading of the sheets corresponding
to the fundamental representation reads
\[\label{eq:Beauty.Grading}
\sgrad_k=(-1,-1,+1,+1,+1,+1,-1,-1).
\]
The sheets of the quasi-momentum are arranged as follows
\[\label{eq:Beauty.Sheets}
p_{1,2,7,8}=\hat p_{1,2,3,4},\qquad
p_{3,4,5,6}=\tilde p_{1,2,3,4}.
\]
%

\subsection{Global Charges}
\label{sec:Beauty.Charges}

The global fillings are defined as
\[\label{eq:Beauty.FillingsSigma}
K_j=
\sum_{a=1}^{A}
\frac{\sqrt{\lambda}}{8\pi^2 i}\oint_{\contour{C}_a} dx\lrbrk{1-\frac{1}{x^2}}
\sum_{k=1}^j \sgrad_kp_k(x).
\]
The global filling $K_j$ essentially measures the total filling of all
$(k,l)$-cuts with $k\leq j< l$.
The global fillings are directly related to the Dynkin labels
$[r_1;r_2;r_3,r_4,r_5;r_6;r_7]$ of a solution.
The Dynkin labels are obtained through the residues at infinity
\[\label{eq:Beauty.DynkinSigma}
\bar\sgrad_j r_j=
\frac{\sqrt{\lambda}}{8\pi^2 i}\oint_\infty dx\, \bigbrk{p_{j}(x)-p_{j+1}(x)},
\]
where $\bar\sgrad_j=[-1;+1;+1,+1,+1;+1;-1]$ are
conventional factors for the definition of the Dynkin labels.
These are given by
\<\label{eq:Beauty.DynkinFilling}
r_1\eq K_2-2K_1,
\nln
r_2\eq K_3-K_1+\half \delta E,
\nln
r_3\eq K_2+K_4-2K_3,
\nln
r_4\eq L-2K_4+K_3+K_5,
\nln
r_5\eq K_4+K_6-2K_5,
\nln
r_6\eq K_5-K_7+\half \delta E,
\nln
r_7\eq K_6-2K_7.
\>
or for short
\[\label{eq:Beauty.DynkinFillingShort}
\bar\sgrad_j r_j=V_j\, L+\bar V_j\,\delta E-M_{j,j'}K_{j'}
\]
The labels $\bar V_j=[0;\half;0,0,0;\half;0]$ indicate the
change of (fermionic) Dynkin labels induced by the energy shift.
Note that the Dynkin labels obey the central charge constraint
\[\label{eq:Beauty.DynkinCentral}
-r_1+2r_2+r_3=r_5+2r_6-r_7.
\]
The inverse relation is given by
\[\label{eq:Beauty.FillingDynkin}
\arraycolsep1.2pt
\begin{array}{rccrcrcrcrcrcrcrcrcrcrcr}
K_1&=&-&\half L&+&\half B &-&\sfrac{1}{4}r^\ast&-&\sfrac{3}{4} r_1&+&\half r_2&+&\half r_3&+&\half r_4&+&\half r_5&+&\half r_6&-&\sfrac{1}{4} r_7&-&\half \delta E,
\\[0.7ex]
K_2&=&-&L&+&B&-&\half r^\ast& -&\half r_1&+&r_2&+&r_3&+&r_4&+&r_5&+&r_6&-&\sfrac{1}{2} r_7&-&\delta E,
\\[0.7ex]
K_3&=&-&\half L&+&\half B&-&\sfrac{1}{4}r^\ast& -& \half r_1 &+& r_2& +&\sfrac{1}{4}r_3 &+&\half r_4&+&\sfrac{3}{4}r_5&+&r_6&-& \half r_7&-&\delta E,
\\[0.7ex]
K_4&=&&&&&&& -& \half r_1 &+& r_2 &+& \half r_3&&& +& \half r_5&+&r_6&-&\half r_7 &-&\delta E,
\\[0.7ex]
K_5&=&-&\half L&-&\half B&+&\sfrac{1}{4}r^\ast&-&\half r_1&+&r_2&+&\sfrac{3}{4}r_3&+&\half r_4&+&\sfrac{1}{4} r_5&+&r_6&-&\sfrac{1}{2} r_7& -& \delta E,
\\[0.7ex]
K_6&=&-&L&-&B &+&\half r^\ast& -&\half r_1&+&r_2&+&r_3&+&r_4&+&r_5&+&r_6&-&\half r_7&-&\delta E ,
\\[0.7ex]
K_7&=&-&\half L&-&\half B &+&\sfrac{1}{4}r^\ast&-&\sfrac{1}{4} r_1&+&\half r_2&+&\half r_3&+&\half r_4&+&\half r_5&+&\half r_6&-&\sfrac{3}{4} r_7&-&\half\delta E.
\\[0.7ex]
\end{array}
\]
The constant $B$ represents the hypercharge of the vacuum.

\subsection{Integral Representation}
\label{sec:Beauty.Integral}

We present the reduction of a full set of resolvents
into seven simple ones $G_j$ for the $AdS_5\times S^5$ superstring.
The easiest way to reduce the expressions is
to drop all but the resolvents between adjacent sheets.
When the remaining resolvents are replaced by the suitably defined
simple resolvents $G_j$
\<\label{eq:Beauty.IntegralSimple}
G_1\eq -\hat G_{21}+\ldots,
\nln
G_2\eq -G^\ast_{12}+\ldots,
\nln
G_3\eq +\tilde G_{12}+\ldots,
\nln
G\indup{mom}=G_4\eq +\tilde G_{23}+\ldots,
\nln
G_5\eq +\tilde G_{34}+\ldots,
\nln
G_6\eq +G^\ast_{43}+\ldots,
\nln
G_7\eq -\hat G_{43}+\ldots,
\>
the original expressions are recovered. The quasi-momenta
in terms of simple resolvents read
\<\label{eq:Beauty.IntegralSheets}
p_1(x)\eq
\mathord{\phantom{H_{1}(u)}}
-H_{1}(x)
+G_{2}(1/x)
-\frac{G_{2}(0)}{1-1/x^2}
+\frac{(c_1+d_1)/x}{1-1/x^2}\,,
\nln
p_2(x)\eq
H_{1}(x)
-H_{2}(x)
+G_{2}(1/x)
-\frac{G_{2}(0)}{1-1/x^2}
+\frac{(c_1+d_1)/x}{1-1/x^2}\,,
\nln
p_3(x)\eq
H_{3}(x)
-H_{2}(x)
+G_{2}(1/x)
-\frac{G_{2}(0)}{1-1/x^2}
+\frac{(c_1+d_1)/x}{1-1/x^2}
-G_{4}(1/x)
+G_{4}(0),
\nln
p_4(x)\eq
H_{4}(x)
-H_{3}(x)
+G_{2}(1/x)
-\frac{G_{2}(0)}{1-1/x^2}
+\frac{(c_1+d_1)/x}{1-1/x^2}
-G_{4}(1/x)
+G_{4}(0),
\nln
p_5(x)\eq
H_{5}(x)
-H_{4}(x)
-G_{6}(1/x)
+\frac{G_{6}(0)}{1-1/x^2}
+\frac{(c_1-d_1)/x}{1-1/x^2}
+G_{4}(1/x)
-G_{4}(0),
\nln
p_6(x)\eq
H_{6}(x)
-H_{5}(x)
-G_{6}(1/x)
+\frac{G_{6}(0)}{1-1/x^2}
+\frac{(c_1-d_1)/x}{1-1/x^2}
+G_{4}(1/x)
-G_{4}(0),
\nln
p_7(x)\eq
H_{6}(x)
-H_{7}(x)
-G_{6}(1/x)
+\frac{G_{6}(0)}{1-1/x^2}
+\frac{(c_1-d_1)/x}{1-1/x^2}\,,
\nln
p_8(x)\eq
H_{7}(x)
\phantom{\mathord{}-H_{7}(u)}
-G_{6}(1/x)
+\frac{G_{6}(0)}{1-1/x^2}
+\frac{(c_1-d_1)/x}{1-1/x^2}
\>
with
$c_1=2\pi B/\sqrt{\lambda}+\half G'_6(0)-\half G'_2(0)$
and
$d_1=2\pi L/\sqrt{\lambda}+G'_{4}(0)-\half G'_{2}(0)-\half G'_{6}(0)$.
The integral equations are given by
\[\label{eq:Beauty.IntegralBethe}
\sheetsl{}_{j+1}(x)-\sheetsl{}_{j}(x)=
-\sum_{j'=1}^7 M_{j,j'}\resolvHsl_{j'}(x)-F_j(x)
=2\pi n_{j,a}\,
\quad\mbox{for }x\in \contour{C}_{j,a}
\]
with $M_{j,j'}$ the Cartan matrix.
Here, the non-zero potentials $F_j(x)$ read
\<\label{eq:Beauty.IntegralPotentials}
F_2(x)=F_6(x)\eq G_{4}(1/x)-G_{4}(0),
\nln
F_4(x)\eq
-2G_{4}(1/x)+2G_{4}(0)
+\frac{2G'_{4}(0)/x}{1-1/x^2}
\nl
+G_{2}(1/x)
-\frac{G_{2}(0)}{1-1/x^2}
-\frac{G'_{2}(0)/x}{1-1/x^2}
\nl
+G_{6}(1/x)
-\frac{G_{6}(0)}{1-1/x^2}
-\frac{G'_{6}(0)/x}{1-1/x^2}
\nl
+\frac{4\pi L}{\sqrt{\lambda}}\frac{1/x}{1-1/x^2}\,.
\>

\bibliography{bksz}

\begin{thebibliography}{10}
\ifx\href\asklfhas\newcommand{\href}[2]{#2}\fi
\raggedright
\small
\parskip 0pt

\bibitem{Maldacena:1998re}
J.~M.~Maldacena,
\textit{``The large N limit of superconformal field theories and
  supergravity''},
\textsf{Adv.~Theor.~Math.~Phys.~2,~231~(1998)},
\href{http://arXiv.org/abs/hep-th/9711200}{\texttt{hep-th/9711200}}.
%
\bibitem{Gubser:1998bc}
S.~S.~Gubser, I.~R.~Klebanov and A.~M.~Polyakov,
\textit{``Gauge theory correlators from non-critical string theory''},
\textsf{Phys.~Lett.~B428,~105~(1998)},
\href{http://arXiv.org/abs/hep-th/9802109}{\texttt{hep-th/9802109}}.
%
\bibitem{Witten:1998qj}
E.~Witten,
\textit{``Anti-de Sitter space and holography''},
\textsf{Adv.~Theor.~Math.~Phys.~2,~253~(1998)},
\href{http://arXiv.org/abs/hep-th/9802150}{\texttt{hep-th/9802150}}.
%
\bibitem{Metsaev:1998it}
R.~R.~Metsaev and A.~A.~Tseytlin,
\textit{``Type IIB superstring action in $AdS_5\times S^5$ background''},
\textsf{Nucl.~Phys.~B533,~109~(1998)},
\href{http://arXiv.org/abs/hep-th/9805028}{\texttt{hep-th/9805028}}.
%
\bibitem{Blau:2001ne}
M.~Blau, J.~Figueroa-O'Farrill, C.~Hull and G.~Papadopoulos,
\textit{``A new maximally supersymmetric background of IIB superstring
  theory''},
\textsf{JHEP~0201,~047~(2002)},
\href{http://arXiv.org/abs/hep-th/0110242}{\texttt{hep-th/0110242}}.
%
\bibitem{Blau:2002dy}
M.~Blau, J.~Figueroa-O'Farrill, C.~Hull and G.~Papadopoulos,
\textit{``Penrose limits and maximal supersymmetry''},
\textsf{Class.~Quant.~Grav.~19,~L87~(2002)},
\href{http://arXiv.org/abs/hep-th/0201081}{\texttt{hep-th/0201081}}.
%
\bibitem{Berenstein:2002jq}
D.~Berenstein, J.~M.~Maldacena and H.~Nastase,
\textit{``Strings in flat space and pp waves from {$\mathcal{N}=\mathord{}$4}
  {Super} {Yang Mills}''},
\textsf{JHEP~0204,~013~(2002)},
\href{http://arXiv.org/abs/hep-th/0202021}{\texttt{hep-th/0202021}}.
%
\bibitem{Metsaev:2001bj}
R.~R.~Metsaev,
\textit{``Type IIB Green-Schwarz superstring in plane wave Ramond- Ramond
  background''},
\textsf{Nucl.~Phys.~B625,~70~(2002)},
\href{http://arXiv.org/abs/hep-th/0112044}{\texttt{hep-th/0112044}}.
%
\bibitem{Metsaev:2002re}
R.~R.~Metsaev and A.~A.~Tseytlin,
\textit{``Exactly solvable model of superstring in plane wave Ramond- Ramond
  background''},
\textsf{Phys.~Rev.~D65,~126004~(2002)},
\href{http://arXiv.org/abs/hep-th/0202109}{\texttt{hep-th/0202109}}.
%
\bibitem{Callan:2003xr}
C.~G.~Callan,~Jr., H.~K.~Lee, T.~McLoughlin, J.~H.~Schwarz, I.~Swanson and
  X.~Wu,
\textit{``Quantizing string theory in $AdS_5\times S^5$: Beyond the pp-wave''},
\textsf{Nucl.~Phys.~B673,~3~(2003)},
\href{http://arXiv.org/abs/hep-th/0307032}{\texttt{hep-th/0307032}}.
%
\bibitem{Parnachev:2002kk}
A.~Parnachev and A.~V.~Ryzhov,
\textit{``Strings in the near plane wave background and AdS/CFT''},
\textsf{JHEP~0210,~066~(2002)},
\href{http://arXiv.org/abs/hep-th/0208010}{\texttt{hep-th/0208010}}.
%
\bibitem{Callan:2004uv}
C.~G.~Callan,~Jr., T.~McLoughlin and I.~Swanson,
\textit{``Holography beyond the Penrose limit''},
\textsf{Nucl.~Phys.~B694,~115~(2004)},
\href{http://arXiv.org/abs/hep-th/0404007}{\texttt{hep-th/0404007}}.
%
\bibitem{Callan:2004ev}
C.~G.~Callan,~Jr., T.~McLoughlin and I.~Swanson,
\textit{``Higher impurity AdS/CFT correspondence in the near-BMN limit''},
\textsf{Nucl.~Phys.~B700,~271~(2004)},
\href{http://arXiv.org/abs/hep-th/0405153}{\texttt{hep-th/0405153}}.
%
\bibitem{McLoughlin:2004dh}
T.~McLoughlin and I.~Swanson,
\textit{``N-impurity superstring spectra near the pp-wave limit''},
\textsf{Nucl.~Phys.~B702,~86~(2004)},
\href{http://arXiv.org/abs/hep-th/0407240}{\texttt{hep-th/0407240}}.
%
\bibitem{Gubser:2002tv}
S.~S.~Gubser, I.~R.~Klebanov and A.~M.~Polyakov,
\textit{``A semi-classical limit of the gauge/string correspondence''},
\textsf{Nucl.~Phys.~B636,~99~(2002)},
\href{http://arXiv.org/abs/hep-th/0204051}{\texttt{hep-th/0204051}}.
%
\bibitem{Frolov:2002av}
S.~Frolov and A.~A.~Tseytlin,
\textit{``Semiclassical quantization of rotating superstring in {$AdS_5 \times
  S^5$}''},
\textsf{JHEP~0206,~007~(2002)},
\href{http://arXiv.org/abs/hep-th/0204226}{\texttt{hep-th/0204226}}.
%
\bibitem{Russo:2002sr}
J.~G.~Russo,
\textit{``Anomalous dimensions in gauge theories from rotating strings in
  {$AdS_5 \times S^5$}''},
\textsf{JHEP~0206,~038~(2002)},
\href{http://arXiv.org/abs/hep-th/0205244}{\texttt{hep-th/0205244}}.
%
\bibitem{Minahan:2002rc}
J.~A.~Minahan,
\textit{``Circular semiclassical string solutions on {$AdS_5\times S^5$}''},
\textsf{Nucl.~Phys.~B648,~203~(2003)},
\href{http://arXiv.org/abs/hep-th/0209047}{\texttt{hep-th/0209047}}.
%
\bibitem{Tseytlin:2002ny}
A.~A.~Tseytlin,
\textit{``Semiclassical quantization of superstrings: {$AdS_5\times S^5$} and
  beyond''},
\textsf{Int.~J.~Mod.~Phys.~A18,~981~(2003)},
\href{http://arXiv.org/abs/hep-th/0209116}{\texttt{hep-th/0209116}}.
%
\bibitem{Frolov:2003qc}
S.~Frolov and A.~A.~Tseytlin,
\textit{``Multi-spin string solutions in {$AdS_5\times S^5$}''},
\textsf{Nucl.~Phys.~B668,~77~(2003)},
\href{http://arXiv.org/abs/hep-th/0304255}{\texttt{hep-th/0304255}}.
%
\bibitem{Beisert:2003xu}
N.~Beisert, J.~A.~Minahan, M.~Staudacher and K.~Zarembo,
\textit{``Stringing Spins and Spinning Strings''},
\textsf{JHEP~0309,~010~(2003)},
\href{http://arXiv.org/abs/hep-th/0306139}{\texttt{hep-th/0306139}}.
%
\bibitem{Beisert:2003ea}
N.~Beisert, S.~Frolov, M.~Staudacher and A.~A.~Tseytlin,
\textit{``Precision Spectroscopy of AdS/CFT''},
\textsf{JHEP~0310,~037~(2003)},
\href{http://arXiv.org/abs/hep-th/0308117}{\texttt{hep-th/0308117}}.
%
\bibitem{Serban:2004jf}
D.~Serban and M.~Staudacher,
\textit{``Planar {$\mathcal{N}=\mathord{}$4} gauge theory and the Inozemtsev
  long range spin chain''},
\textsf{JHEP~0406,~001~(2004)},
\href{http://arXiv.org/abs/hep-th/0401057}{\texttt{hep-th/0401057}}.
%
\bibitem{Tseytlin:2003ii}
A.~A.~Tseytlin,
\textit{``Spinning strings and AdS/CFT duality''},
\href{http://arXiv.org/abs/hep-th/0311139}{\texttt{hep-th/0311139}},
in: \textit{``From Fields to Stings: Circumnavigating Theoretical Physics''},
Ian Kogan Memorial Volume,
ed.: M.~Shifman, A.~Vainshtein and J.~Wheater,
World Scientific (2005),
Singapore.
%
\bibitem{Tseytlin:2004cj}
A.~A.~Tseytlin,
\textit{``Semiclassical strings in $AdS_5\times S^5$ and scalar operators in
  {$\mathcal{N}=\mathord{}$4} SYM theory''},
\textsf{Comptes~Rendus~Physique~5,~1049~(2004)},
\href{http://arXiv.org/abs/hep-th/0407218}{\texttt{hep-th/0407218}}.
%
\bibitem{Beisert:2004ry}
N.~Beisert,
\textit{``The Dilatation Operator of {$\mathcal{N}=\mathord{}$4} Super
  Yang-Mills Theory and Integrability''},
\textsf{Phys.~Rept.~405,~1~(2005)},
\href{http://arXiv.org/abs/hep-th/0407277}{\texttt{hep-th/0407277}}.
%
\bibitem{Beisert:2004yq}
N.~Beisert,
\textit{``Higher-loop integrability in {$\mathcal{N}=\mathord{}$4} gauge
  theory''},
\textsf{Comptes~Rendus~Physique~5,~1039~(2004)},
\href{http://arXiv.org/abs/hep-th/0409147}{\texttt{hep-th/0409147}}.
%
\bibitem{Zarembo:2004hp}
K.~Zarembo,
\textit{``Semiclassical Bethe ansatz and AdS/CFT''},
\textsf{Comptes~Rendus~Physique~5,~1081~(2004)},
\href{http://arXiv.org/abs/hep-th/0411191}{\texttt{hep-th/0411191}}.
%
\bibitem{Kazakov:2004qf}
V.~A.~Kazakov, A.~Marshakov, J.~A.~Minahan and K.~Zarembo,
\textit{``Classical/quantum integrability in AdS/CFT''},
\textsf{JHEP~0405,~024~(2004)},
\href{http://arXiv.org/abs/hep-th/0402207}{\texttt{hep-th/0402207}}.
%
\bibitem{Arutyunov:2003uj}
G.~Arutyunov, S.~Frolov, J.~Russo and A.~A.~Tseytlin,
\textit{``Spinning strings in $AdS_5\times S^5$ and integrable systems''},
\textsf{Nucl.~Phys.~B671,~3~(2003)},
\href{http://arXiv.org/abs/hep-th/0307191}{\texttt{hep-th/0307191}}.
%
\bibitem{Arutyunov:2003za}
G.~Arutyunov, J.~Russo and A.~A.~Tseytlin,
\textit{``Spinning strings in {$AdS_5\times S^5$}: New integrable system
  relations''},
\textsf{Phys.~Rev.~D69,~086009~(2004)},
\href{http://arXiv.org/abs/hep-th/0311004}{\texttt{hep-th/0311004}}.
%
\bibitem{Zakharov:1973pp}
V.~E.~Zakharov and A.~V.~Mikhailov,
\textit{``Relativistically invariant two-dimensional models in field theory
  integrable by the inverse problem technique''},
\textsf{Sov.~Phys.~JETP~47,~1017~(1978)},
in russian.
%
\bibitem{Pohlmeyer:1975nb}
K.~Pohlmeyer,
\textit{``Integrable Hamiltonian systems and interactions through quadratic
  constraints''},
\textsf{Commun.~Math.~Phys.~46,~207~(1976)}.
%
\bibitem{Luscher:1977rq}
M.~L{\"u}scher and K.~Pohlmeyer,
\textit{``Scattering of massless lumps and nonlocal charges in the
  two-dimensional classical nonlinear sigma model''},
\textsf{Nucl.~Phys.~B137,~46~(1978)}.
%
\bibitem{Brezin:1979am}
E.~Brezin, C.~Itzykson, J.~Zinn-Justin and J.~B.~Zuber,
\textit{``Remarks about the existence of nonlocal charges in two-dimensional
  models''},
\textsf{Phys.~Lett.~B82,~442~(1979)}.
%
\bibitem{Eichenherr:1981sk}
H.~Eichenherr and M.~Forger,
\textit{``Higher local conservation laws for nonlinear sigma models on
  symmetric spaces''},
\textsf{Commun.~Math.~Phys.~82,~227~(1981)}.
%
\bibitem{Minahan:2002ve}
J.~A.~Minahan and K.~Zarembo,
\textit{``The Bethe-ansatz for {$\mathcal{N}=\mathord{}$4} super Yang-Mills''},
\textsf{JHEP~0303,~013~(2003)},
\href{http://arXiv.org/abs/hep-th/0212208}{\texttt{hep-th/0212208}}.
%
\bibitem{Beisert:2003yb}
N.~Beisert and M.~Staudacher,
\textit{``The {$\mathcal{N}=\mathord{}$4} SYM Integrable Super Spin Chain''},
\textsf{Nucl.~Phys.~B670,~439~(2003)},
\href{http://arXiv.org/abs/hep-th/0307042}{\texttt{hep-th/0307042}}.
%
\bibitem{Beisert:2003tq}
N.~Beisert, C.~Kristjansen and M.~Staudacher,
\textit{``The dilatation operator of {$\mathcal{N}=\mathord{}$4} conformal
  super Yang-Mills theory''},
\textsf{Nucl.~Phys.~B664,~131~(2003)},
\href{http://arXiv.org/abs/hep-th/0303060}{\texttt{hep-th/0303060}}.
%
\bibitem{Beisert:2003ys}
N.~Beisert,
\textit{``The su(2$/$3) dynamic spin chain''},
\textsf{Nucl.~Phys.~B682,~487~(2004)},
\href{http://arXiv.org/abs/hep-th/0310252}{\texttt{hep-th/0310252}}.
%
\bibitem{Staudacher:2004tk}
M.~Staudacher,
\textit{``The factorized S-matrix of CFT/AdS''},
\textsf{JHEP~0505,~054~(2005)},
\href{http://arXiv.org/abs/hep-th/0412188}{\texttt{hep-th/0412188}}.
%
\bibitem{Beisert:2004hm}
N.~Beisert, V.~Dippel and M.~Staudacher,
\textit{``A Novel Long Range Spin Chain and Planar {$\mathcal{N}=\mathord{}$4}
  Super Yang-Mills''},
\textsf{JHEP~0407,~075~(2004)},
\href{http://arXiv.org/abs/hep-th/0405001}{\texttt{hep-th/0405001}}.
%
\bibitem{Sutherland:1995aa}
B.~Sutherland,
\textit{``Low-Lying Eigenstates of the One-Dimensional Heisenberg Ferromagnet
  for any Magnetization and Momentum''},
\textsf{Phys.~Rev.~Lett.~74,~816~(1995)}.
%
\bibitem{Kruczenski:2003gt}
M.~Kruczenski,
\textit{``Spin chains and string theory''},
\textsf{Phys.~Rev.~Lett.~93,~161602~(2004)},
\href{http://arXiv.org/abs/hep-th/0311203}{\texttt{hep-th/0311203}}.
%
\bibitem{Kruczenski:2004kw}
M.~Kruczenski, A.~V.~Ryzhov and A.~A.~Tseytlin,
\textit{``Large spin limit of $AdS_5\times S^5$ string theory and low energy
  expansion of ferromagnetic spin chains''},
\textsf{Nucl.~Phys.~B692,~3~(2004)},
\href{http://arXiv.org/abs/hep-th/0403120}{\texttt{hep-th/0403120}}.
%
\bibitem{Kazakov:2004nh}
V.~A.~Kazakov and K.~Zarembo,
\textit{``Classical/quantum integrability in non-compact sector of AdS/CFT''},
\textsf{JHEP~0410,~060~(2004)},
\href{http://arXiv.org/abs/hep-th/0410105}{\texttt{hep-th/0410105}}.
%
\bibitem{Beisert:2004ag}
N.~Beisert, V.~A.~Kazakov and K.~Sakai,
\textit{``Algebraic curve for the SO(6) sector of AdS/CFT''},
\href{http://arXiv.org/abs/hep-th/0410253}{\texttt{hep-th/0410253}},
to appear in Comm.~Math.~Phys..
%
\bibitem{Schafer-Nameki:2004ik}
S.~Sch{\"a}fer-Nameki,
\textit{``The algebraic curve of 1-loop planar {$\mathcal{N}=\mathord{}$4}
  SYM''},
\textsf{Nucl.~Phys.~B714,~3~(2005)},
\href{http://arXiv.org/abs/hep-th/0412254}{\texttt{hep-th/0412254}}.
%
\bibitem{Arutyunov:2004yx}
G.~Arutyunov and S.~Frolov,
\textit{``Integrable Hamiltonian for classical strings on $AdS_5\times S^5$''},
\textsf{JHEP~0502,~059~(2005)},
\href{http://arXiv.org/abs/hep-th/0411089}{\texttt{hep-th/0411089}}.
%
\bibitem{Mikhailov:2004xw}
A.~Mikhailov,
\textit{``Supersymmetric null-surfaces''},
\textsf{JHEP~0409,~068~(2004)},
\href{http://arXiv.org/abs/hep-th/0404173}{\texttt{hep-th/0404173}}.
%
\bibitem{Hernandez:2004kr}
R.~Hern\'andez and E.~L\'opez,
\textit{``Spin chain sigma models with fermions''},
\textsf{JHEP~0411,~079~(2004)},
\href{http://arXiv.org/abs/hep-th/0410022}{\texttt{hep-th/0410022}}.
%
\bibitem{Bena:2003wd}
I.~Bena, J.~Polchinski and R.~Roiban,
\textit{``Hidden symmetries of the {$AdS_5\times S^5$} superstring''},
\textsf{Phys.~Rev.~D69,~046002~(2004)},
\href{http://arXiv.org/abs/hep-th/0305116}{\texttt{hep-th/0305116}}.
%
\bibitem{Beisert:2003jj}
N.~Beisert,
\textit{``The Complete One-Loop Dilatation Operator of
  {$\mathcal{N}=\mathord{}$4} Super Yang-Mills Theory''},
\textsf{Nucl.~Phys.~B676,~3~(2004)},
\href{http://arXiv.org/abs/hep-th/0307015}{\texttt{hep-th/0307015}}.
%
\bibitem{Beisert:2005di}
N.~Beisert, V.~A.~Kazakov, K.~Sakai and K.~Zarembo,
\textit{``Complete Spectrum of Long Operators in {$\mathcal{N}=\mathord{}$4}
  SYM at One Loop''},
\textsf{JHEP~0507,~030~(2005)},
\href{http://arXiv.org/abs/hep-th/0503200}{\texttt{hep-th/0503200}}.
%
\bibitem{Hatsuda:2004it}
M.~Hatsuda and K.~Yoshida,
\textit{``Classical integrability and super Yangian of superstring on
  $AdS_5\times S^5$''},
\textsf{Int.~J.~Mod.~Phys.~A19,~4715~(2004)},
\href{http://arXiv.org/abs/hep-th/0407044}{\texttt{hep-th/0407044}}.
%
\bibitem{Das:2004hy}
A.~Das, J.~Maharana, A.~Melikyan and M.~Sato,
\textit{``The algebra of transition matrices for the $AdS_5\times S^5$
  superstring''},
\textsf{JHEP~0412,~055~(2004)},
\href{http://arXiv.org/abs/hep-th/0411200}{\texttt{hep-th/0411200}}.
%
\bibitem{Kallosh:1998zx}
R.~Kallosh, J.~Rahmfeld and A.~Rajaraman,
\textit{``Near horizon superspace''},
\textsf{JHEP~9809,~002~(1998)},
\href{http://arXiv.org/abs/hep-th/9805217}{\texttt{hep-th/9805217}}.
%
\bibitem{Berkovits:1999zq}
N.~Berkovits, M.~Bershadsky, T.~Hauer, S.~Zhukov and B.~Zwiebach,
\textit{``Superstring theory on $AdS_2\times S^2$ as a coset supermanifold''},
\textsf{Nucl.~Phys.~B567,~61~(2000)},
\href{http://arXiv.org/abs/hep-th/9907200}{\texttt{hep-th/9907200}}.
%
\bibitem{Roiban:2000yy}
R.~Roiban and W.~Siegel,
\textit{``Superstrings on $AdS_5\times S^5$ supertwistor space''},
\textsf{JHEP~0011,~024~(2000)},
\href{http://arXiv.org/abs/hep-th/0010104}{\texttt{hep-th/0010104}}.
%
\bibitem{Berkovits:2004jw}
N.~Berkovits,
\textit{``BRST cohomology and nonlocal conserved charges''},
\textsf{JHEP~0502,~060~(2005)},
\href{http://arXiv.org/abs/hep-th/0409159}{\texttt{hep-th/0409159}}.
%
\bibitem{Berkovits:2004xu}
N.~Berkovits,
\textit{``Quantum consistency of the superstring in $AdS_5\times S^5$
  background''},
\textsf{JHEP~0503,~041~(2005)},
\href{http://arXiv.org/abs/hep-th/0411170}{\texttt{hep-th/0411170}}.
%
\bibitem{Mikhailov:2004au}
A.~Mikhailov,
\textit{``Notes on fast moving strings''},
\href{http://arXiv.org/abs/hep-th/0409040}{\texttt{hep-th/0409040}}.
%
\bibitem{Mikhailov:2005wn}
A.~Mikhailov,
\textit{``Plane wave limit of local conserved charges''},
\href{http://arXiv.org/abs/hep-th/0502097}{\texttt{hep-th/0502097}}.
%
\bibitem{Berkovits:1999im}
N.~Berkovits, C.~Vafa and E.~Witten,
\textit{``Conformal field theory of AdS background with Ramond-Ramond flux''},
\textsf{JHEP~9903,~018~(1999)},
\href{http://arXiv.org/abs/hep-th/9902098}{\texttt{hep-th/9902098}}.
%
\bibitem{Dolan:1999dc}
L.~Dolan and E.~Witten,
\textit{``Vertex operators for $AdS_3$ background with Ramond-Ramond flux''},
\textsf{JHEP~9911,~003~(1999)},
\href{http://arXiv.org/abs/hep-th/9910205}{\texttt{hep-th/9910205}}.
%
\bibitem{Metsaev:2000mv}
R.~R.~Metsaev and A.~A.~Tseytlin,
\textit{``Superparticle and superstring in $AdS_3\times S^3$ Ramond-Ramond
  background in light-cone gauge''},
\textsf{J.~Math.~Phys.~42,~2987~(2001)},
\href{http://arXiv.org/abs/hep-th/0011191}{\texttt{hep-th/0011191}}.
%
\bibitem{Arutyunov:2003rg}
G.~Arutyunov and M.~Staudacher,
\textit{``Matching Higher Conserved Charges for Strings and Spins''},
\textsf{JHEP~0403,~004~(2004)},
\href{http://arXiv.org/abs/hep-th/0310182}{\texttt{hep-th/0310182}}.
%
\bibitem{Arutyunov:2004xy}
G.~Arutyunov and M.~Staudacher,
\textit{``Two-loop commuting charges and the string/gauge duality''},
\href{http://arXiv.org/abs/hep-th/0403077}{\texttt{hep-th/0403077}},
in: \textit{``Lie Theory and its Applications in Physics V''},
Proceedings of the Fifth International Workshop, Varna, Bulgaria, 16-22 June
  2003,
ed.: H.-D.~Doebner and V.~K.~Dobrev,
World Scientific (2004),
Singapore.
%
\bibitem{Arutyunov:2004vx}
G.~Arutyunov, S.~Frolov and M.~Staudacher,
\textit{``Bethe ansatz for quantum strings''},
\textsf{JHEP~0410,~016~(2004)},
\href{http://arXiv.org/abs/hep-th/0406256}{\texttt{hep-th/0406256}}.
%
\bibitem{Beisert:2004jw}
N.~Beisert,
\textit{``Spin chain for quantum strings''},
\textsf{Fortsch.~Phys.~53,~852~(2005)},
\href{http://arXiv.org/abs/hep-th/0409054}{\texttt{hep-th/0409054}}.
%
\bibitem{Swanson:2004qa}
I.~Swanson,
\textit{``Quantum string integrability and AdS/CFT''},
\textsf{Nucl.~Phys.~B709,~443~(2005)},
\href{http://arXiv.org/abs/hep-th/0410282}{\texttt{hep-th/0410282}}.
%
\bibitem{Beisert:2005mq}
N.~Beisert, A.~A.~Tseytlin and K.~Zarembo,
\textit{``Matching quantum strings to quantum spins: one-loop vs.~finite-size
  corrections''},
\textsf{Nucl.~Phys.~B715,~190~(2005)},
\href{http://arXiv.org/abs/hep-th/0502173}{\texttt{hep-th/0502173}}.
%
\bibitem{Hernandez:2005nf}
R.~Hern\'andez, E.~L\'opez, A.~Peri\'a\~nez and G.~Sierra,
\textit{``Finite size effects in ferromagnetic spin chains and quantum
  corrections to classical strings''},
\textsf{JHEP~0506,~011~(2005)},
\href{http://arXiv.org/abs/hep-th/0502188}{\texttt{hep-th/0502188}}.
%
\bibitem{Polyakov:1983tt}
A.~M.~Polyakov and P.~B.~Wiegmann,
\textit{``Theory of nonabelian Goldstone bosons in two dimensions''},
\textsf{Phys.~Lett.~B131,~121~(1983)}.
%
\bibitem{Polyakov:1984et}
A.~M.~Polyakov and P.~B.~Wiegmann,
\textit{``Goldstone fields in two-dimensions with multivalued actions''},
\textsf{Phys.~Lett.~B141,~223~(1984)}.
%
\bibitem{Faddeev:1985qu}
L.~D.~Faddeev and N.~Y.~Reshetikhin,
\textit{``Integrability of the principal chiral field model in
  $(1+1)$-dimension''},
\textsf{Ann.~Phys.~167,~227~(1986)}.
%
\bibitem{Ogievetsky:1987vv}
E.~Ogievetsky, P.~Wiegmann and N.~Reshetikhin,
\textit{``The principal chiral field in two-dimensions on classical Lie
  algebras: The Bethe ansatz solution and factorized theory of scattering''},
\textsf{Nucl.~Phys.~B280,~45~(1987)}.
%
\bibitem{Novikov:1984id}
S.~Novikov, S.~V.~Manakov, L.~P.~Pitaevsky and V.~E.~Zakharov,
\textit{``Theory of Solitons. The Inverse Scattering Method''},
Consultants Bureau (1984),
New York, USA,
276p,
Contemporary Soviet Mathematics.
%
\bibitem{Its:1975aa}
A.~R.~Its and V.~B.~Matveev,
\textit{``Schr{\"o}dinger operators with finite-gap spectrum and N-soliton
  solutions of the Korteweg-de Vries equation''},
\textsf{Theor.~Math.~Phys.~23,~343~(1975)}.
%
\bibitem{Dubrovin:1976xx}
B.~A.~Dubrovin, M.~V.~B. and S.~P.~Novikov,
\textit{``Non-linear equations of Korteweg-de Vries type, finite zone linear
  operators, and Abelian varieties''},
\textsf{Russ.~Math.~Surveys~31,~59~(1976)}.
%
\bibitem{Krichever:1980aa}
I.~M.~Krichever,
\textit{``Elliptic solutions of KP equations and integrable systems of
  particles''},
\textsf{Funk.~Anal.~App.~14,~282~(1980)}.
%
\end{thebibliography}
\bibliographystyle{nb}

\end{document}